\documentclass{eccomas}

\usepackage{graphicx}
\usepackage{subfigure}
\usepackage{longtable}
\usepackage{fancyhdr}
\usepackage{multirow,bigstrut, booktabs}
\usepackage[english]{babel}
\usepackage[latin1]{inputenc}
\usepackage[T1]{fontenc}
\usepackage{indentfirst}
\usepackage{amsmath,amsfonts,amssymb,amsthm}
\usepackage{wasysym}
\usepackage{latexsym}
\usepackage{appendix}
\usepackage{eqnarray}
\usepackage{colortbl}
\usepackage{dsfont}
\usepackage{xtab}
\usepackage{pifont}
\usepackage[strict]{chngpage}
\usepackage{float}
\usepackage{rotating}
\usepackage[centerlast,small,bf]{caption}
\usepackage{multicol}
\usepackage{transparent}
\usepackage{epsfig}
\usepackage{epstopdf}
\usepackage{longtable}
\usepackage[hyperindex]{hyperref}
\usepackage[final]{listings}
\usepackage{cite}

\usepackage{geometry}
\usepackage{array}
\newcolumntype{C}[1]{p{#1}}
\usepackage{icomma}

\title{A second-generation URANS model (STRUCT$-\epsilon$) applied to a Generic Side Mirror and its Impact on Sound Generation}
\author{J. Munoz-Paniagua$^{1,\dag}$, J. Garc\'ia$^1$, E. Latorre-Iglesias$^2$}
\heading{-}

\address{
$^1$ Dpto. Ingenier\'ia Energ\'etica, Universidad Polit\'ecnica de Madrid\\
C/ Jos\'e Guti\'errez Abascal 2, 28006, Madrid, Spain\\
e-mail: le.munoz@upm.es\\

$^2$ Dpto. Ingenier\'ia Audiovisual y Comunicaciones, Universidad Polit\'ecnica de Madrid\\
C/ Nikolas Tesla, 28031, Madrid, Spain}

\abstract{A generic side mirror can be approximated to the combination of a half cylinder topped with a quarter of sphere. The flow structure in the wake of the side mirror is highly transient and the turbulence plays an important role affecting aeroacoustics through pressure fluctuation. Thus, this geometry is one of the test cases object of several numerical studies in recent years to assess the aerodynamic and aeroacoustic capabilities of the turbulence models. In this context, this study presents how the second-generation URANS close STRUCT$-\epsilon$ is able to properly predict the expected stagnation, flow separation and vortex shedding phenomena. Besides, the predictive accuracy for the noise generation mechanism is evaluated by comparing the spectra of the sound pressure level measured at several static pressure sensors with the numerical results obtained with the STRUCT$-\epsilon$. The response of this turbulence model has overachieved other hybrid methods and is in good agreement with the results from Large-Eddy Simulations or the experiments. To conclude the paper, the applicability of STRUCT$-\epsilon$ to construct a Spectral Proper Orthogonal Decomposition method that helps identifying the most energetic modes to appropriately capture the dominant flow structures is also introduced.}

\begin{document}
%\maketitle

\newpage

\section{Introduction}

As electric cars become more and more popular, wind noise reduction results into an issue of great importance. Indeed, when the engine noise is removed from a car, all the other noise sources gain a more relevant role, \cite{tec-otto-n-et-al-1999-electric-cars-noise-sources}. Exposed components such as side mirrors or underfloor details generate flow structures which are the primary sources of noise generation around the vehicle above approximately $120\,\mathrm{km\,h^{-1}}$, \cite{pap-ask-j-and-davidson-l-2009-les-aeroacoustics-generic-side-mirror}. The flow structure in the wake of the side mirror is highly transient and will generate strong pressure fluctuation on the door panels and windows. This unsteady pressure fluctuation ultimately propagates into the carriage and exterior as noise, \cite{pap-lokhande-b-et-al-2003-les-generic-side-mirror}. Therefore, it is evident the interest to properly simulate the flow around the side mirror with accurate, robust and computationally affordable turbulence models for a potential shape optimization. While the contribution of the side mirror to total aerodynamic drag for modern passenger cars amounts up to $5\%$, \cite{ibo-gilhaus-a-and-hoffmann-r-1998-contribution-side-mirror-drag-vehicle}, its relative position to the vehicle A-pillars makes it a significant contributor of generated aerodynamic noise, and this feature has been the objective function in the optimization proposed in \cite{cnf-grahs-t-and-othmer-c-2006-sas-and-des-generic-side-mirror-optimization} or \cite{pap-papoutsiskiachagias-e-et-al-2015-optimization-side-mirror}.

A generic side mirror (GSM) can be approximated to the combination of a half cylinder topped with a quarter of sphere. Even though such a simple geometry, it is well known that the flow around a cylinder is complex, and turbulence plays an important role affecting aeroacoustics through pressure fluctuation. Among other flow phenomena, it is expected flow separation, Kelvin-Helmholtz instabilities and vortex shedding. Besides, the drag crisis observed around the critical Reynolds number for both cylinders and sphere is expected as well in the GSM, associated to the laminar-to-turbulent boundary layer transition. These features highlight the challenge of an accurate prediction in both the near and far downstream regions.

Experiments of the flow around the GSM run by Daimler-Chrysler are presented in \cite{cnf-hold-r-et-al-1999-experiments-side-mirror-acoustics-basic-analysis} and \cite{cnf-siegert-r-et-al-1999-experiments-generic-side-mirror} for Reynolds number Re$_D = 7.2\times 10^5$, and for Re$_D = 5.2\times 10^5$ in \cite{cnf-rung-th-et-al-2002-comparison-urans-des-side-mirror}, where $D$ is the diameter of the cylinder of the GSM. Similar experiments have also been conducted and reported in \cite{pap-kato-c-et-al-2007-generic-side-mirror-reynolds-effect-yaw-angle} for Re$_D = 1.4-2.4\times 10^5$. Based on these works, the GSM became one of the test cases object of several numerical studies in recent years, focusing on the turbulence of the flow around the body. Different turbulence models have been considered in this benchmark case, from incompressible Large-Eddy Simulation (LES) in \cite{pap-lokhande-b-et-al-2003-les-generic-side-mirror,pap-ask-j-and-davidson-l-2009-les-aeroacoustics-generic-side-mirror} and compressible LES in \cite{pap-yao-hd-and-davidson-l-2018-generic-side-mirror-interior-cavity-noise} to a wide variety of hybrid RANS/LES methods. Among the latter, we can highlight the (SST-)Delayed Detached-Eddy Simulation (DDES) in \cite{pap-yu-l-et-al-2021-fast-transient-fractional-step-method-scheme-generic-side-mirror}, the (SST-)Improved Delayed Detached-Eddy Simulation (IDDES) in \cite{cnf-tosh-a-et-al-2018-des-generic-side-mirror}, the eXtra Large Eddy Simulation (XLES) of \cite{pap-capizzano-f-et-al-2019-xles-generic-side-mirror} and the recently developed Stress-Blended Eddy Simulation (SBES) in \cite{pap-chode-k-et-al-2021-sbes-generic-side-mirror}. Apart from them, the second-generation RANS model Scale-Adaptive Simulation (SAS), \cite{pap-egorov-y-et-al-2010-sas-generic-side-mirror} has also been tested in this case. A second-generation URANS turbulence model, named STRUCT$-\epsilon$ model \cite{phd-xu-l-2020-struct}, is considered in this paper as an alternative to the aforementioned approaches. The model adopts as its baseline URANS a $k-\epsilon$ anisotropic non-linear eddy-viscosity model (NLEVM) with a cubic stress-strain relation while enabling controlled scale-resolution inside flow regions where the scale separation assumption of URANS is not satisfied, \cite{pap-lenci-g-and-baglietto-e-2021-struct}. This turbulence model has been validated in train aerodynamic studies in \cite{pap-garcia-j-et-al-2020-struct-freight-trains}. As it is well known that unsteady vortex shedding is expected in the case considered in this paper, which is better resolved using hybrid RANS/LES models as opposed to RANS, \cite{pap-robertson-e-et-al-2015-cfd-aeroacustical-spectral-analysis-spherocylinder}, it would be interesting to compare the performance of STRUCT$-\epsilon$ with other LES or hybrid models for both aerodynamic and aeroacoustic applications. 

The assessment against both experimental and numerical data available in literature demonstrates promising results, in terms of accuracy and computational cost, \cite{pap-garcia-j-et-al-2020-struct-freight-trains}, making the model suitable for aerodynamic optimization studies. Particularly, the use of optimization methods which may use an extensive Design of Experiments (DoE) like Genetic Algorithms (GA) \cite{pap-munozpaniagua-j-and-garcia-j-2020-ga-optimization-hst}, Particle Swarm Optimization (PSO), \cite{pap-goncalves-w-j-and-margnat-f-2020-shape-optimization-pso-noise-bluff-bodies}, or fractional factorial design \cite{pap-beigmoradi-s-and-vahdati-m-2019-fractional-factorial-design-hatchback-rear-vehicle-aerodynamic-optimization} in vehicle aerodynamics and aeroacoustics requires a turbulence model with great efficiency, in order to support such large number of numerical simulations. Besides, accurate resolution of the influence of flow unsteadiness is necessary to predict both the aerodynamic and aeroacoustic response. It is in this context that the new STRUCT$-\epsilon$ model can provide accuracy and speed advantages in comparison to existing hybrid models.

\subsection{Scope of the study}

\begin{itemize}
%\item The flow past a spherocylinder body is a standard validation case in fluid dynamics. Thus, the first objective of the paper is to analyze the performance of IDDES and STRUCT turbulence models in this particular case.
\item The flow past a spherocylinder body is a standard validation case in fluid dynamics. Thus, the first objective of the paper is to analyze the performance of the STRUCT$-\epsilon$ turbulence model in this particular case.
%\item There is a notable disparity in the bibliography on the size of the computational domain where the GSM is placed. Here, two different computational domains are considered and the effect the boundary conditions may have on the pressure distribution over the mirror surface and the pressure fluctuations measured in different sensors in the far field is analyzed. Besides, the dependency of the sound pressure spectra on the Reynolds number of the flow is discussed.
\item A Spectral Proper-Orthogonal Decomposition (SPOD) analysis is included in the paper to further describe the flow field observed around the GSM. The capability of the STRUCT$-\epsilon$ turbulence model to let developing such a reduced order model is presented in this work.
\end{itemize}

% from Technical Report: Development and analysis of turbulence models
% for flows with strong curvature and rotation
% Olof Grundestam
% The purpose of this study is partly to investigate how far the
% modeling used in EARSM can be taken in terms of using models
% nonlinear in the strain and rotation rate and/or the Reynolds stress
% anisotropy tensor.

%Comparison of DES, RANS and LES for the separated flow around a flat
%plate at high incidence, M. Breuer et al.
% One of the objectives was to compare the resources (CPU time,
% memory) required for all four techniques, while at the same time to
% investigate and evaluate the quality of the predicted results for
% steady and unsteady simulations. For this purpose, a test case which
% is favourable to the [...] was chosen.
% % mio!
% Since not so much experimental information is available for this
% test case, a further comparison was performed among the turbulence
% models chosen in this paper. For this purpose, the best turbulence
% model in terms of pressure coefficient value and distribution
% prediction served as a reference case.
% 
% %Comparison of DES, RANS and LES for the separated flow around a flat
% %plate at high incidence, M. Breuer et al.
% For all four techniques, a grid sensitivity study was carried out,
% and when converged, predictions were carried out and compared with
% each other based on important integral parameters (Strouhal number,
% mean drag and side-force coefficients and their standard
% deviations), the instantaneous and time-averaged flow structures,
% and higher-order statistics.

The paper is organized as follows. In Section \ref{sec-turbulence-models-formulation}, the formulation of the STRUCT$-\epsilon$ turbulence model considered in our paper is described. Section \ref{sec:methodology} is devoted to the methodology followed in the paper. This encompasses the computational domain and the numerical set-up. The discussion of the results is presented in Section \ref{sec:results}. Finally, the conclusions are given in Section \ref{sec:summary}.

\section{Turbulence model formulation. STRUCT$-\epsilon$ model}
\label{sec-turbulence-models-formulation}

A new STRUCT$-\epsilon$ model \cite{phd-xu-l-2020-struct} of a fully self-adaptive STRUCT model proposed by \cite{phd-lenci-g-2016-struct} and \cite{pap-lenci-g-and-baglietto-e-2021-struct} is considered in this paper and described in this section. This model is implemented in the commercial code \textsf{Star-CCM$+$}. The original STRUCT model addressed the inadequacy of URANS models by introducing a nonlinear eddy viscosity formulation (NLEVM) and the inapplicability of the scale separation assumption in rapidly varying flows through locally resolving a significant portion of the turbulent fluctuations in those regions. The resolved time scale is defined based on the second invariant of the resolved velocity gradient tensor with several advantages. The soundness of the STRUCT$-\epsilon$ approach has been demonstrated through its application to a variety of flow cases, including configurations that had not been addressed successfully by other hybrid models, \cite{phd-xu-l-2020-struct}. In the original STRUCT approach, the hybridization is implemented in a straightforward way by reducing the overall eddy viscosity through a reduction parameter depending on the ratio of the resolved time scale and the modeled flow time scale. The new revised version reduces the eddy viscosity implicitly by adding a source term  $C_{\mathit{\epsilon 3}}k \left| \overline{\mathit{II}}\right|$, where $\left| \overline{\mathit{II}}\right|$ is the second invariant of the resolved gradient tensor, also referred to as $Q$-criterion, in the rate of dissipation of turbulence energy ($\epsilon$) transport equation of the standard $k-\epsilon$ model with the form by \cite{pap-jones-w-and-launder-b-1972-standard-k-epsilon} and coefficients by \cite{pap-launder-b-and-sharma-b-1974-standard-k-epsilon}, similarly as the SAS-SST model:

\begin{equation}
\nu _t = C_{\mu }\frac{k^2}{\epsilon},
\label{eq-nut}
\end{equation}

\begin{equation}
\frac{{\partial}k}{{\partial}t}+ \frac{{\partial}\overline{u}_j k}{{\partial}x_j}=\frac{{\partial}}{{\partial}x_j}[(\nu +\frac{\nu _t}{\sigma _k})\frac{{\partial}k}{{\partial}x_j}]+P_k-\epsilon,
\end{equation}

\begin{equation}
\frac{{\partial}\epsilon }{{\partial}t}+\overline u_j\frac{{\partial}\epsilon
}{{\partial}x_j}=\frac{{\partial}}{{\partial}x_j}[(\nu +\frac{\nu _t}{\sigma _k})\frac{{\partial}\epsilon
}{{\partial}x_j}]+C_{\mathit{\epsilon 1}}\frac{\epsilon }{k}P_k-C_{\mathit{\epsilon
2}}\frac{\epsilon ^2}{k}+C_{\mathit{\epsilon 3}}k\left| \overline{\mathit{II}}\right|; 
\end{equation}

with $P_k=-\overline{u_iu_j}\frac{{\partial}\overline u_i}{{\partial}x_j}$, and the coefficients $C_{\epsilon 1}=1.44$, $C_{\epsilon 2}=1.92$, $\sigma _k=1.0$ and $\sigma
_{\epsilon }=1.30$, respectively. The new value of  $C_{\epsilon 3}$  is selected through sensitivity study on several test cases and may be subjected to further improvements in the further. With this new STRUCT$-\epsilon$ model, the hybridization region no longer depends on the inlet turbulence. In addition, the reduced version is consistent with the original STRUCT idea implying the comparison of  $\left|\overline{\mathit{II}}\right|^{1/2}$ and  $\epsilon /k$, i.e. the modification of the $\epsilon$  equation would only become noticeable when  $\left|\overline{\mathit{II}}\right|^{1/2}$  is larger than  $\epsilon /k$. The baseline URANS model is a cubic NLEVM formulation proposed by \cite{pap-baglietto-e-and-ninokata-h-2007-struct, pap-lenci-g-and-baglietto-e-2021-struct}.

The Reynolds stress tensor is calculated as follows,

\begin{eqnarray}
 \overline{u_iu_j}=\frac 2 3k\delta _{\mathit{ij}}-2\nu _t\overline{S_{\mathit{ij}}}+4C_1\nu _t\frac k{\epsilon}\left[\overline{S_{\mathit{ik}}}\overline{S_{\mathit{kj}}}-\frac 1 3\delta_{\mathit{ij}}\overline{S_{\mathit{kl}}}\overline{S_{\mathit{kl}}}\right]+\nonumber \\
4C_2\nu _t\frac k{\epsilon }\left[\overline{\Omega_{\mathit{ik}}}\overline{S_{\mathit{kj}}}+ 
\overline{\Omega _{\mathit{jk}}}\overline{S_{\mathit{ki}}}\right]+4C_3\nu _t\frac k{\epsilon }[\overline{\Omega _{\mathit{ik}}}\overline{\Omega _{\mathit{kj}}}-\frac 1 3\delta _{\mathit{ij}}\overline{\Omega_{\mathit{kl}}}\overline{\Omega _{\mathit{kl}}}]+\nonumber \\
8C_4\nu _t\frac{k^2}{\epsilon ^2}[\overline{S_{\mathit{ki}}}\overline{\Omega
_{\mathit{lj}}}+\overline{S_{\mathit{kj}}}\overline{\Omega _{\mathit{li}}}]\overline{S_{\mathit{kl}}}+8C_5\nu
_t\frac{k^2}{\epsilon ^2}[\overline{S_{\mathit{kl}}}\overline{S_{\mathit{kl}}}-\overline{\Omega _{\mathit{kl}}}\overline{\Omega
_{\mathit{kl}}}]\overline{S_{\mathit{ij}}}.
\label{eq-rst}
\end{eqnarray}

The coefficients in the equations \ref{eq-nut} and \ref{eq-rst} are

\begin{equation}
C_{\mu}=\frac{0.667}{3.9+S}, \quad S=\frac{k_m}{\epsilon}\sqrt{2\overline{S_{\mathit{ij}}}\overline{S_{\mathit{ij}}}}
\end{equation}

and 

\begin{eqnarray}
C_1=\frac{0.8}{(1000+S^3)C_{\mu }}, \quad C_2=\frac{11}{(1000+S^3)C_{\mu }},\nonumber \\
C_3=\frac{4.5}{(1000+S^3)C_{\mu }}, \quad C_4=-5C_{\mu }^2, \quad C_5=-4.5C_{\mu }^2.
\end{eqnarray}
 
This new STRUCT$-\epsilon$ model has been tested on a variety of simple flow cases including all configurations that original STRUCT was evaluated on and has exhibited similar behaviour in wall-bounded flows. Additional challenging test cases include the Ahmed body, flow over periodic hills and natural transition on the back of a hydrofoil, \cite{phd-xu-l-2020-struct}, and freight train aerodynamics, \cite{pap-garcia-j-et-al-2020-struct-freight-trains}. 

\section{Methodology}
\label{sec:methodology}

\subsection{Computational domain and boundary conditions}

\begin{figure}[!ht]
\begin{center}
        \subfigure[]{%
           \includegraphics[width=0.45\textwidth]{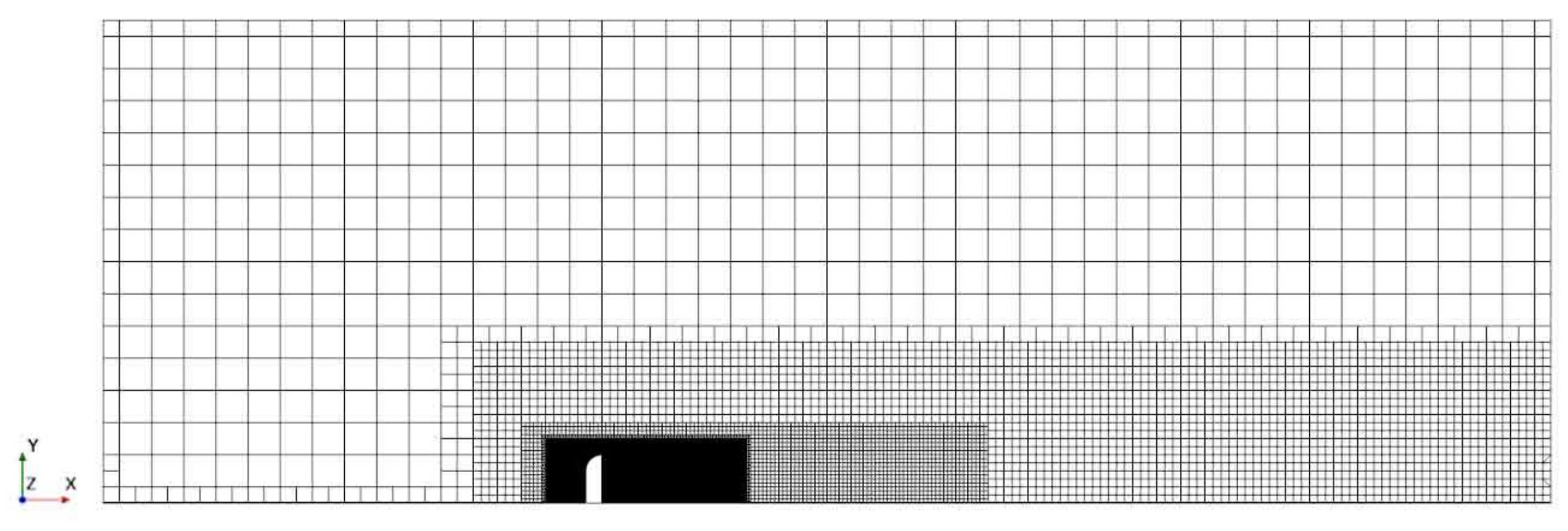}
        }%
        \subfigure[]{%
           \includegraphics[width=0.45\textwidth]{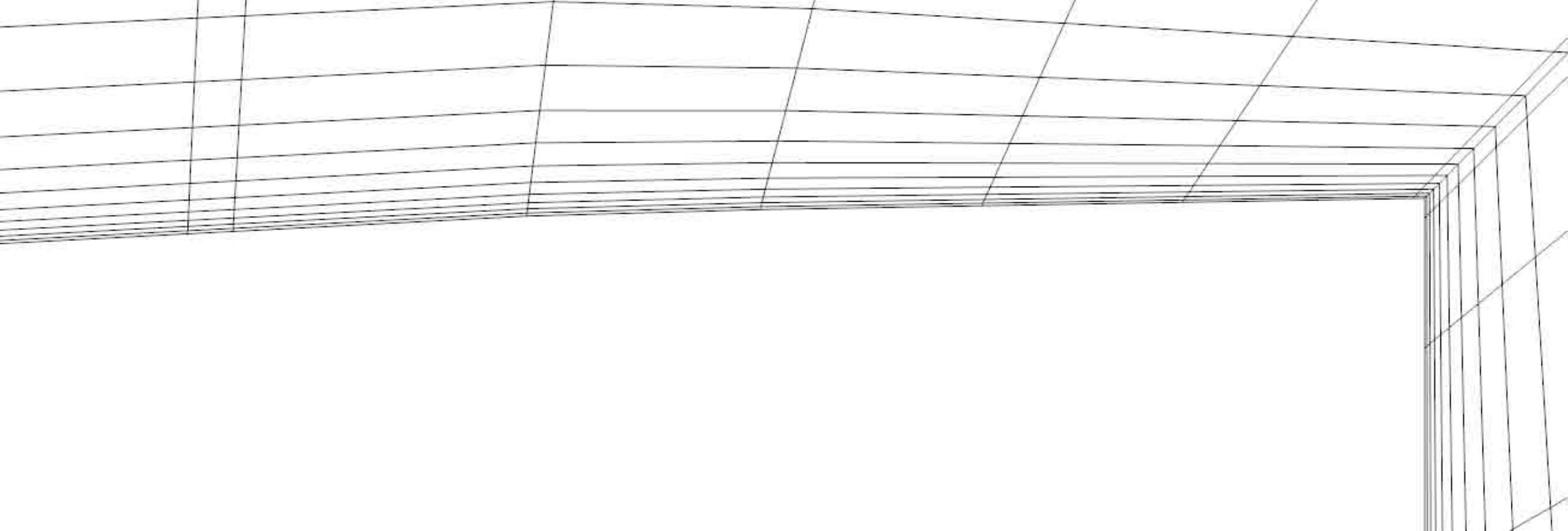}
        }\\
        \subfigure[]{%
           \includegraphics[width=0.45\textwidth]{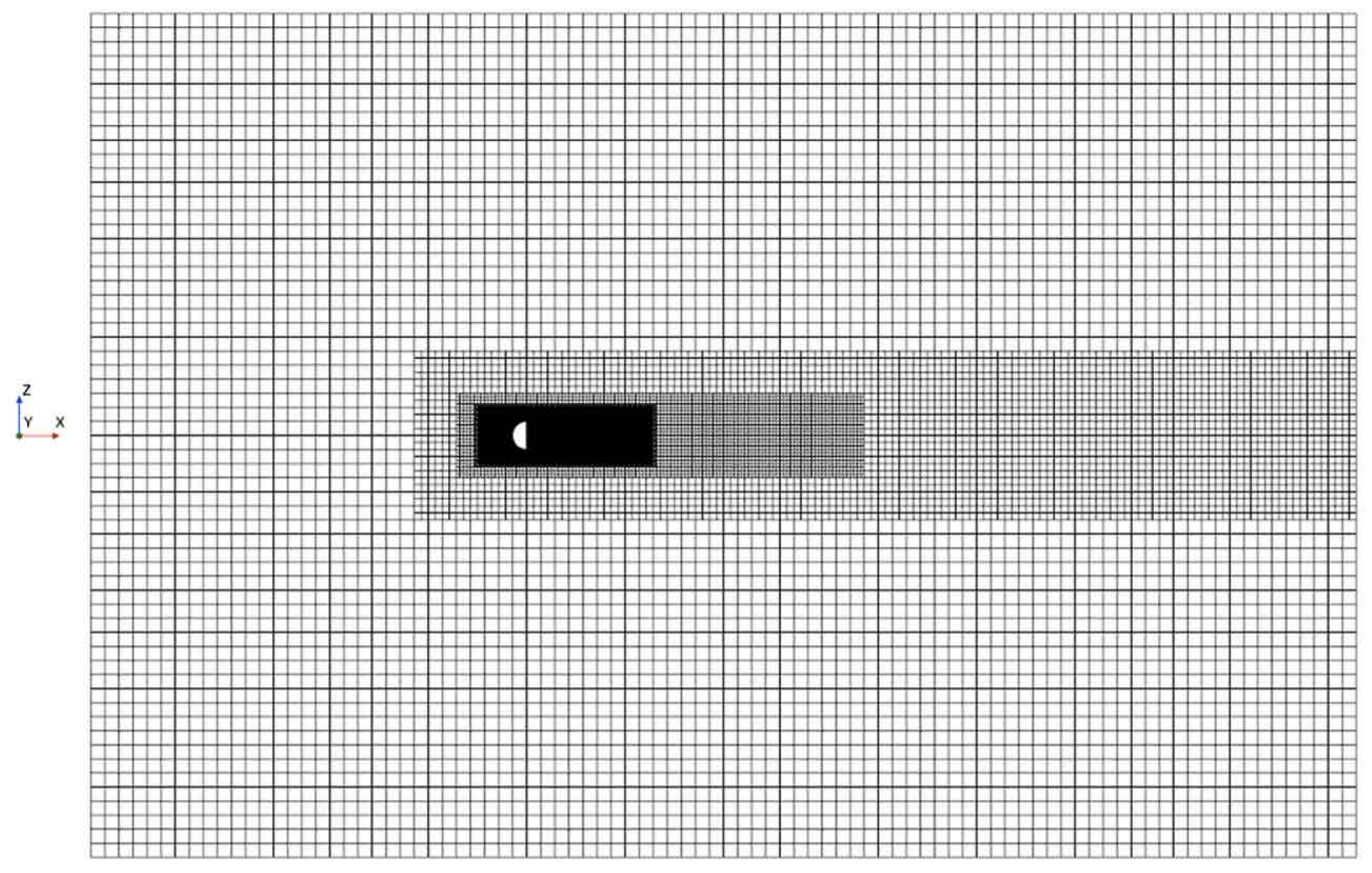}
        }%
        \subfigure[]{%
           \includegraphics[width=0.45\textwidth]{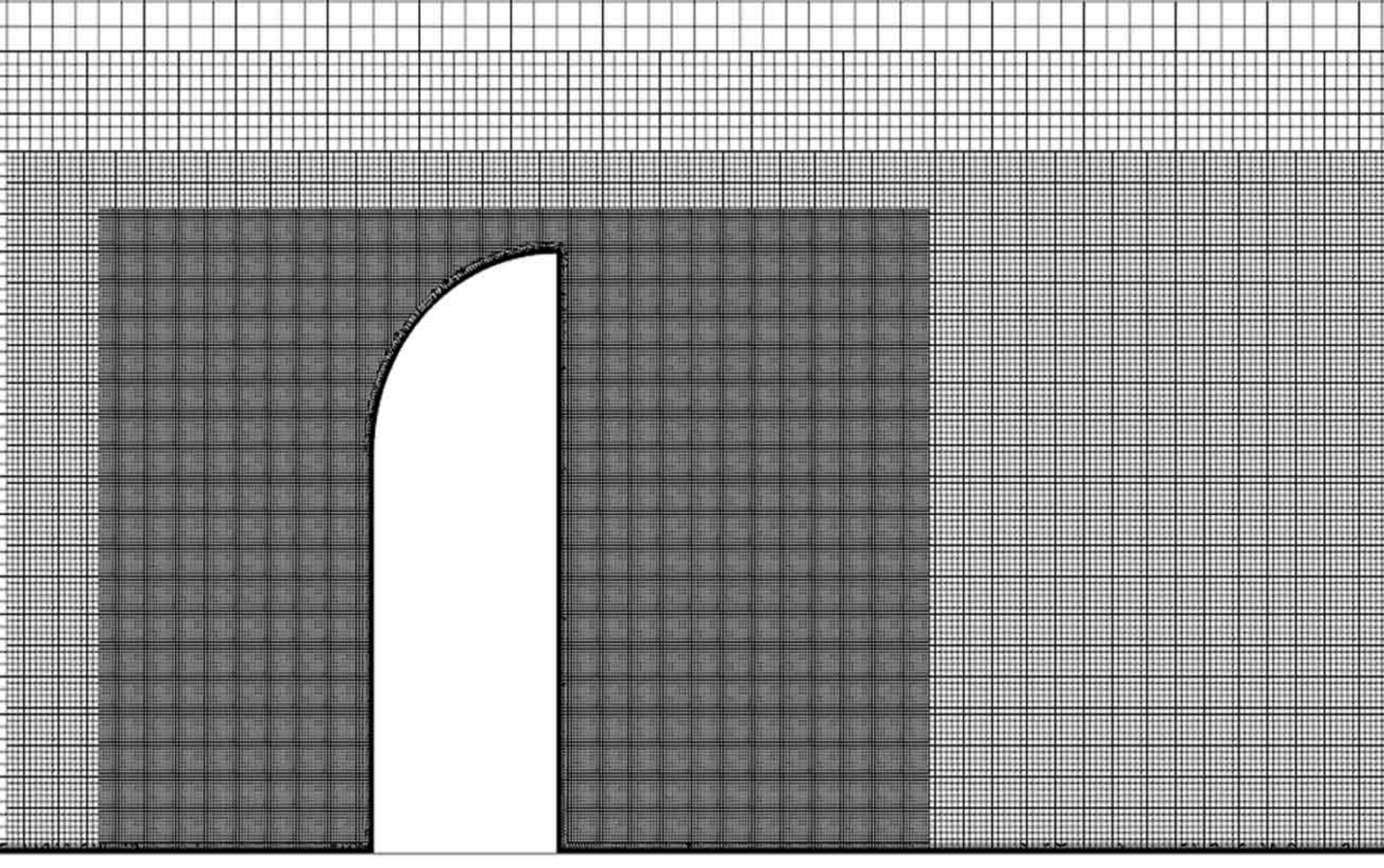}
        }\\
\caption{\small{Computational domain. Mesh corresponding to the STRUCTFi case. From left to right, and from top to bottom, general side view, detail of the boundary layer mesh, top view and detail view of refinement volumes around the GSM, respectively.}}
\label{fig-computational-domain}
\end{center}
\end{figure}

The computational domain is based on the experiment conducted at Daimler-Chrysler Aerospace and published in \cite{cnf-hold-r-et-al-1999-experiments-side-mirror-acoustics-basic-analysis} and \cite{cnf-siegert-r-et-al-1999-experiments-generic-side-mirror}. The body consists in a half cylinder with diameter $D=0.2\,\mathrm{m}$, topped by a quarter of a sphere with the same diameter which gives a total height of $H=0.3\,\mathrm{m}$. The computational domain is an hexahedral box inside which the model is arranged as indicated in Fig. \ref{fig-computational-domain}. The domain has an extension of $43D$ in the streamwise direction, a width of $30D$ and the height reads $15D$. The GSM is placed $15D$ downstream of the inlet boundary. 

The Reynolds number based on the GSM diameter is $\mathrm{Re}_D = 5.2\times 10^5$, which gives a freestream velocity of $39\,\mathrm{m\,s^{-1}}$ that is imposed at the inlet boundary, considering also a turbulence intensity of $0.1\%$ and a turbulent length scale of $0.1\,\mathrm{m}$. Besides, to study the Reynolds number effect, a second freestream velocity is considered. Following the study of \cite{pap-kato-c-et-al-2007-generic-side-mirror-reynolds-effect-yaw-angle}, simulations have been run with $\mathrm{Re}_D = 1.9\times 10^5$, which corresponds to a velocity of $12.56\,\mathrm{m\,s^{-1}}$.

\subsection{Numerical set-up} 
\label{sec:numericalSetUp}

% % from Cokljat et al, 'Embedded LES Methodology for General-Purpose CFD Solvers'
% The general purpose CFD code \textsf{ANSYS-FLUENT} has been used to solve the Navier-Stokes equations considering the IDDES hybrid RANS/LES turbulence model. The bounded central difference scheme is used for the momentum equation and the Least Square Method (LSM) is used for the gradients. For the latter, this choice allows a better representation of the second derivative of the velocity field that is required for the model formulation, \cite{tec-menter-f-2012-best-practices-sas}. The bounded second order implicit Euler scheme is used for the transient terms. 

The general purpose CFD code \textsf{Star-CCM$+$} has been used to solve the Navier-Stokes equations considering the STRUCT turbulence model. The baseline model is the standard low-Re with a cubic NLEVM. For the STRUCT$-\epsilon$ simulations, the corresponding constants have been modified according to those proposed by \cite{pap-baglietto-e-and-ninokata-h-2007-struct}. The spatial discretization schemes are second-order upwind for the convective terms while for the time integration a second-order accurate, three time level implicit scheme was used. Concerning the upwind scheme, its selection is based on the fact that high-order central schemes typically used in LES, as the bounded central difference (BCD) scheme applied in \cite{pap-ask-j-and-davidson-l-2009-les-aeroacoustics-generic-side-mirror}, often suffer from spurious oscillations with coarse grids at the far-field boundary and near the wall, \cite{pap-xiao-z-et-al-2012-numerical-dissipation-massive-separation-tandem-cylinders}. A blended model implemented in \textsf{Star-CCM$+$} (all  wall-treatment) is chosen to calculate the near-wall turbulence quantities. 

\begin{figure}[!htp]
\begin{center}
\includegraphics[width=0.95\textwidth]{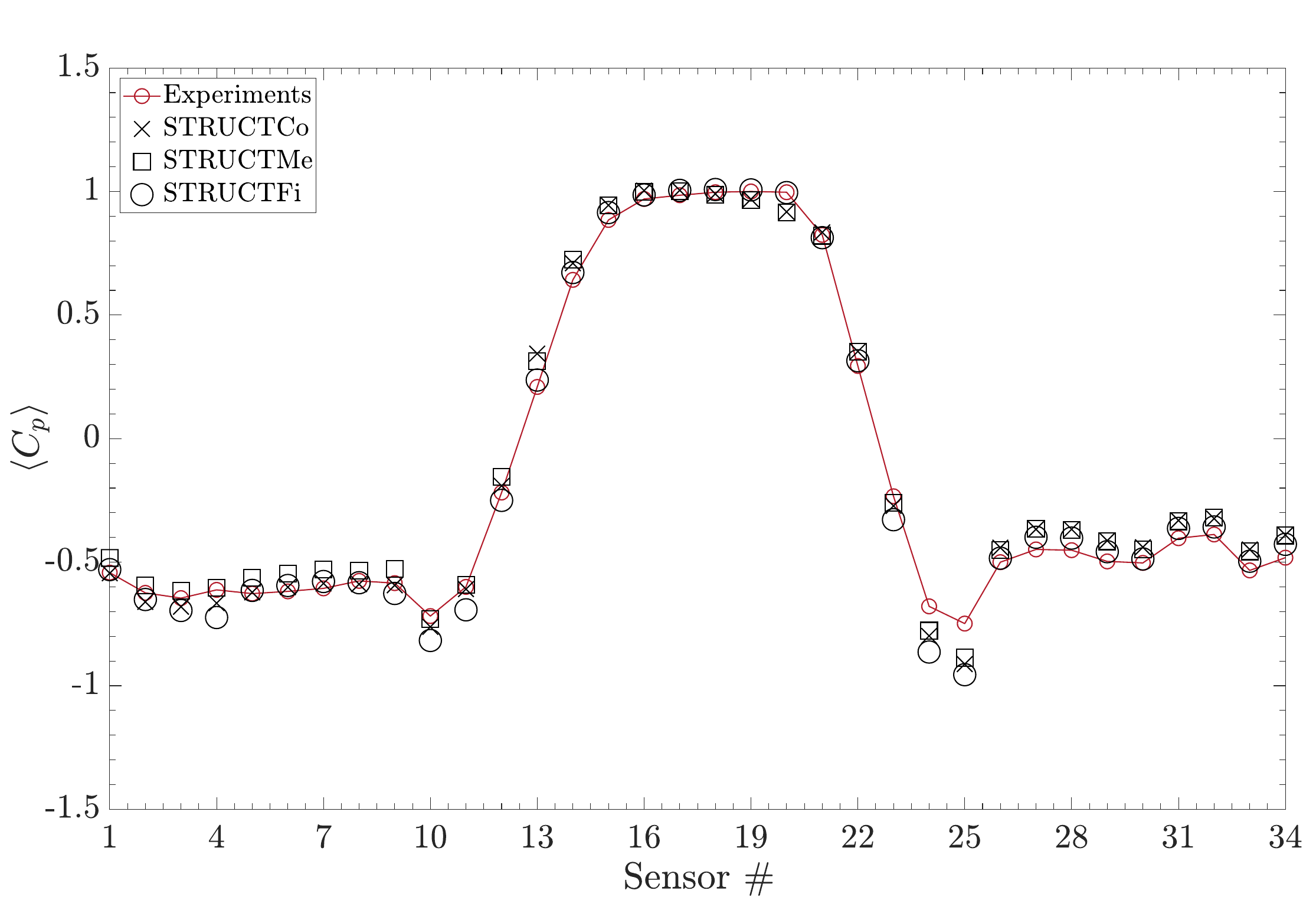}
\caption{\small{Mean pressure distribution along the side mirror surface sensors as a function of the mesh size. Experiments from \cite{cnf-rung-th-et-al-2002-comparison-urans-des-side-mirror}.}}
\label{fig-mesh-sensitivity-cp}
\end{center}
\end{figure}

Several meshes have been used to assure the results are grid independent. The STRUCT$-\epsilon$ turbulence model has been tested with a coarse (STRUCTCo), a medium (STRUCTMe) and a fine (STRUCTFi) mesh of $4.97$, $9.20$ and $12.92\times 10^6$ cells, respectively. All the grids are generated with the same topology. As it is observed in fig. \ref{fig-computational-domain}, the computational domain is divided into five control volumes, whose size is kept constant for the two coarser meshes and for which a refinement ratio of $0.206$ is set. The finest one still considers the same five control volumes, but the two ones closer to the GSM are extended so that the refinement around the body and in the wake is larger. A trimmed hexahedral mesh with prismatic cell layers near the walls has been considered. A fine grid area is defined so as to follow the footprint of the expected shedding vortices from the cylindrical part and the crest of the quarter sphere. Since we are interested in computing the aeroacoustic noise sources, we have to guarantee an appropriate resolution of the vortex structure of the flow field. To ensure an accurate resolution of highest frequencies, for the STRUCTFi case a fixed time step of $2\times\,10^{-5}\,\mathrm{s}$ was used for the time integration. This time step corresponds to approximately $8\times\,10^{-5}$ times the time taken for the fluid to travel the computational domain length, and is in good agreement with the time-step values of \cite{pap-ask-j-and-davidson-l-2009-les-aeroacoustics-generic-side-mirror}. For the STRUCTMe and STRUCTCo cases, the time step has been increased consistently with the cell size in order to maintain a similar CFL. All cases cover $0.9\,\mathrm{s}$ of physical time. Table \ref{tab-gci} indicates the size of the three cases considered in the mesh sensitivity analysis and the corresponding drag coefficient on the GSM. For comparison, it is included also information of other numerical simulations from available publications.  Unfortunately, there is not available information of the $\langle C_D\rangle$ measured in the experiments so as to validate the numerical results. Nevertheless, the time-averaged static pressure distribution (in terms of pressure coefficient $\langle C_p\rangle$) reported from the experiments in \cite{cnf-rung-th-et-al-2002-comparison-urans-des-side-mirror} let us discuss the results of $\langle C_D\rangle$ as both coefficients are somehow related. The higher drag force coefficient obtained from the STRUCTFi simulation compared to the LES or hybrid methods cases is probably due to a lower pressure recovery over the mirror rear side as it is identified in fig. \ref{fig-mean-cp-sensors}. All the turbulence models overpredict that pressure recovery when compared to the experiments. An extensive discussion of our results compared to published results is presented in the next sections. Concerning only our simulations, it is the STRUCTFi simulation the one that approximates better to the experimental data. Figure \ref{fig-mesh-sensitivity-cp} shows the pressure coefficient $\langle C_p\rangle$ on the side mirror model, measured with the pressure sensors, which are located according to fig. \ref{fig-position-sensors}, for the three cases, namely STRUCTCo, STRUCTMe and STRUCTFi. 

%The Grid Convergence Index ($GCI$) is used to estimate the relative error of the computed value with respect to the asymptotic numerical value. To compute the GCI, a mean cell size $h_i$ is defined as $h_i = \sqrt{V/N_i}$, where $V$ is the computational domain volume and $N_i$ is the total number of cells of the mesh. The grid refinement ratio $r_i$ is set to X, as it is shown in table \ref{tab-gci}
%The coarse grids are about 70\% grid points in all three directions compared with [...]

\begin{table}[!ht]
\begin{center}
\begin{tabular}{l r c | l r c}
\hline\hline
Case & $N_i$ ($\times 10^6$) & $\langle C_D\rangle$ & Case & $N_i$ ($\times 10^6$) & $\langle C_D\rangle$\\ 
\hline
STRUCTCo & 4.97 & 0.522 & LES & 31.10 & 0.444\\
STRUCTMe & 9.20 & 0.513 & DDES & 14.00 & 0.445\\
STRUCTFi & 12.92 & 0.497 & SBES & 6.86 & 0.472\\
\hline\hline
\end{tabular}
\end{center}
\caption{\small{General characteristics of meshes used in this paper, indicating the number of cells $N_i$ of each one, as well as the resulting time-averaged drag coefficient $\langle C_D\rangle$. For the sake of comparison, the same information of other numerical simulations is given: incompressible LES from \cite{pap-ask-j-and-davidson-l-2009-les-aeroacoustics-generic-side-mirror}, DDES from \cite{pap-yu-l-et-al-2021-fast-transient-fractional-step-method-scheme-generic-side-mirror} and SBES from \cite{pap-chode-k-et-al-2021-sbes-generic-side-mirror}.}}
\label{tab-gci}
\end{table}

\begin{figure}[!htp]
\begin{center}
\includegraphics[width=0.95\textwidth]{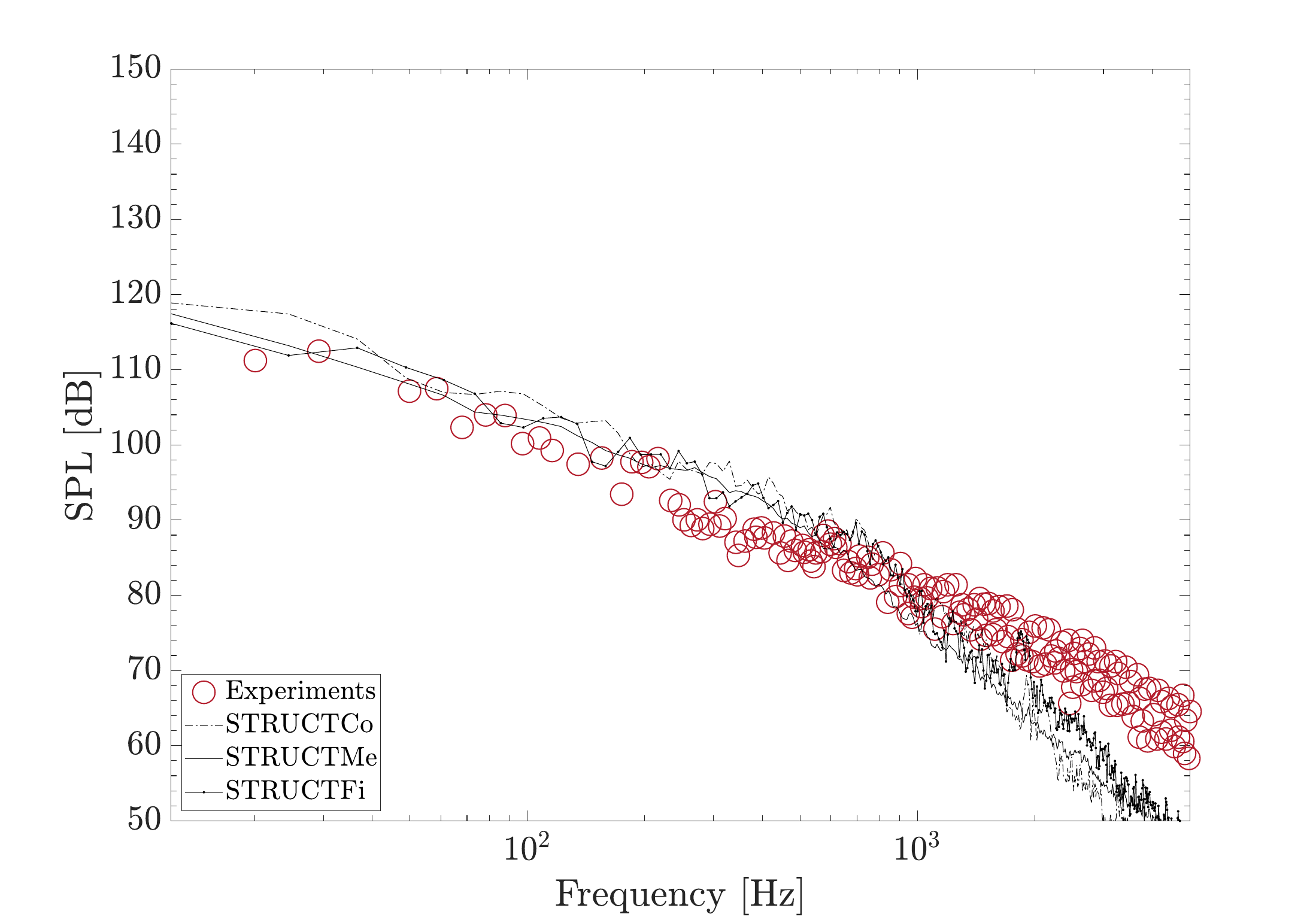}
\caption{\small{SPL at surface sensor $\#113$ as a function of the turbulence model. Experiments from \cite{cnf-rung-th-et-al-2002-comparison-urans-des-side-mirror}.}}
\label{fig-mesh-sensitivity-spectra}
\end{center}
\end{figure}

Figure \ref{fig-mesh-sensitivity-spectra} plots the sound pressure level (SPL) obtained for sensor $\#113$ (table \ref{tab-dynamic-pressure-sensors} indicates also the position of this sensor at the plate) for the three cases previously mentioned. The SPL is computed as 

\begin{equation}
 \mathrm{SPL} = 20\log_{10}\frac{\hat{p}}{p_0},
 \label{eq-spl}
\end{equation}

where $p_0 = 2\times 10^{-5}\,\mathrm{Pa}$ is the reference pressure and $\hat{p}$ is the filtered Fourier transformed results of the fluctuating pressure measured and calculated at the corresponding sensor. The time sequences were resampled and split into windows containing $2^{12}$ samples. The sampling frequency is $50\,\mathrm{kHz}$. A Hanning filter is applied to each window with an overlap of $50\%$. The differences are not significant as all the cases show a similar decay of the spectra and approximate roughly similar to the experiments up to $1000\,\mathrm{Hz}$. However, the combination of the response of the three cases let us conclude that the differences between the fine and medium meshes are not negligable, so hereafter the fine mesh STRUCTFi is considered for all the simulations.

\section{Discussion of the results}
\label{sec:results}

\subsection{Aerodynamic force coefficients and pressure distribution}

\begin{figure}[!ht]
\begin{center}
\includegraphics[width=0.95\textwidth]{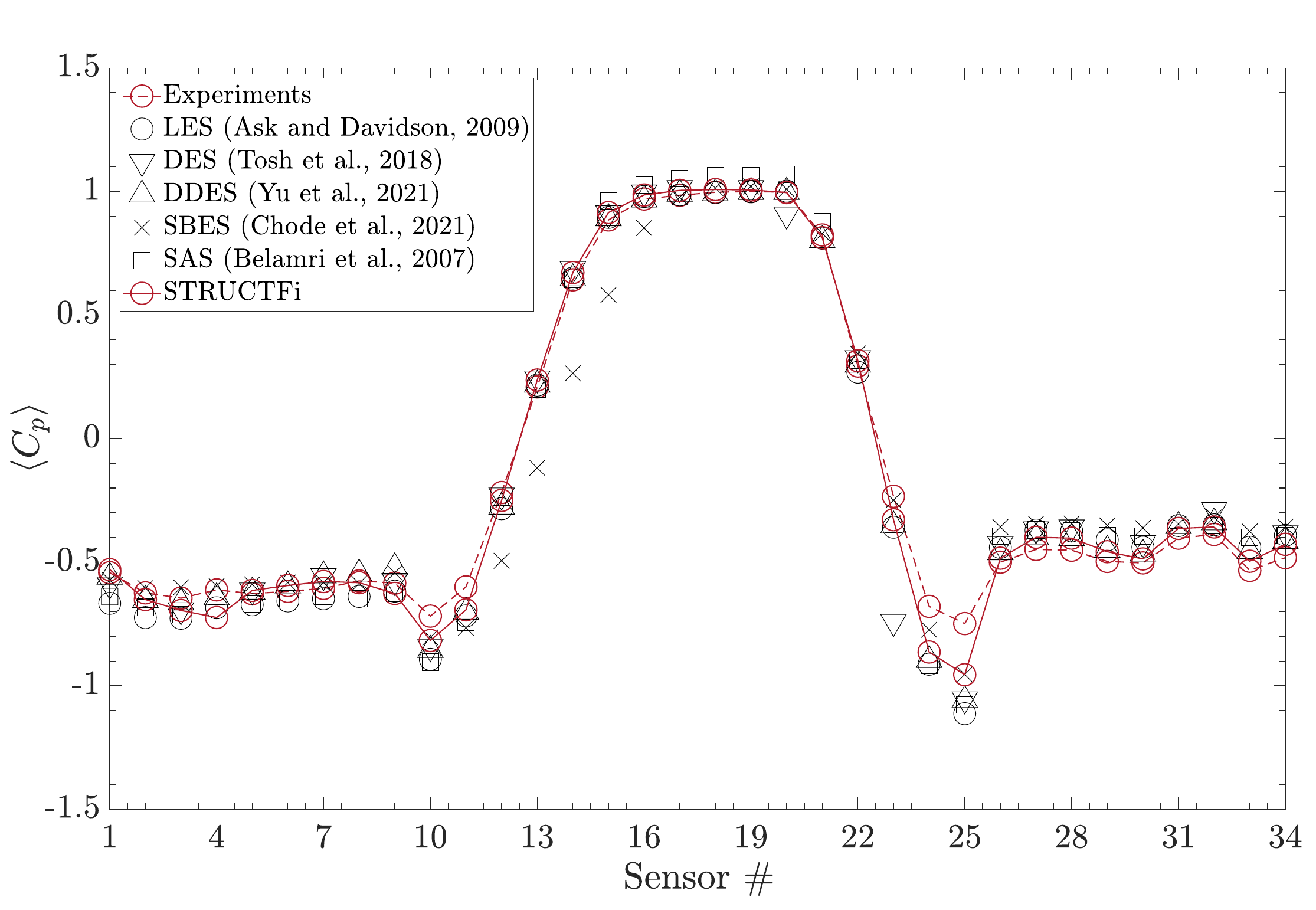}
\caption{\small{Mean pressure distribution along the side mirror surface sensors as a function of the turbulence model.}}
\label{fig-mean-cp-sensors}
\end{center}
\end{figure}

Once a grid has been chosen as our reference for validation of the turbulence model, the results obtained with the STRUCT$-\epsilon$ model are compared with different published results involving alternative turbulence models. Figure \ref{fig-mean-cp-sensors} shows the $\langle C_p\rangle$ values on the side mirror model obtained with the STRUCTFi simulation compared with the experiments published in \cite{cnf-rung-th-et-al-2002-comparison-urans-des-side-mirror}, the LES simulations from \cite{pap-ask-j-and-davidson-l-2009-les-aeroacoustics-generic-side-mirror} and several hybrid RANS/LES methods, including the SST-DES \cite{cnf-tosh-a-et-al-2018-des-generic-side-mirror}, the SST-DDES \cite{pap-yu-l-et-al-2021-fast-transient-fractional-step-method-scheme-generic-side-mirror} and the SBES results from \cite{pap-chode-k-et-al-2021-sbes-generic-side-mirror}, as well as the SST-SAS \cite{pap-egorov-y-et-al-2010-sas-generic-side-mirror}.

Sensors \#1 to \#9 are located 6 mm upstream of the mirror rear side, which means an angle $\theta = 86.6\,\mathrm{deg}$ measured from the front stagnation point, see fig. \ref{fig-position-sensors}. Sensors \#1 to \#4 are placed in the cylindrical part of the mirror, while sensors \#5 to \#9 are found in the spherical region of the body. While most of the turbulence models underpredict the pressure in the cylindrical part (including our simulation with STRUCT$-\epsilon$), differences in the behaviour of the aforementioned turbulence models are observed for the spherical region. Here, again the LES in \cite{pap-ask-j-and-davidson-l-2009-les-aeroacoustics-generic-side-mirror} and the SST-SAS from \cite{pap-egorov-y-et-al-2010-sas-generic-side-mirror} result into lower pressure values than that from the experiments. However, both SST-DES and SST-DDES of \cite{cnf-tosh-a-et-al-2018-des-generic-side-mirror} and \cite{pap-yu-l-et-al-2021-fast-transient-fractional-step-method-scheme-generic-side-mirror}, respectively, as well as our STRUCTFi simulation, slightly overpredict the pressure coefficient in sensors \#5 to \#8. In these sensors, the average relative error of the STRUCT when compared to the experiments is of $2.94\%$. Meanwhile, sensor \#9, located in the symmetry plane of the GSM, is underpredicted with the STRUCT$-\epsilon$ turbulence model, indicating the flow is attached rather than detached as it is observed in the experiments. These variations in the response of the turbulence model, under and overpredicting the pressure for the same azimuthal angle $\theta$ might be explained by the different critical Re number at which the flow transites from laminar to turbulent in a cylinder and a sphere, respectively. This issue is addressed afterwards in this paper.

Sensors \#10 to \#20 are located along the symmetry plane of the GSM. The agreement of our STRUCTFi simulation with the experiments and the other turbulence is notably satisfactory for all the sensors but for sensors \#10 and \#11, which are in the topward region. Particularly, the pressure corresponding to the sensors located in the cylindrical part (\#16 to \#20) is predicted with an average relative error of $1.49\%$. When the STRUCT$-\epsilon$ results are compared with that of the second-generation URANS turbulence models (SST-SAS) or advanced hybrid models like the SBES of \cite{pap-chode-k-et-al-2021-sbes-generic-side-mirror}, the response of STRUCT$-\epsilon$ overachieves even for the sensors placed in the spherical region. It is interesting to remark that sensors \#16 and \#20 are close to the upper and lower limits of the cylindrical part of the mirror (the former near the transition to the spherical shape and the latter close to the plate), and in these two sensors the value of $\langle C_p\rangle$ is a little less than 1.00, meaning that the mean flow was not totally stagnant there but moving upwards and downwards, respectively.

Sensors \#21 to \#25 (as well as \#17 and \#2) are placed along a horizontal curve in the front side of the body at $y/D = 0.667$. Therefore, they are found in the cylindrical part of the side mirror. As expected, the $\langle C_p\rangle$ profile shows a peak at the leading edge stagnation point at azimuthal angle $\theta = 0^\circ$ (sensor \#17). Further away, pressure decreases as $\theta$ (absolute value) increases. A minimum value of $C_p$ is obtained for sensor \#25, which corresponds to $\theta = 75\,\mathrm{deg}$. The STRUCTFi simulation, as well as all the turbulence models here studied, underpredicts the pressure when compared to the experiments, although it is relevant to highlight that it is able to get closer to the experiments than any other turbulence model considered in the analysis. This lower pressure coefficient is explained by the resulting delayed separation of the boundary layer, which takes place in the experiments at $\theta \approx 72.5\,\mathrm{deg}$ and is observed in the STRUCTFi simulation at $\theta \approx 73.7\,\mathrm{deg}$, fig. \ref{fig-wall-shear-stress}. This larger value of the flow separation angle points out to a premature turbulent transition in the boundary layer. Such behaviour has also been observed in \cite{pap-pereira-fs-et-al-2019-cfd-circular-cylinder-re-140000} for typical scale-resolving simulation approaches like DES, DDES or SAS simulating the flow around a smooth circular cylinder at sub-critical regime. The errors in separation location prediction are probably due to inability of SST models to predict transition because of the absence of damping functions, \cite{pap-robertson-e-et-al-2015-cfd-aeroacustical-spectral-analysis-spherocylinder}. To conclude the analysis of this set of sensors, it is also interesting to highlight that once the flow is detached from the side mirror (both experiments and any simulation enumerated before show how at sensor \#2 the flow is already detached), the STRUCTFi simulation predicts the pressure at this point with a relative error (with respect to the experiments) of $4.30\%$.

% from robertson.e et al. 2015
%The SST model implementation does not include damping functions in the $k$ and $\omega$, which are sometimes necessary to simulate low-Re effects (such as transition). These functions modify the production of the transport equations to properly represent the behaviour of turbulence near the wall.

Sensors \#26 to \#34 are located at the rear side of the mirror. Particularly, sensors \#26 to \#31 are found close to the trailing edge, while sensors \#32 to \#34 are located in the centerline of the rear side, see fig. \ref{fig-position-sensors}. A very good agreement with the experiments (relative error of $3.20\%$) is observed for the predicted pressure coefficient at sensors \#26 and \#30, which are placed symmetrically at a height $y/D = 0.425$. Any other turbulence model considered in this analysis overpredicts $\langle C_p\rangle$ at these sensors with a minimal relative error of $7.40\%$ for the SST-DDES of \cite{pap-yu-l-et-al-2021-fast-transient-fractional-step-method-scheme-generic-side-mirror} and of $11.96\%$ for the LES of \cite{pap-ask-j-and-davidson-l-2009-les-aeroacoustics-generic-side-mirror}. The values obtained for the time-averaged $C_p$ at these sensors confirm that the recirculating vortices detached from the trailing edge, and so the wake flow, is symmetric, which is typical of sub-critical regime, \cite{pap-demartino-c-and-ricciardelli-f-2017-theory-flow-cylinders}. For the rest of the sensors of this subset, all the turbulence models, even our simulation STRUCTFi, overpredict $\langle C_p\rangle$ when compared to the experiments.

To conclude the subsection, taking advantage of the tabulated information available in the literature that allows a more accurate evaluation of the prediction capability of STRUCT$-\epsilon$, table \ref{tab-comparison-cp} presents a comparison of $\langle C_p\rangle$ obtained at several sensors from our STRUCTFi simulation and those published in \cite{pap-yao-hd-and-davidson-l-2018-generic-side-mirror-interior-cavity-noise}. Our results outdo those from the compressible LES (the second best one) for sensors \#5 and \#10. Meanwhile, the differences between the best one (again the compressible LES simulation of \cite{pap-yao-hd-and-davidson-l-2018-generic-side-mirror-interior-cavity-noise}) and STRUCTFi simulation for sensors \#15, \#20, \#25 and \#30 are negligable (lower than $2.50\%$ between them). Only at the sensor \#34 STRUCT$-\epsilon$ gives a notable different prediction of $\langle C_p\rangle$ when compared with the experiments or the compressible LES simulation. 

\begin{table}[!ht]
\begin{center}
\begin{tabular}{l r r r r r r r}
\hline\hline
Case / Sensors & \#5 & \#10 & \#15 & \#20 & \#25 & \#30 & \#34\\ 
\hline
Experiments & $-0.629$ & $-0.725$ & $0.886$ & $0.991$ & $-0.753$ & $-0.507$ & $-0.484$\\
Compressible LES & $-0.457$ & $-0.592$ & $0.879$ & $0.991$ & $-0.557$ & $-0.498$ & $-0.472$\\
Incompressible DES & $-0.668$ & $-0.896$ & $0.866$ & $0.956$ & $-1.112$ & $-0.453$ & $-0.451$\\
Incompressible LES & $-0.727$ & $-0.898$ & $0.898$ & $1.000$ & $-1.102$ & $-0.477$ & $-0.443$\\
STRUCTFi & $-0.614$ & $-0.817$ & $0.915$ & $0.996$ & $-0.995$ & $-0.487$ & $-0.428$\\
\hline\hline
\end{tabular}
\end{center}
\caption{\small{Comparison of $\langle C_p\rangle$ at several sensors on the mirror surfaces obtained from the experiments \cite{cnf-hold-r-et-al-1999-experiments-side-mirror-acoustics-basic-analysis}, incompressible DES and LES from \cite{pap-ask-j-and-davidson-l-2009-les-aeroacoustics-generic-side-mirror} and compressible LES from \cite{pap-yao-hd-and-davidson-l-2018-generic-side-mirror-interior-cavity-noise}.}}
\label{tab-comparison-cp}
\end{table}

\begin{figure}[!ht]
\begin{center}
\includegraphics[width=1\textwidth]{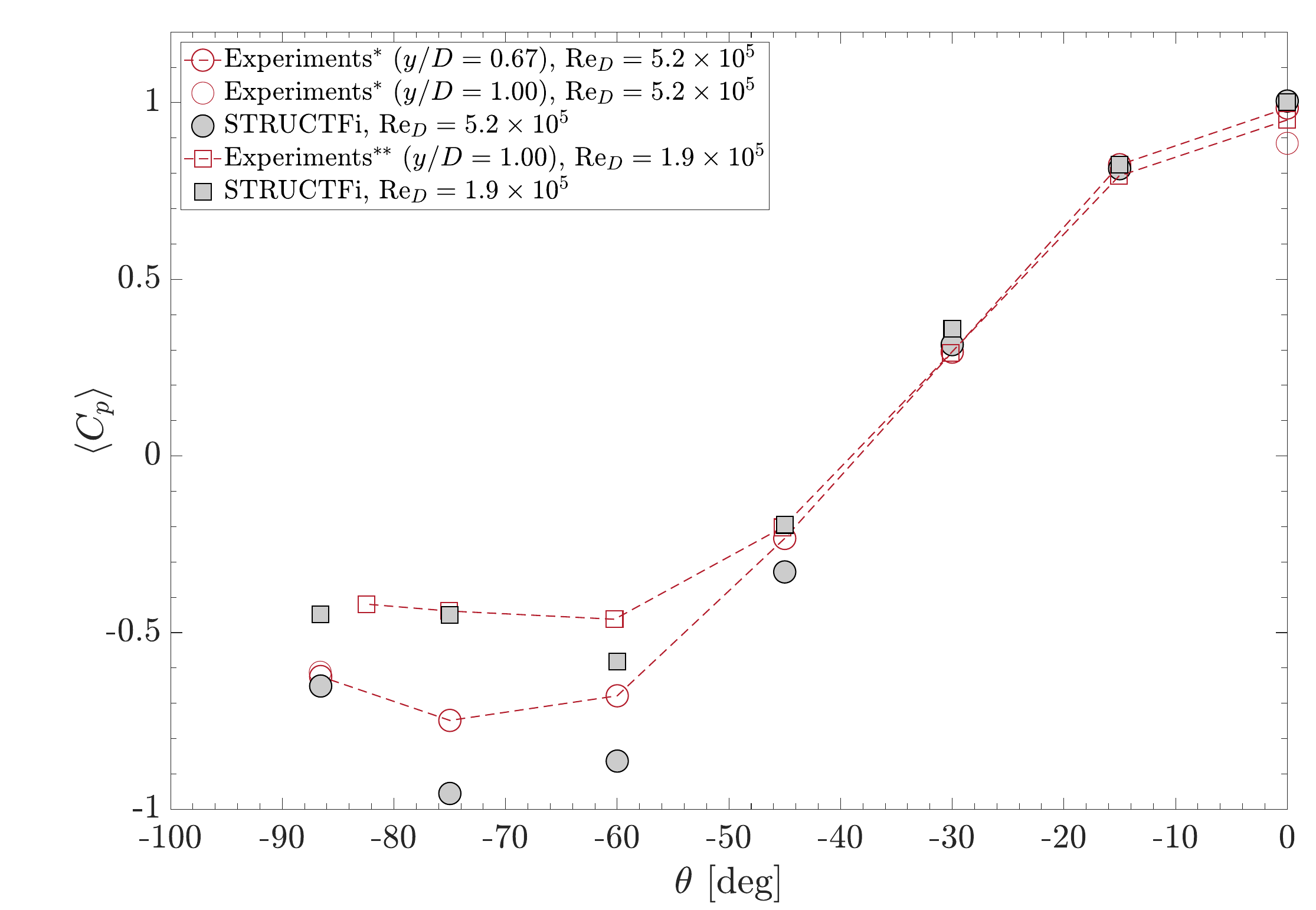}
\caption{\small{Mean pressure coefficient along the cylinder surface $\langle C_p\rangle(\theta)$ as a function of the Reynolds number. Experiments$^*$ refers to \cite{cnf-rung-th-et-al-2002-comparison-urans-des-side-mirror}, while Experiments$^{**}$ refers to \cite{pap-kato-c-et-al-2007-generic-side-mirror-reynolds-effect-yaw-angle}.}}
\label{fig-mean-cp-angle}
\end{center}
\end{figure}

\subsection{Reynolds number effect}

The experiments for this benchmark case that have been published involve Reynolds numbers which might correspond to different flow regimes. In \cite{cnf-hold-r-et-al-1999-experiments-side-mirror-acoustics-basic-analysis} and \cite{cnf-siegert-r-et-al-1999-experiments-generic-side-mirror}, $\mathrm{Re}_D = 7.01\times 10^5$, while in \cite{cnf-rung-th-et-al-2002-comparison-urans-des-side-mirror} the experiments are run for $\mathrm{Re}_D =5.2\times 10^5$. Meanwhile, in \cite{pap-kato-c-et-al-2007-generic-side-mirror-reynolds-effect-yaw-angle} the Reynolds number varies from $1.4\times 10^5$ to $2.4\times 10^5$. These Reynolds numbers are found to be \emph{controversial} as it is not well defined how the transition range is set for the GSM, which involves half a cylinder and a quarter of sphere. \cite{pap-roshko-a-1961-supercritical-reynolds-single-cylinder} indicates that, for a cylinder, the whole range from Re $= 2\,\times\,10^5$ to $3.5\,\times\,10^6$ could be called the transition range, although other authors define such region in the range Re $= 2\,\times\,10^5$ to $5\,\times\,10^5$, labelled in \cite{pap-roshko-a-1961-supercritical-reynolds-single-cylinder} as lower or critical transition. The Reynolds number considered in this paper is Re$_D$ $= 5.2\,\times\,10^5$ (identical to most of the aforementioned numerical publications analyzing the GSM). This value, for a circular cylinder, is close, if not over, to the critical value. However, as mentioned in \cite{pap-ask-j-and-davidson-l-2009-les-aeroacoustics-generic-side-mirror}, the experimentalists observed a sub-critical flow state in their tests. This observation is in good agreement with the conclusions reported in \cite{pap-kato-c-et-al-2007-generic-side-mirror-reynolds-effect-yaw-angle}, but the latter considered a lower Reynolds number. Thus, for the sake of comparison, fig. \ref{fig-mean-cp-angle} shows the pressure distribution around the GSM for Re$_D$ $= 1.9\,\times\,10^5$ and Re$_D$ $= 5.2\,\times\,10^5$ reported in \cite{pap-kato-c-et-al-2007-generic-side-mirror-reynolds-effect-yaw-angle} and \cite{cnf-rung-th-et-al-2002-comparison-urans-des-side-mirror}, respectively. The former only presents results of $\langle C_p\rangle$ measured at $y/D = 1.00$, while the latter gives the pressure distribution along the cylindrical side of the side mirror, measured at $y/D = 0.67$ (sensors $\#2$, $\#17$ and $\#21$ to $\#25$) and at $y/D = 1.00$ (sensors $\#4$ and $\#15$). These two sensors let us compare the influence of the relative height of the location of the sensors. While it has a small influence in the stagnation point (i.e. $\#15$ versus $\#17$), it is negligible for sensors $\#2$ and $\#4$, as the flow is already detached at these points. This feature is evinced when we compare the results of \cite{pap-kato-c-et-al-2007-generic-side-mirror-reynolds-effect-yaw-angle} with the results of the STRUCTFi simulation at Re$_D$ $= 1.9\,\times\,10^5$, as a smaller value of $\langle C_p\rangle$ is reported in the experiments at $y/D = 1.00$ than that obtained in the simulation at $y/D = 0.67$, while no difference is observed when the flow is already detached (at $\theta = 75\,\mathrm{deg}$). 

%The influence of the Reynolds number is further presented in the following sections, involving.

%The $\langle C_p\rangle$ of a circular cylinder for a sub-critical Reynolds number is also plotted in fig. \ref{fig-mean-cp-angle}. 

%\cite{pap-zdravkovich-1997-smooth-cylinder-turbulent-transition} remarks that, for a smooth circular cylinder, the regime where turbulent transition occurs in the free shear-layer is observed in the Reynolds number range of $3.5\times 10^2$ to $2.0\times 10^5$. This regime 

\subsection{Time-averaged flow fields}

\begin{figure}[!htp]
\begin{center}
        \subfigure[]{%
           \includegraphics[height=0.297\textwidth]{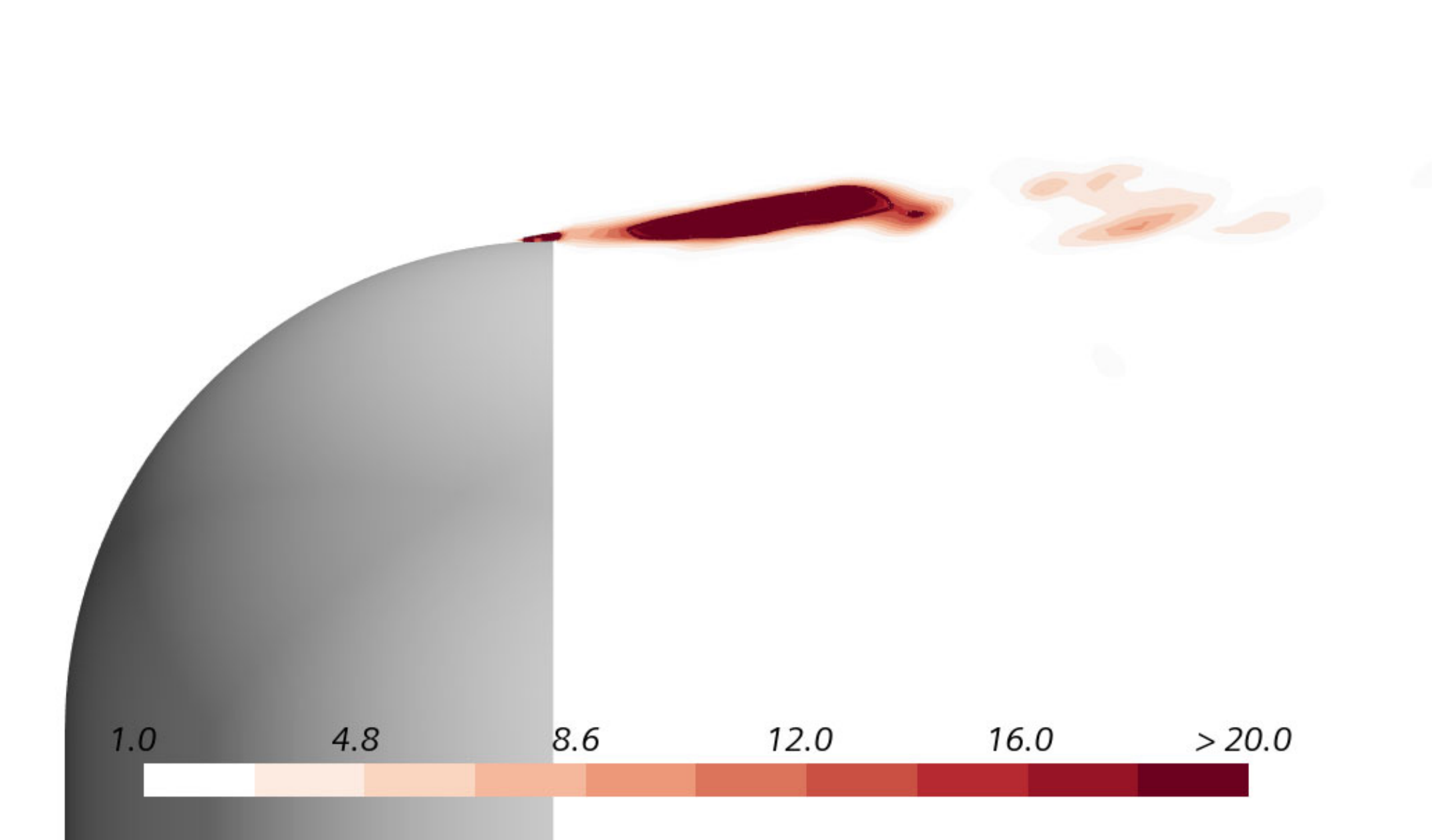}
        }%
        \subfigure[]{%
           \includegraphics[height=0.297\textwidth]{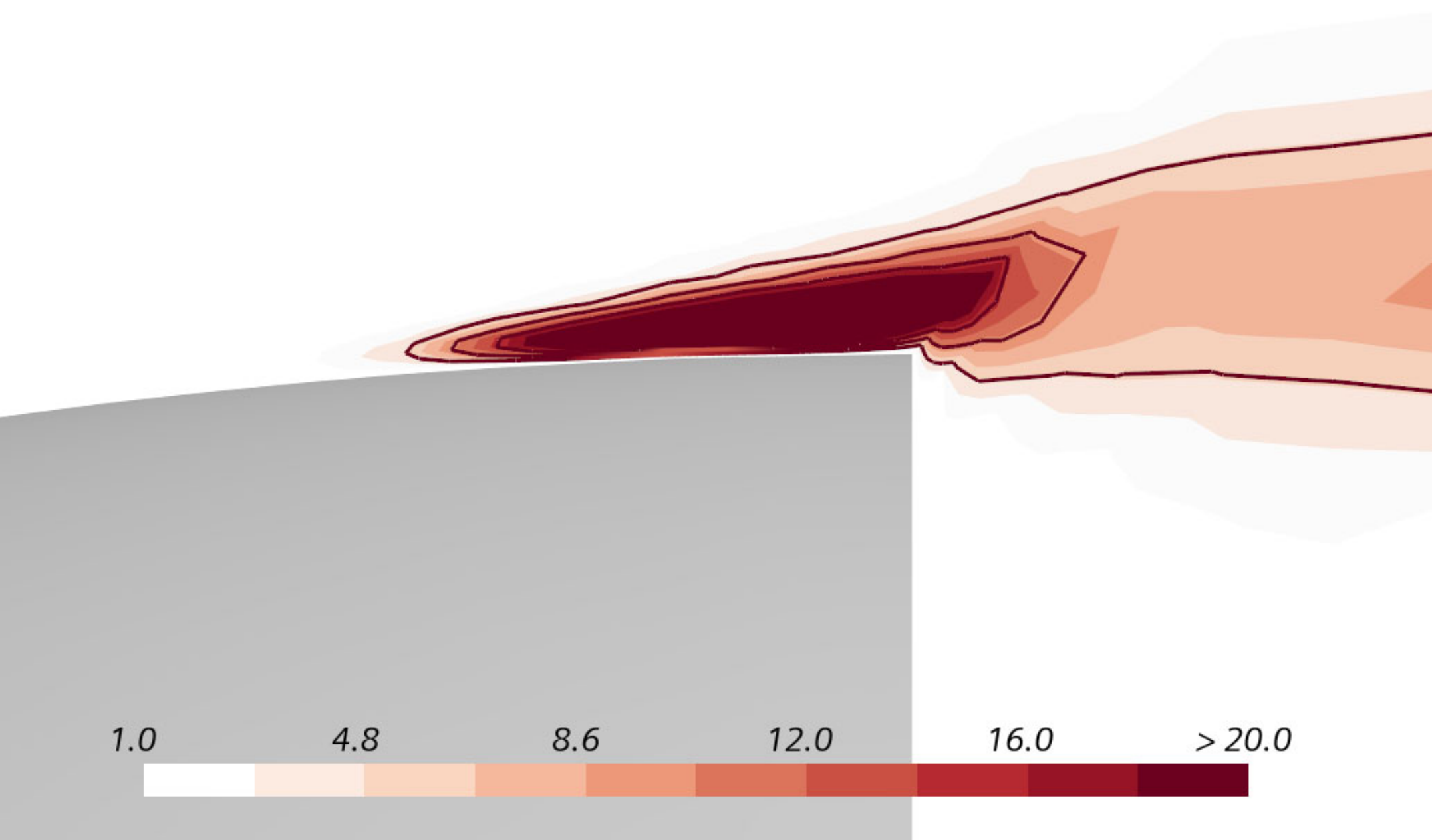}
        }\\
\caption{\small{Eddy viscosity ratio over plane $z = 0$ for $\mathrm{Re}_D = 5.2\times 10^5$, including a zoom view of the activated zone.}}
\label{fig-eddy-viscosity}
\end{center}
\end{figure}

Very significant differences are found in the field of the turbulent viscosity normalized by the molecular viscosity around the top of the side mirror. Figure \ref{fig-eddy-viscosity} (a) and (b) show the viscosity ratio field and a zoomed detail of it. The contours are limited up to $\frac{\nu_t}{\nu} = 20$ for this case to be compared with the results obtained considering the DES in \cite{pap-ask-j-and-davidson-l-2009-les-aeroacoustics-generic-side-mirror} (which are further detailed in \cite{cnf-ask-j-and-davidson-l-2006-des-aeroacoustics-generic-side-mirror}) and the SBES in \cite{pap-chode-k-et-al-2021-sbes-generic-side-mirror}. The aforementioned references predict a delayed separation when compared to the experiments of \cite{cnf-rung-th-et-al-2002-comparison-urans-des-side-mirror}, which obtains a separation line $0.15D$ upstream of the trailing edge of the body. The SBES simulations presented in \cite{pap-chode-k-et-al-2021-sbes-generic-side-mirror} result into a separation line $0.0625D$ from the trailing edge, while not any specific length is given in \cite{pap-ask-j-and-davidson-l-2009-les-aeroacoustics-generic-side-mirror} for the identification of the separation line, but it is reported that all simulations conducted indicate a delayed separated region. In particular for the DES simulation, the separation line is not even detected. As it is pointed out in these references, an excessive production of turbulent viscosity might be responsible for delaying or preventing the ocurrence of the separation. For the sake of clarity, table \ref{t-comparison-eddy-viscosity} is given in this section, indicating at the symmetry plane (i.e., $z = 0$) which is the distance from the trailing edge of three reference values of $\frac{\nu_t}{\nu}$, namely 5, 10 and 15. When compared with the results of \cite{cnf-ask-j-and-davidson-l-2006-des-aeroacoustics-generic-side-mirror}, a good agreement is found for $\frac{\nu_t}{\nu} = 15$, while the values of $5$ and $10$ are obtained notably downstream than the location given in \cite{cnf-ask-j-and-davidson-l-2006-des-aeroacoustics-generic-side-mirror}. Consequently, a very low viscosity ratio is observed around the separation line location reported in \cite{cnf-rung-th-et-al-2002-comparison-urans-des-side-mirror}, what helps predicting more accurately the flow detachment. More evident are even the differences when compared to the results published in \cite{pap-chode-k-et-al-2021-sbes-generic-side-mirror}, as a viscosity ratio of 11.44 is reported at $0.13D$ from the trailing edge, while the STRUCTFi simulation brings a $\frac{\nu_t}{\nu} < 2$ at that location. Thus, unlike it might be expected from standard URANS turbulence model simulations, \cite{pap-nakayama-a-and-miyashita-k-2001-urans-smooth-topography}, not very large values of $\nu_t$ are observed outside the shear region.

\begin{table}[!ht]
\begin{center}
\begin{tabular}{l r r r}
\hline\hline
Turbulence model & $\frac{\nu_t}{\nu} = 5$ & $\frac{\nu_t}{\nu} = 10$ & $\frac{\nu_t}{\nu} = 15$ \\ 
\hline
Incompressible DES & $0.250D$ & $0.130D$ & $0.025D$ \\
STRUCTFi  & $0.047D$ & $0.043D$ & $0.040D$\\
\hline\hline
\end{tabular}
\end{center}
\caption{\small{Comparison of turbulent viscosity ratio values at different locations in the symmetry plane for the incompressible DES of \cite{pap-ask-j-and-davidson-l-2009-les-aeroacoustics-generic-side-mirror} and the present simulations considering the STRUCT$-\epsilon$ turbulence model at $\mathrm{Re}_D = 5.2 \times 10^5$.}}
\label{t-comparison-eddy-viscosity}
\end{table}

\begin{figure}[!htp]
\begin{center}
\includegraphics[width = 1.0\textwidth]{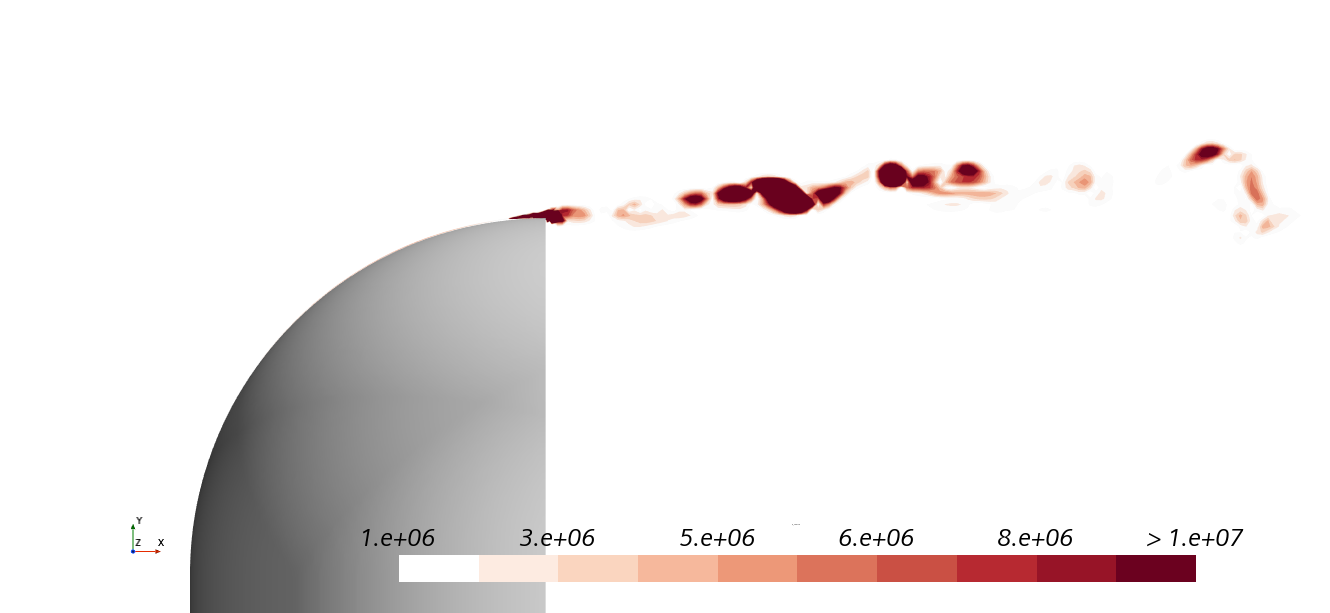}
\caption{\small{Activation source term field at $z = 0$ for $\mathrm{Re}_D = 5.2\times 10^5$.}}
\label{fig-activation-e-source-term}
\end{center}
\end{figure}

The viscosity ratio here can be understood as a criterion to measure the amount of modeled turbulence and, consequently, to detect the RANS and the LES zone. Figure \ref{fig-activation-e-source-term} shows the hybrid activation regions based on the field of the source term  $C_{\mathit{\epsilon 3}}k \left| \overline{\mathit{II}}\right|$. Indeed, a key aspect of the STRUCT-$\epsilon$ model is its ability to considerably improve the description of the flow structures, while activating the source term in the $\epsilon-$equation only in very limited regions, as can be seen in Fig. \ref{fig-activation-e-source-term}. The hybridization is mostly activated in the proximity of the trailing edge. The source term has a high value for the separation regions, while the term is almost negligible away from these regions. So, it is consistent with the STRUCT concept of increasing flow resolution in areas with strong deformation of rapidly varying flows, \cite{phd-xu-l-2020-struct}. Due to the model activation, the STRUCT$-\epsilon$ model provides a significantly increased resolution of the turbulent structures, as they can be observed in section \ref{sec:instantaneous-flow-structures}. This increased resolution of unsteady flow structures is extremely helpful for properly modelling the aeroacoustic behaviour of the side mirror.

\begin{figure}[!htp]
\begin{center}
        \subfigure[]{%
           \includegraphics[width=0.5\textwidth]{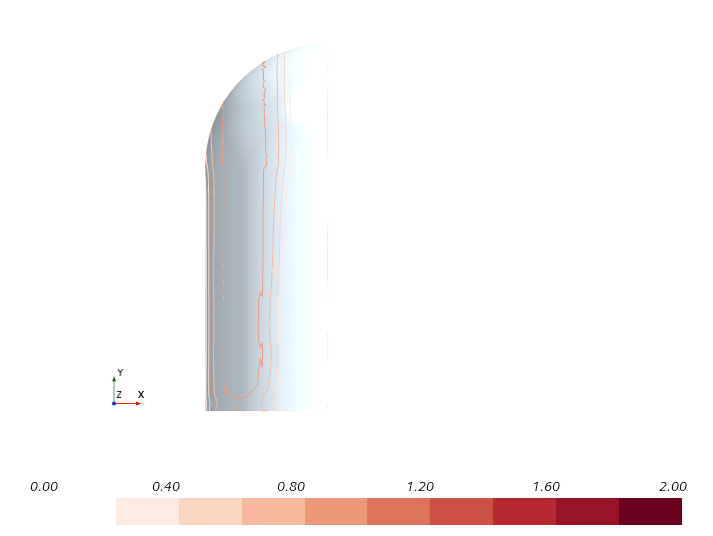}
        }%
        \subfigure[]{%
           \includegraphics[width=0.5\textwidth]{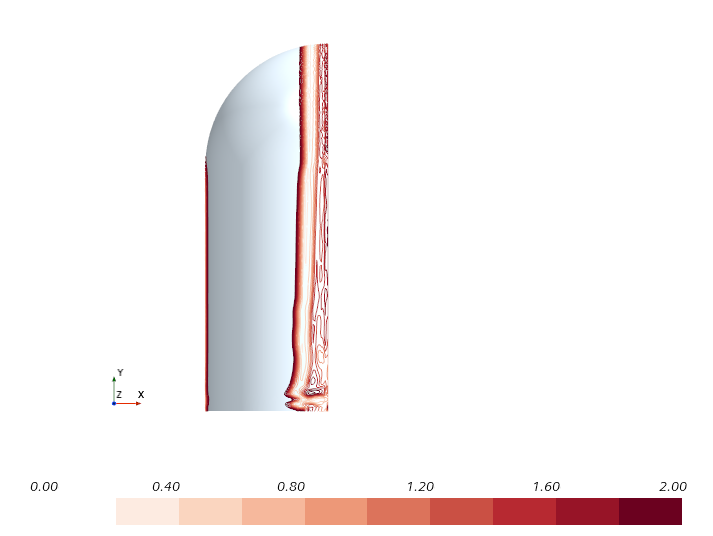}
        }\\
\caption{\small{Wall-shear stress for (a) $\mathrm{Re}_D = 1.9\times 10^5$ and (b) $\mathrm{Re}_D = 5.2\times 10^5$ over the side mirror.}}
\label{fig-wall-shear-stress}
\end{center}
\end{figure}

The experiments run by Daimler Chrysler Research and Technology included oil visualization to identify the separation line, and these results are reported in \cite{pap-ask-j-and-davidson-l-2009-les-aeroacoustics-generic-side-mirror}, what let us compare the wall-shear stress $\tau_{wall}$ predicted on the surface of the mirror within the STRUCTFi simulations and the available published results. Figure \ref{fig-wall-shear-stress} shows this flow field variable in the range $0 \leq \tau_{wall} \leq 2\,\mathrm{N\,m^{-2}}$. As it has been mentioned in previous sections, the separation of the boundary layer takes place in the experiment at $\theta \approx 72.5\,\mathrm{deg}$, which is also in good agreement with the results reported in \cite{pap-kato-c-et-al-2007-generic-side-mirror-reynolds-effect-yaw-angle} for $\mathrm{Re}_D = 1.4 - 2.5 \times 10^5$ ($\theta \approx 70\,\mathrm{deg}$). Figure \ref{fig-wall-shear-stress} (b) shows that, in the present study, $\tau_{wall} = 2$ is found at $\approx 0.14D$ from the trailing edge, in very good agreement with the experiments of \cite{cnf-rung-th-et-al-2002-comparison-urans-des-side-mirror}, and notably upstream that the results considering the DES simulation in \cite{pap-ask-j-and-davidson-l-2009-les-aeroacoustics-generic-side-mirror} or the SBES results presented in \cite{pap-chode-k-et-al-2021-sbes-generic-side-mirror}, which reports a location of $0.05D$ from the trailing edge.  

\begin{figure}[!ht]
\begin{center}
\includegraphics[width = 1.0\textwidth]{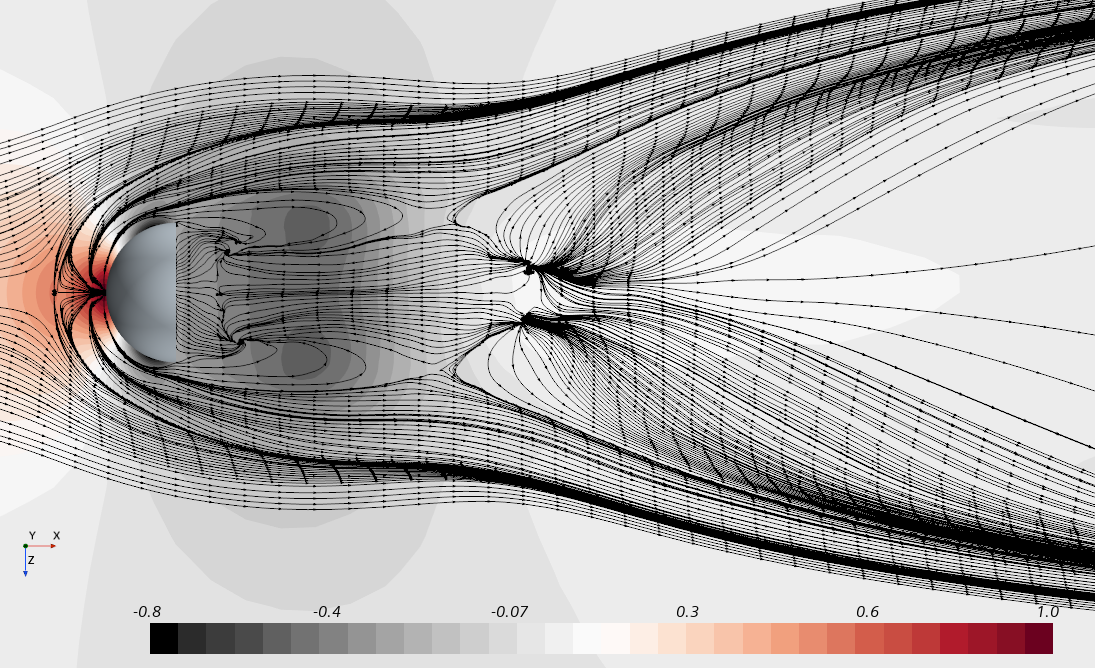}
\caption{\small{Streamlines projected on the floor (i.e. $y = 0$), surface colored by time averaged pressure coefficient $\langle C_p\rangle$ for $\mathrm{Re}_D = 5.2\times 10^5$.}}
\label{fig-pathlines-re-520}
\end{center}
\end{figure}

\begin{figure}[!htp]
\begin{center}
        \subfigure[]{%
           \includegraphics[width=0.5\textwidth]{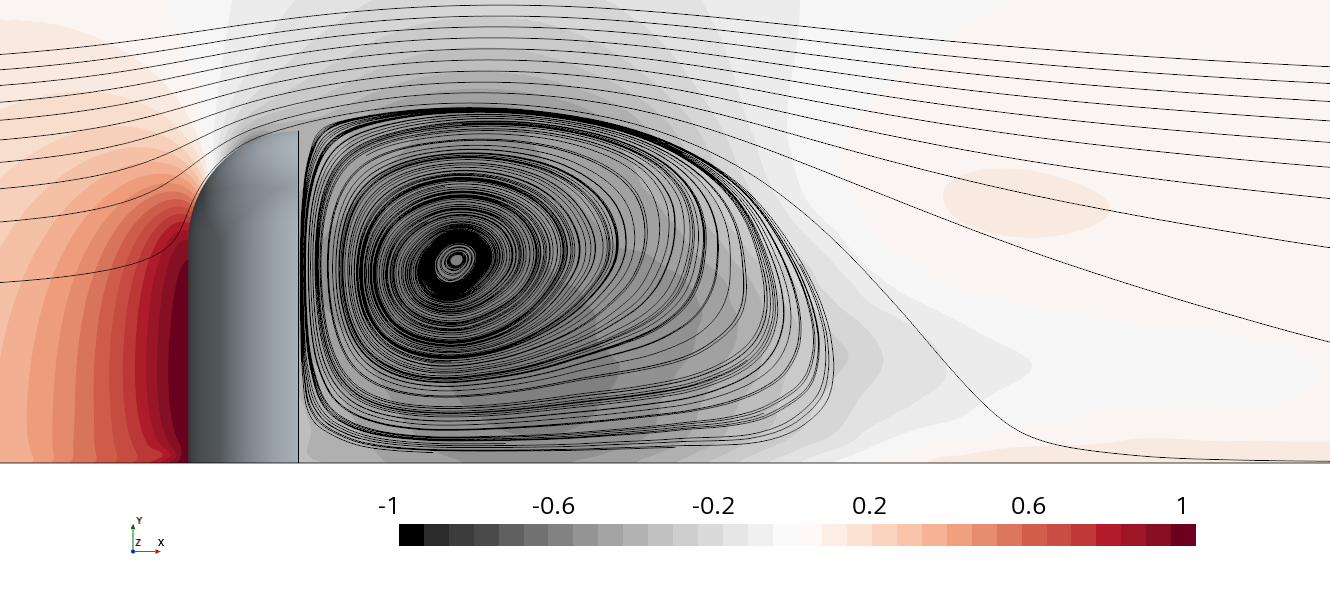}
        }%
        \subfigure[]{%
           \includegraphics[width=0.5\textwidth]{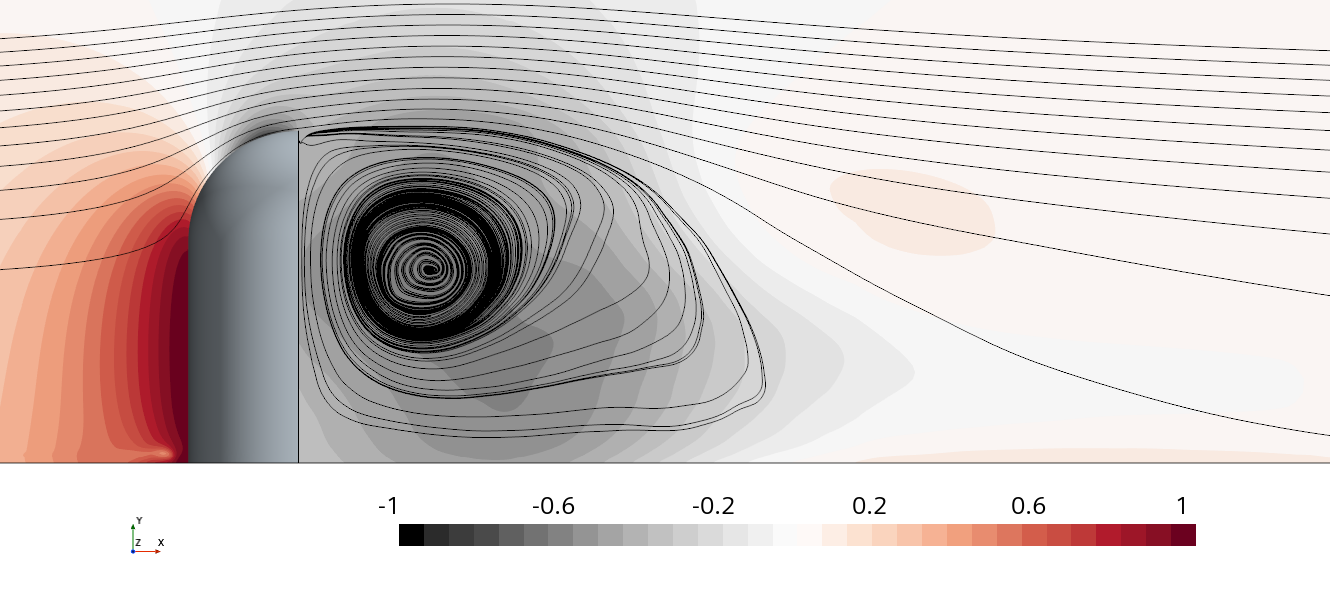}
        }\\
\caption{\small{Streamlines for (a) $\mathrm{Re}_D = 1.9\times 10^5$ and (b) $\mathrm{Re}_D = 5.2\times 10^5$ projected on the symmetry plane (i.e. $z = 0$),  surface colored by time averaged pressure coefficient $\langle C_p\rangle$.}}
\label{fig-vortex-core-detection}
\end{center}
\end{figure}

To conclude the time-averaged flow field study, this subsection is devoted to the vortex core detection based on the analysis of the streamlines projected onto different planes, namely the floor in fig. \ref{fig-pathlines-re-520} and the symmetry plane ($z = 0$) in fig. \ref{fig-vortex-core-detection}, respectively. For the sake of brevity and comparison with available published results, this analysis is limited to $\mathrm{Re}_D = 5.2\times 10^5$. The flow approaching the side mirror experiences a strong adverse pressure gradient and undergoes three-dimensional separation forming a horseshoe vortex. The work of \cite{pap-devenport-w-and-simpson-r-1990-bimodal-horseshoe-vortex} demonstrated the bimodal behaviour of this horseshoe vortex for the flow past a wall-mounted wing at $\mathrm{Re} = 1.15\times 10^5$, and this feature is also observed in this paper for the side mirror at $\mathrm{Re}_D = 5.2\times 10^5$. This phenomenon has been thoroughly studied by \cite{pap-paik-j-et-al-2007-bimodal-dynamics-horseshoe-vortex} or \cite{pap-kirkil-g-and-constantinescu-g-2015-effect-reynolds-number-horseshoe-vortex-system} among others. The simulation captures the presence of several reference lines associated with the horseshoe vortices that are better depicted in fig. \ref{fig-instantaneous-fields}(b) and fig. \ref{fig-visualization-lambda-2}(b). The outer line away from the side mirror is a line of separation which originates at the saddle point located at the plane of symmetry. This point defines the distance $L_{hs}$. In the inner region, an attachment line and a second separation line are observed. The latter is related to the secondary horseshoe vortex that is embraced by the primary horseshoe vortex, as it is better shown in fig. \ref{fig-visualization-lambda-2}(b). The separation line from the saddle point has been considered as a reference to define the length $L_{hz}$ in several publications, which are detailed in table \ref{tab-reference-lengths}. These works report a third length labelled as $L_{ws}$ that apparently is defined from the trailing edge of the side mirror to the unstable nodes that indicate the stagnation points at which the flow reattaches to the floor. We observe two symmetrical nodes that are expected from the interaction of the upper shear layer and the two symmetrical side shear layers. Between them, a middle recirculation is detected, which is also mentioned in \cite{pap-yao-hd-and-davidson-l-2018-generic-side-mirror-interior-cavity-noise}. These three lengths are compared with the LES simulations of \cite{pap-ask-j-and-davidson-l-2009-les-aeroacoustics-generic-side-mirror} and \cite{pap-yao-hd-and-davidson-l-2018-generic-side-mirror-interior-cavity-noise}, the DES and DDES results from \cite{pap-chen-x-and-li-m-2019-ddes-generic-side-mirror}, the SAS results from \cite{cnf-belamri-t-and-menter-f-2007-sas-generic-side-mirror} and the SBES results of \cite{pap-chode-k-et-al-2021-sbes-generic-side-mirror}. A very good agreement is found between our simulation and the results of the LES simulations. For $L_{hz}$ all the turbulence models achieve similar results, but the STRUCTFi simulation is closer to the incompressible LES and DDES results than to any other one. Unfortunately there is no information about this reference length for the compressible LES so as to compare with. Concerning $L_{ws}$, our results are similar to that from \cite{pap-yao-hd-and-davidson-l-2018-generic-side-mirror-interior-cavity-noise}, being in any case the smallest size of recirculation bubble when compared to the rest of the hybrid methods. It is observed that the overprediction of the turbulent viscosity ratio in the proximity of the separation line in the side mirror causes a larger recirculation as the detachment is delayed when compared to the experiments. Therefore, the fact that our simulation get closer to the experimental location of the separation line might explain this smaller size of the recirculation bubble. Finally, when we look at $L_{hx}$, we can observe that the STRUCTFi simulation gives the largest value. Here, as previously mentioned, we have defined this length by the position of the saddle point from which the separation line upstream the side mirror is generated. This criterion is not uniform in all the references, as \cite{pap-chode-k-et-al-2021-sbes-generic-side-mirror} does also define $L_{hx}$ based on the position of the aforementioned saddle point, while \cite{pap-ask-j-and-davidson-l-2009-les-aeroacoustics-generic-side-mirror} indicates this length is defined by the center of the horseshoe vortex and \cite{pap-chen-x-and-li-m-2019-ddes-generic-side-mirror} does not detail how this length is defined in their case. Being aware that a negative bifurcation line on no-slip boundaries indicates separation of the flow from the surface and it is upstream of the associated horseshoe vortex in this situation, it is therefore understandable to have a larger value of $L_{hx}$ than that reported by \cite{pap-ask-j-and-davidson-l-2009-les-aeroacoustics-generic-side-mirror}.

\begin{table}[!ht]
\begin{center}
\begin{tabular}{l c c c}
\hline\hline
Turbulence model & $L_{hx}$ & $L_{hz}$ & $L_{ws}$ \\ 
\hline
Incompressible LES & $0.26D$ & $0.45D$ & $2.58D$ \\
Compressible LES & $-$ & $-$ & $2.50D$ \\
DES & $0.27D$ & $0.40D$ & $3.25D$\\
DDES & $0.27D$ & $0.43D$ & $2.65D$ \\
SAS & $-$ & $-$ & $2.00-4.00D$ \\
SBES & $0.30D$ & $0.42D$ & $2.59D$ \\
STRUCTFi  & $0.35D$ & $0.44D$ & $2.50D$\\
\hline\hline
\end{tabular}
\end{center}
\caption{\small{Comparison of reference lengths of the time-averaged streamlines projected on the floor. Values for the incompressible LES of \cite{pap-ask-j-and-davidson-l-2009-les-aeroacoustics-generic-side-mirror}, the compressible LES from \cite{pap-yao-hd-and-davidson-l-2018-generic-side-mirror-interior-cavity-noise}, the DES and DDES results from \cite{pap-chen-x-and-li-m-2019-ddes-generic-side-mirror}, the SAS from \cite{cnf-belamri-t-and-menter-f-2007-sas-generic-side-mirror}, the SBES results of \cite{pap-chode-k-et-al-2021-sbes-generic-side-mirror} and the present simulations considering the STRUCT$-\epsilon$ turbulence model at $\mathrm{Re}_D = 5.2 \times 10^5$.}}
\label{tab-reference-lengths}
\end{table}

As it is pointed out in \cite{pap-yao-hd-and-davidson-l-2018-generic-side-mirror-interior-cavity-noise}, three main regions are observed, namely a free shear layer, a recirculation bubble and two far-downstream wake branches. The former two are better visualized in fig. \ref{fig-vortex-core-detection}. Flow separation dictates vortex generation and vortex shedding, and a clear vortex is observed in the rear side of the mirror. The center of this vortex varies with the Reynolds number, moving closer to the base face as the Reynolds number increases. The reattachment location does also varies with the Reynolds number, as it is observed in fig.\ref{fig-vortex-core-detection}. The separation of the flow in the side mirror is discussed in fig. \ref{fig-mean-cp-angle} considering the Reynolds numbers of $\mathrm{Re}_D = 1.9\times 10^5$ and $\mathrm{Re}_D = 5.2\times 10^5$, and it is found, as expected, that the separation occurs before for the former than the latter. Thus the wake is wider and so the reattachment occurs later, approximately a $14\%$ larger the reciculation bubble than that from  
$\mathrm{Re}_D = 5.2\times 10^5$.

\subsection{Instantaneous flow structures}
\label{sec:instantaneous-flow-structures}

\begin{figure}[!htp]
\begin{center}
        \subfigure[]{%
           \includegraphics[width=0.5\textwidth]{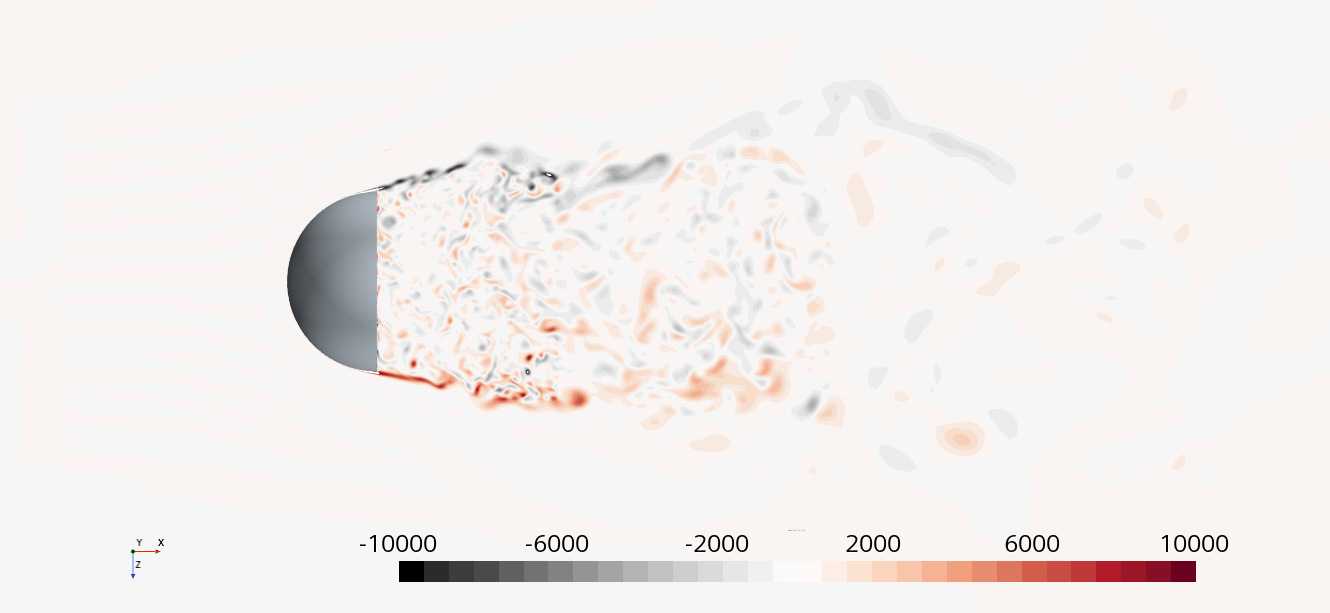}
        }%
        \subfigure[]{%
           \includegraphics[width=0.5\textwidth]{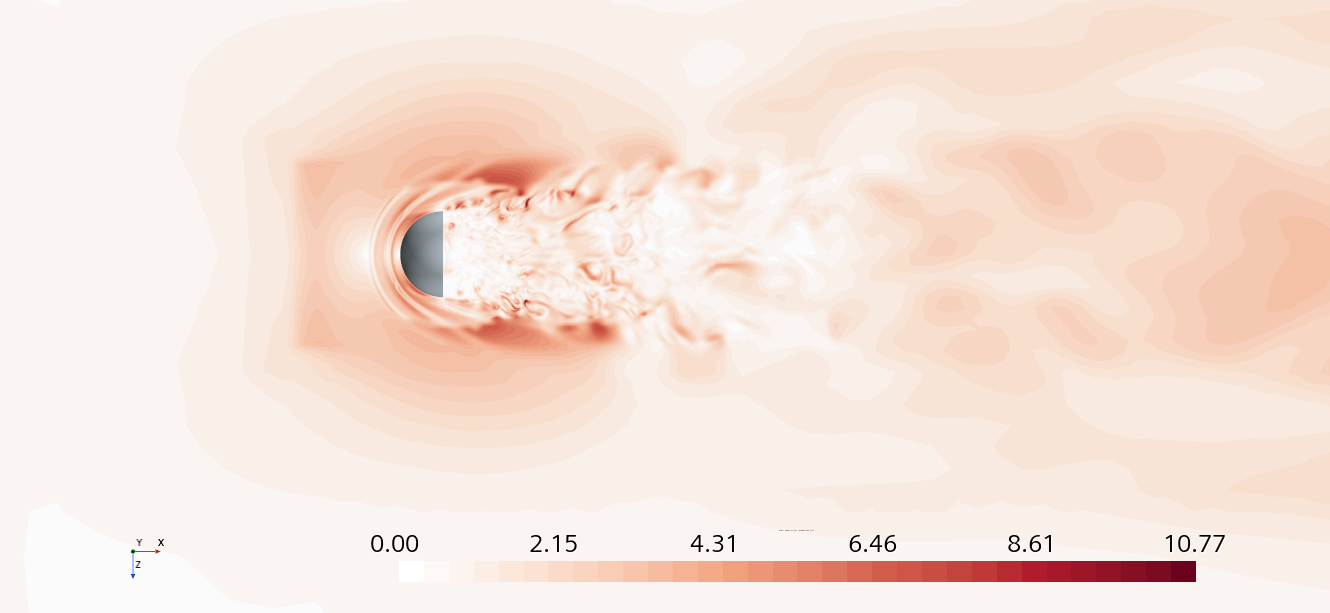}
        }\\
\caption{\small{(a) Instantaneous vorticity field at $y = 0.1D$ and (b) instantaneous wall-shear stress field over the plate (i.e., $y = 0$) for $\mathrm{Re}_D = 5.2\times 10^5$.}}
\label{fig-instantaneous-fields}
\end{center}
\end{figure}

A more thorough analysis of the flow field can be achieved if the instantaneous flow fields are considered. In this section, the instantaneous flow structures obtained from the STRUCT$-\epsilon$ simulations are presented. Figure \ref{fig-instantaneous-fields} (a) and (b) show instantaneous vorticity contours and wall-shear-stress, respectively, illustrating the turbulence source and wall impingement mechanisms. As it has been introduced in fig. \ref{fig-pathlines-re-520}, the impingement of the incoming flow on the side mirror results in a horseshoe vortex that is well identified in fig. \ref{fig-instantaneous-fields} (b), showing a periodicity that has an impact on the SPL spectra measured at sensor \#116 (located upstream of the body). The flow around the side mirror is characterized by large-scale von K\'arm\'an vortices shed from the half cylinder that mix with the upper shear layer from the top of the body. The former are well captured in fig. \ref{fig-instantaneous-fields}, highlighting the capability of the STRUCT$-\epsilon$ turbulence model to reproduce the unsteady flow structures.

\begin{figure}[!htp]
\begin{center}
        \subfigure[]{%
           \includegraphics[width=0.5\textwidth]{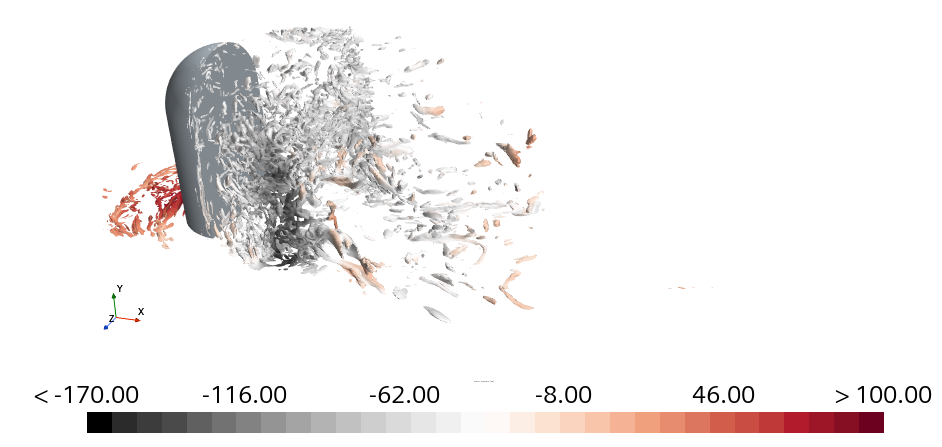}
        }%
        \subfigure[]{%
           \includegraphics[width=0.5\textwidth]{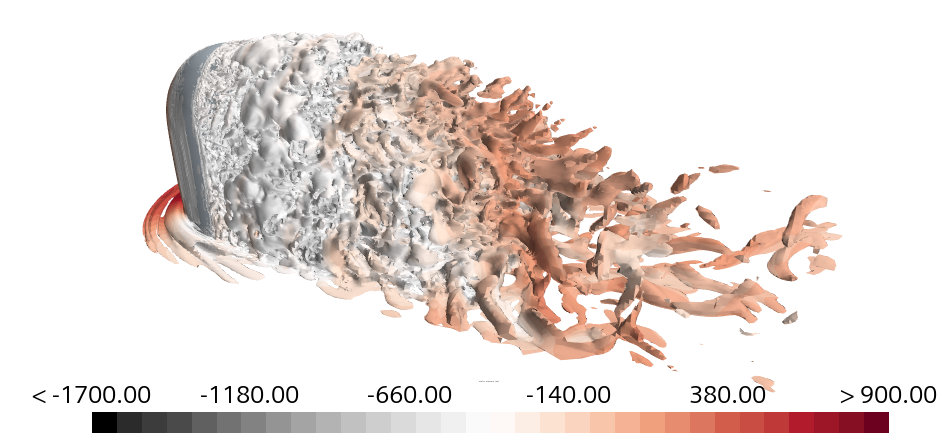}
        }\\
\caption{\small{Iso-surface of $\lambda_2 = 2 \times 10^5\,\mathrm{s^{-2}}$ colored by static pressure (in Pa) for (a) $\mathrm{Re}_D = 1.9\times 10^5$ and (b) $\mathrm{Re}_D = 5.2\times 10^5$, respectively.}}
\label{fig-visualization-lambda-2}
\end{center}
\end{figure}

Figure \ref{fig-visualization-lambda-2} plots the instantaneous flow structures visualized by iso-surfaces of $\lambda_2$. As it is expected, the flow involves three-dimensional separation all along the rear edge of the side mirror, showing how the side and upper shear layers interact and determine the size of the recirculation bubble downstream the body. The two wake branches that are formed downstream the recirculation bubble, as it is introduced in \cite{pap-yao-hd-and-davidson-l-2018-generic-side-mirror-interior-cavity-noise} are slightly observed in fig. \ref{fig-visualization-lambda-2} for both Reynolds numbers. It is clear that all these large eddies captured in fig. \ref{fig-visualization-lambda-2} contribute to the broadband component of the spectrum associated with the predicted transient pressure, \cite{cnf-belamri-t-and-menter-f-2007-sas-generic-side-mirror}, what is presented in the following section. 

\subsection{Sound pressure levels}
\label{sec-sound-pressure-levels}

\begin{figure}[!htp]
\begin{center}
        \subfigure[]{%
           \includegraphics[width=0.4755\textwidth]{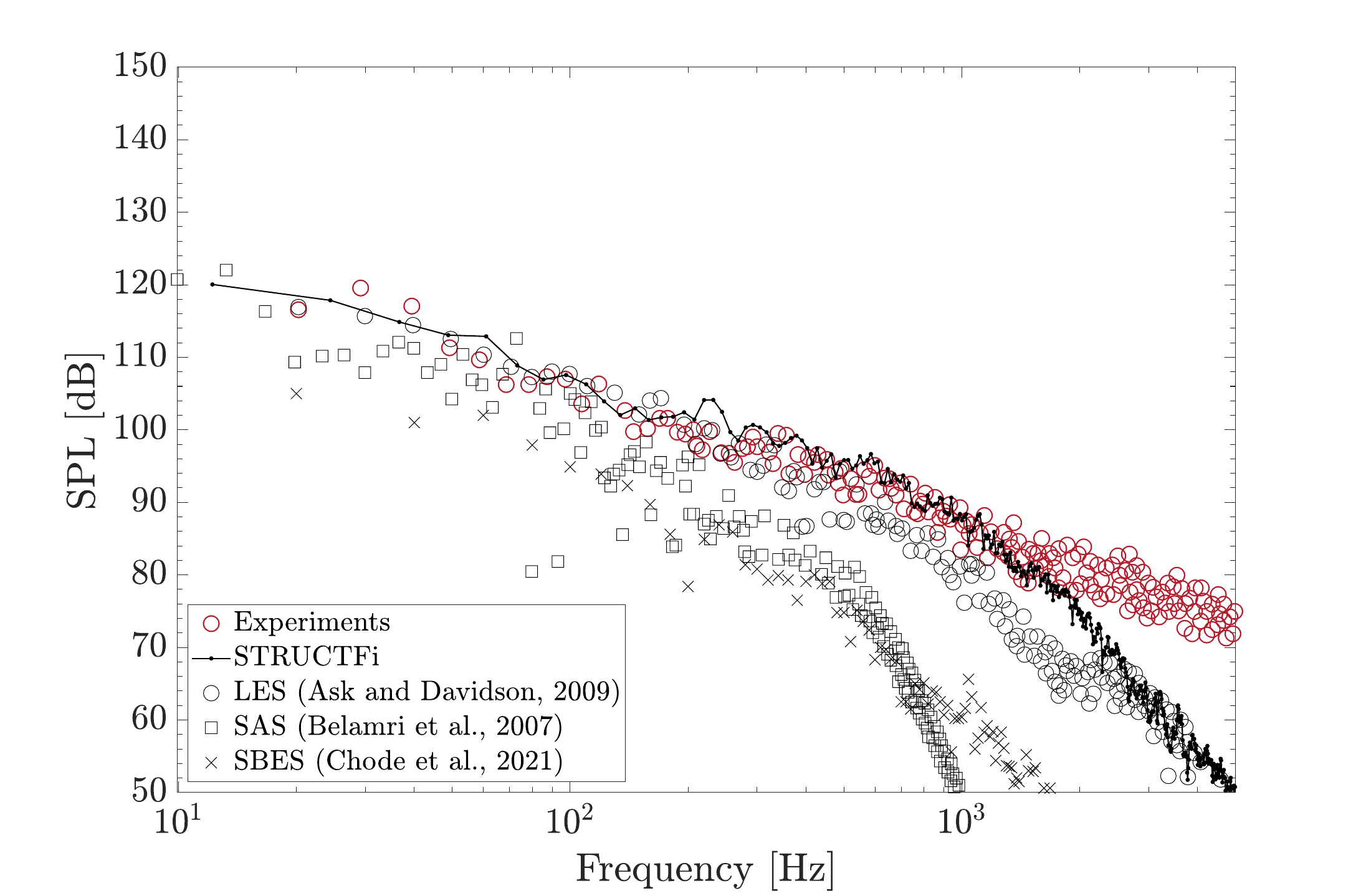}
        }%
        \subfigure[]{%
           \includegraphics[width=0.4755\textwidth]{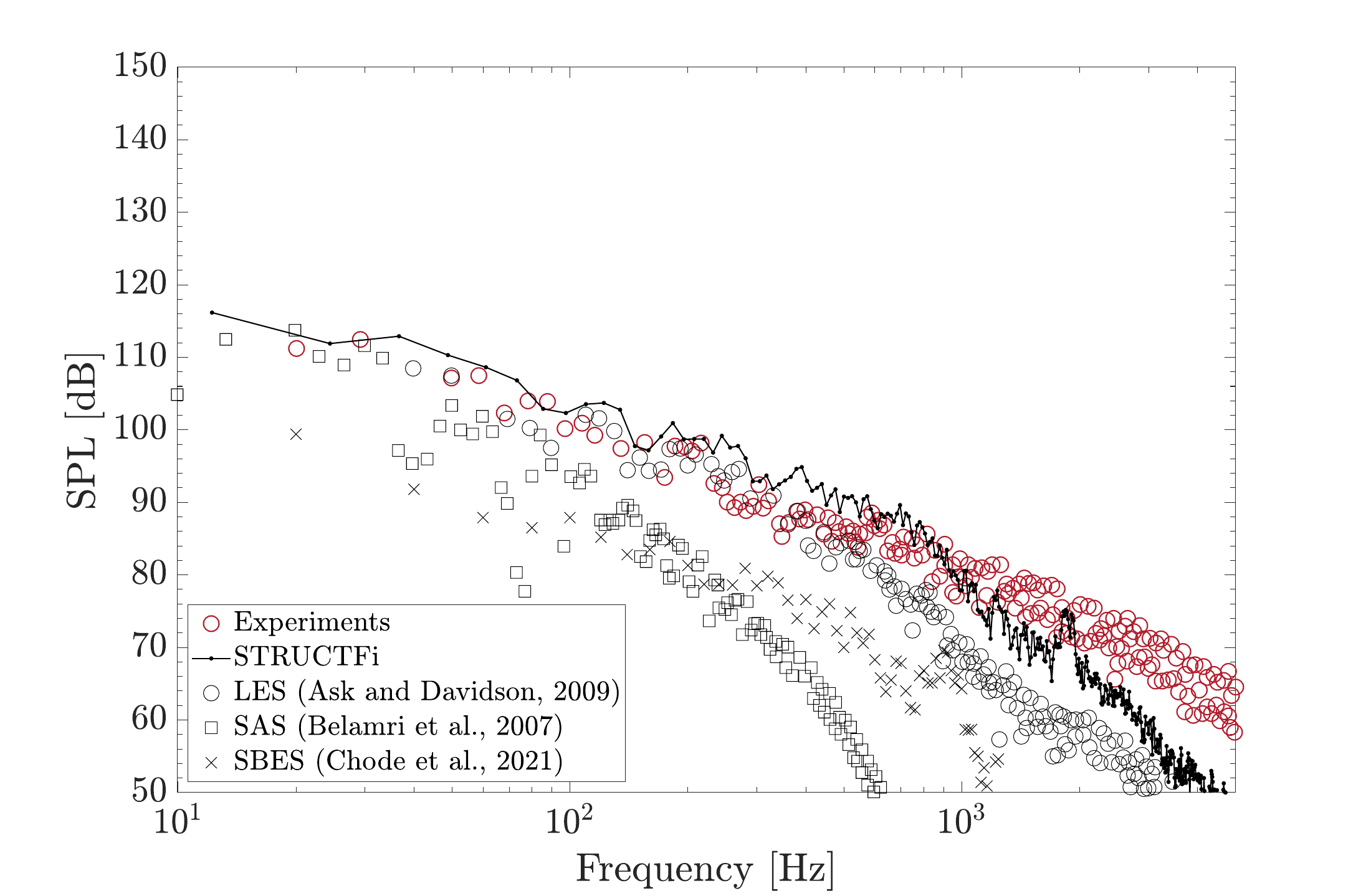}
        }\\
        \subfigure[]{%
           \includegraphics[width=0.4755\textwidth]{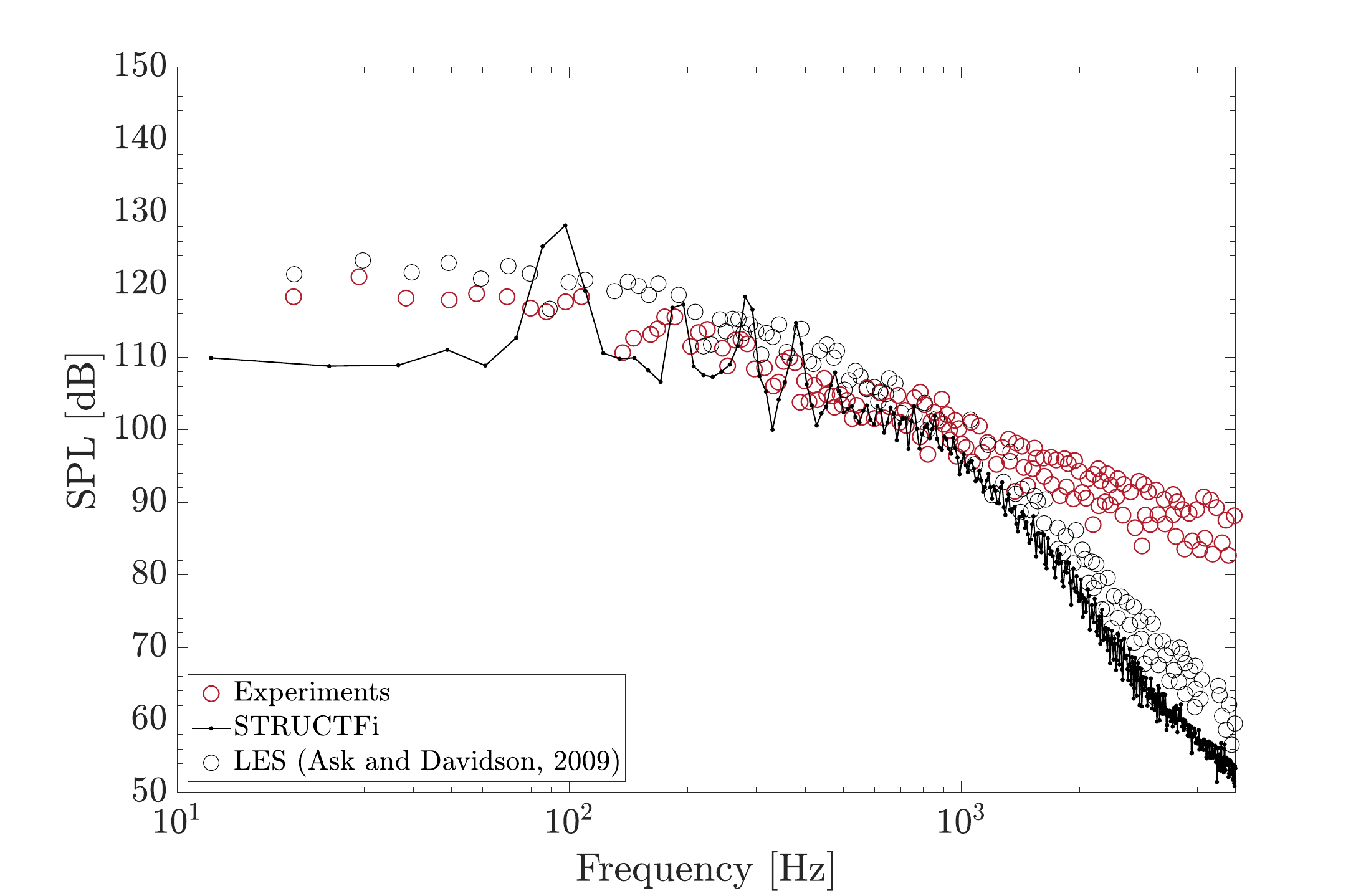}
        }%
        \subfigure[]{%
           \includegraphics[width=0.4755\textwidth]{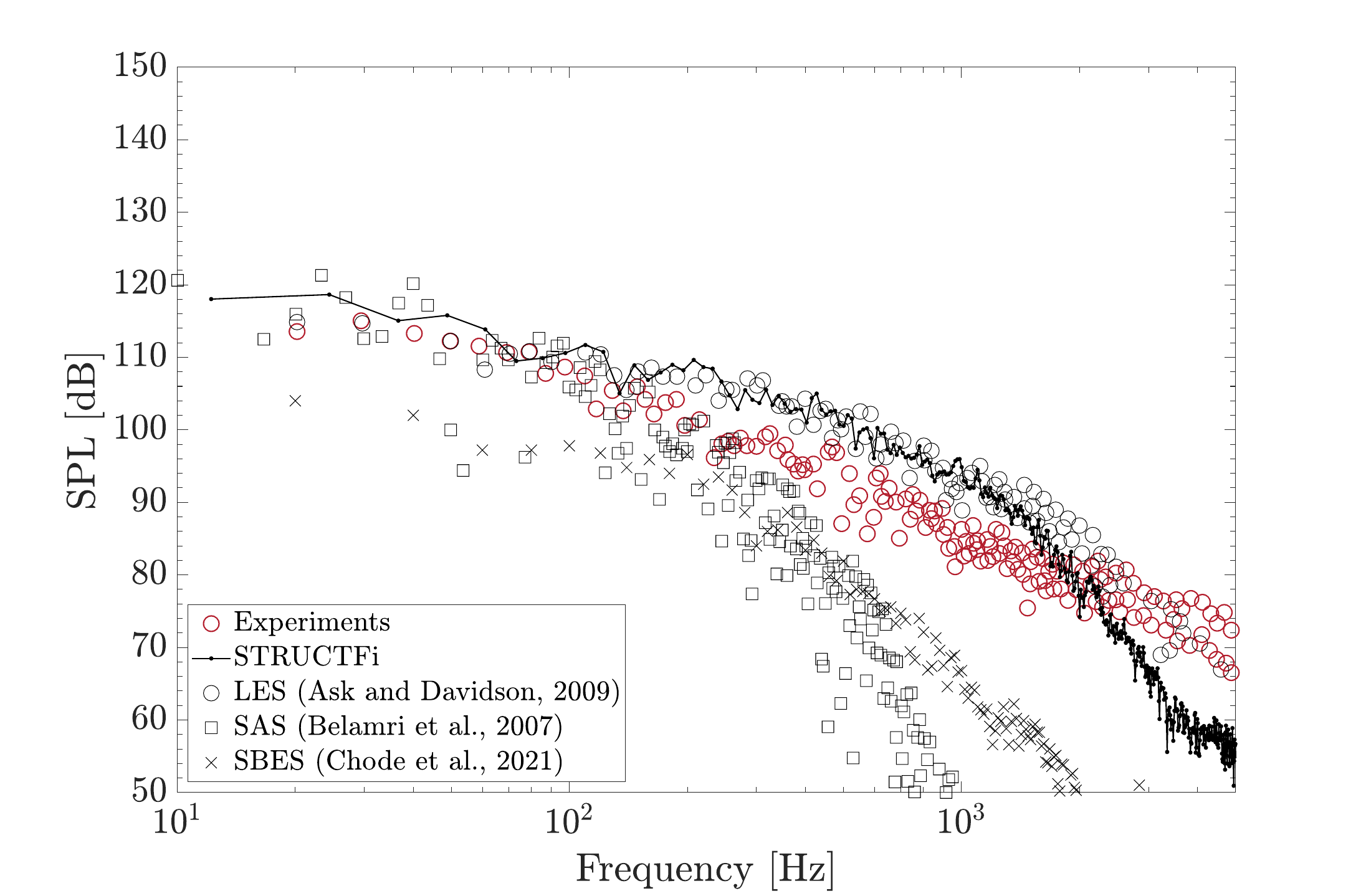}
        }\\
        \subfigure[]{%
           \includegraphics[width=0.4755\textwidth]{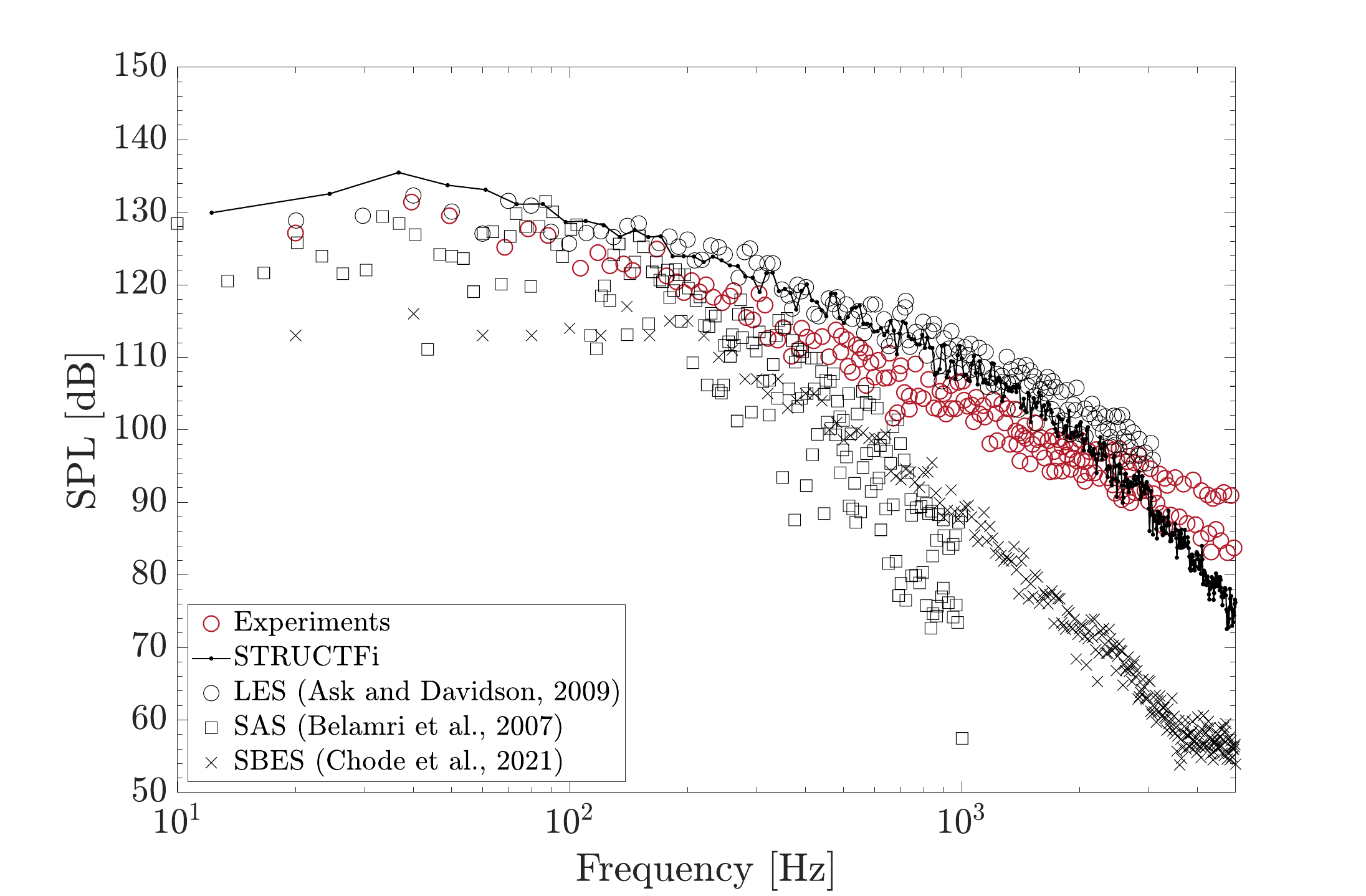}
        }%
        \subfigure[]{%
           \includegraphics[width=0.4755\textwidth]{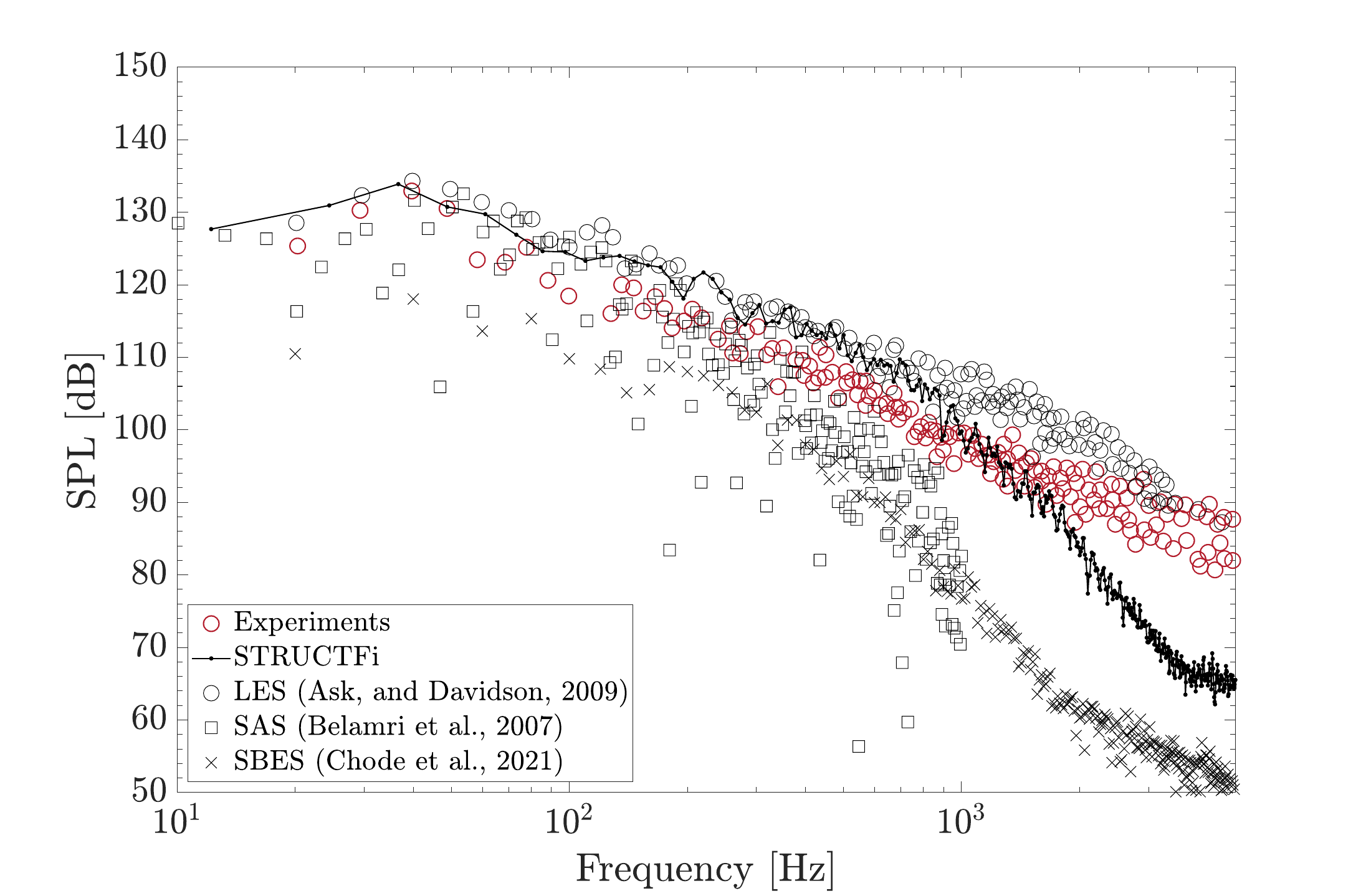}
        }\\
\caption{\small{SPL at surface sensors (a) \#111, (b) \#113, (c) \#116, (d) \#119, (e) \#121 and (f) \#123, respectively. Comparison with the experiments from \cite{cnf-rung-th-et-al-2002-comparison-urans-des-side-mirror}, the LES results from \cite{pap-ask-j-and-davidson-l-2009-les-aeroacoustics-generic-side-mirror}, the SST-SAS from \cite{cnf-belamri-t-and-menter-f-2007-sas-generic-side-mirror} and the SBES from \cite{pap-chode-k-et-al-2021-sbes-generic-side-mirror}.}}
\label{fig-spl}
\end{center}
\end{figure}

The predictive accuracy for the noise generation mechanism depends on the resolution of turbulence/transient interaction since no distinct spectral gap between mean-transient and residual-turbulence phenomena is observed in industrial aeroacoustics, where the relevant frequency range usually spans 5-6 octaves, \cite{cnf-rung-th-et-al-2002-comparison-urans-des-side-mirror}. Considering this particular case, the practically relevant noise frequency goes up to 4 kHz, \cite{cnf-grahs-t-and-othmer-c-2006-sas-and-des-generic-side-mirror-optimization}, so the spectra plotted in figure \ref{fig-spl} are given up to the limit of 5 kHz. Six sensors are presented, namely \#111, \#113, \#116, \#119, \#121 and \#123. These sensors are located in five different regions around the side mirror, namely upstream, back side of the mirror and near, medium and far downstream of the body. The exact position of any of these sensors is given in table \ref{tab-dynamic-pressure-sensors}. The sound pressure level (SPL) is computed as it is given in equation \ref{eq-spl}. The spectra measured at each sensor is compared with the experiments from \cite{cnf-rung-th-et-al-2002-comparison-urans-des-side-mirror} and the numerical results considering the LES turbulence model. 

%but the surface pressure spectra are here plotted only up to the grid-relevant limit of 1 kHz, similarly to \cite{pap-egorov-y-et-al-2010-sas-generic-side-mirror}.
%As it is pointed out in \cite{pap-egorov-y-et-al-2010-sas-generic-side-mirror}, it is not clear the weighted filter applied to the raw experimental data to obtain the spectra. 

Sensors \#111 and \#113 are located at the rear side of the mirror. The SPL measured in these sensors are plotted in figure \ref{fig-spl} (a) and (b), respectively. The latter is found in the upper region while the former is placed at $y/D = 0.5D$. Therefore, the resulting SPL is affected by the observed flow structures, namely upper and side shear layers, \cite{pap-yao-hd-and-davidson-l-2018-generic-side-mirror-interior-cavity-noise}. The side flow structure is well captured in the STRUCTFi simulation, as it has already been evinced in fig. \ref{fig-mean-cp-sensors} for sensor \#26, the nearest static pressure sensor to sensor \#111. The spectra obtained for this sensor is in good agreement with the experiments up to a frequency of $2000\,\mathrm{Hz}$. This behaviour is notably better than the results obtained with SBES in \cite{pap-chode-k-et-al-2021-sbes-generic-side-mirror} or with SAS in \cite{cnf-belamri-t-and-menter-f-2007-sas-generic-side-mirror}, if our turbulence model is compared with equivalent hybrid or second-generation RANS models; and of the same order of accuracy to that from the LES simulations of \cite{pap-ask-j-and-davidson-l-2009-les-aeroacoustics-generic-side-mirror} at a minor computational cost, \cite{pap-garcia-j-et-al-2020-struct-freight-trains}. Meanwhile, the spectra for sensor \#113 obtained in the STRUCTFi simulation overachieves that presented in \cite{pap-ask-j-and-davidson-l-2009-les-aeroacoustics-generic-side-mirror} when compared to the experiments. Therefore, the decay of the SPL is delayed at medium frequencies (in the range of $400$ to $1000\,\mathrm{Hz}$).

Sensor \#116 is located upstream of the mirror along its symmetry plane. Figure \ref{fig-spl} (c) only shows the SPL prediction from the STRUCT$-\epsilon$ turbulence model simulations, the experiments and the LES simulation as there are no further published results considering this sensor. While the prediction of the SPL from the STRUCTFi simulation is comparable to that from the LES case in the frequency range of 400 to 5000 Hz, notable differences are observed in the low frequency range. The location of this sensor is set to capture the horseshoe vortex resulting from the impingement of the incoming flow in the side mirror, and this vortex is responsible for fluctuation levels of pressure larger than the maximum values at the mirror back side. Consequently, the SPL obtained from the STRUCTFi simulation results into large peaks at frequencies up to $400\,\mathrm{Hz}$. Nevertheless, as it is reported in \cite{pap-ask-j-and-davidson-l-2009-les-aeroacoustics-generic-side-mirror}, a flatter spectrum is observed at low frequencies for this sensor \#116 compared to the sensors located downstream of the side mirror. Besides, the decay of the fluctuations is larger at this sensor than in any other reported sensor in this work, which is also highlighted in the work of \cite{pap-ask-j-and-davidson-l-2009-les-aeroacoustics-generic-side-mirror}. Thus, the discrepancies at low frequencies should not invalidate the STRUCT$-\epsilon$ turbulence model performance. It is worth noting that the region upstream of the mirror is of less importance compared to the downstream region for aeroacoustics purposes, as the unsteady flow interacts with the side windows of the car, \cite{cnf-belamri-t-and-menter-f-2007-sas-generic-side-mirror}.

%The shape of the horseshoe is influenced by the incoming boundary layer \cite{pap-martinuzzi-r-and-tropea-c-1993-horseshoe-vortex}.

Sensor \#119 is located downstream of the mirror along its symmetry plane. The agreement with the experimental data is reasonable at lower frequencies, where the deviation is about $2-4\,\mathrm{dB}$ for the STRUCTFi simulation. Meanwhile, at higher frequencies, the deviation is of the order of that presented in \cite{pap-ask-j-and-davidson-l-2009-les-aeroacoustics-generic-side-mirror}, overpredicting the SPL. 

% \begin{figure}[!ht]
% \begin{center}
% \includegraphics[width = 0.7\textwidth]{figures/fig-validation-re-19.pdf}
% \caption{\small{Dimensions of the train model considered in this paper.}}
% \label{fig-sensors-positions}
% \end{center}
% \end{figure}
% 
% \begin{figure}[!ht]
% \begin{center}
% \includegraphics[width = 0.7\textwidth]{figures/fig-validation-re-19_v2.pdf}
% \caption{\small{Dimensions of the train model considered in this paper.}}
% \label{fig-sensors-positions}
% \end{center}
% \end{figure}

% \begin{figure}[!ht]
% \begin{center}
% \includegraphics[width = 1.05\textwidth]{figures/fig-spl-s111-s114.png}
% \caption{\small{RMS of wall pressure (Pa) over the plate for (a) \cite{pap-ask-j-and-davidson-l-2009-les-aeroacoustics-generic-side-mirror}, (b) STRUCTCo case and (c) STRUCTFi case, respectively.}}
% \label{fig-spl-s111-s114}
% \end{center}
% \end{figure}
% 
% \begin{figure}[!ht]
% \begin{center}
% \includegraphics[width = 1.05\textwidth]{figures/fig-spl-s120-s121.png}
% \caption{\small{RMS of wall pressure (Pa) over the plate for (a) \cite{pap-ask-j-and-davidson-l-2009-les-aeroacoustics-generic-side-mirror}, (b) STRUCTCo case and (c) STRUCTFi case, respectively.}}
% \label{fig-spl-s120-s121}
% \end{center}
% \end{figure}

The sensors located at the footprint of the vortex detached from the cylindrical part of the side mirror and presented in this paper are the sensors \#121 and \#123. These correspond to medium- and far-downstream zones (of the side mirror), respectively. The local refinement considered in this region helps resolving the turbulent structures and so the noise generated from this. Figures \ref{fig-spl} (e) and (f) present the SPL obtained for sensors \#121 and \#123, respectively. Again, there is a good agreement with the results obtained from the LES simulations, improving the response from SAS or SBES turbulence models. Indeed, the SPL decay occurs far later than in any of the latter turbulence models (about $1000\,\mathrm{Hz}$ compared to $500\,\mathrm{Hz}$ approximately from SAS or SBES). A peak at low frequency is observed, apparently due to large-scale periodic shedding of vortices, \cite{pap-kato-c-et-al-2007-generic-side-mirror-reynolds-effect-yaw-angle}. It results into a Strouhal number, based on the GSM diameter and the freestream velocity, of $\mathrm{St} = 0.187$. This result is in very good agreement with the Strouhal number obtained in \cite{pap-ask-j-and-davidson-l-2009-les-aeroacoustics-generic-side-mirror} of $0.19$, and similar to usual von K\'arm\'an vortices ($\mathrm{St} = 0.18 - 0.20$) for this Reynolds number range.

\subsection{Spectral Proper-Orthogonal Decomposition}

\begin{figure}[!htp]
\begin{center}
        \subfigure[]{%
           \includegraphics[height=0.325\textwidth]{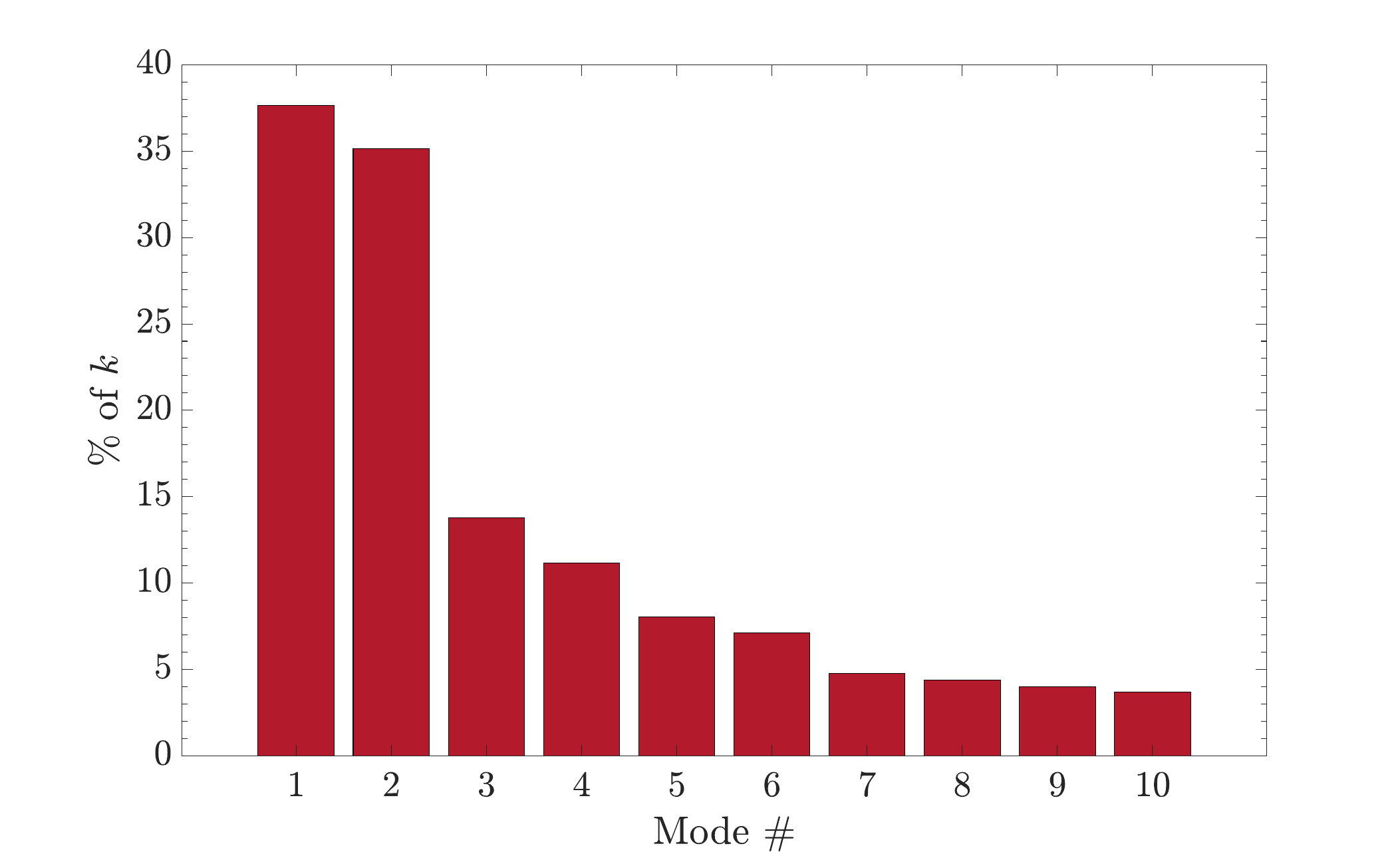}
        }%
        \hspace{-0.5cm}
        \subfigure[]{%
           \includegraphics[height=0.325\textwidth]{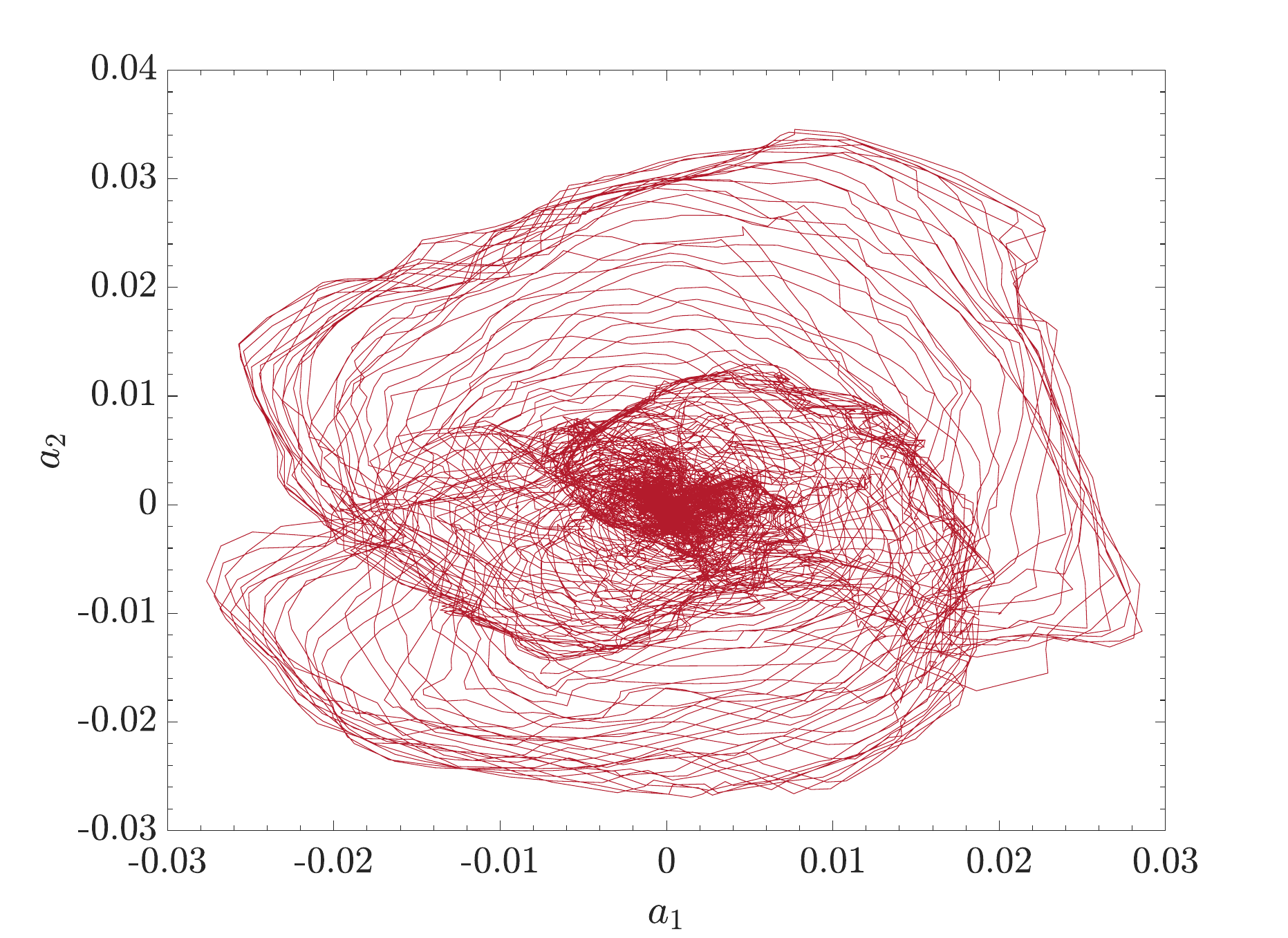}
        }\\

\caption{\small{(a) Contribution (in \%) of each mode to the turbulent kinetic energy of the flow field, and (b) phase portraits (Lissajous figures) of the temporal coefficients of the two most energetic SPOD modes.}}
\label{fig-pod-analysis}
\end{center}
\end{figure}

\begin{figure}[!htp]
\begin{center}
        \subfigure[]{%
           \includegraphics[width=0.40495\textwidth]{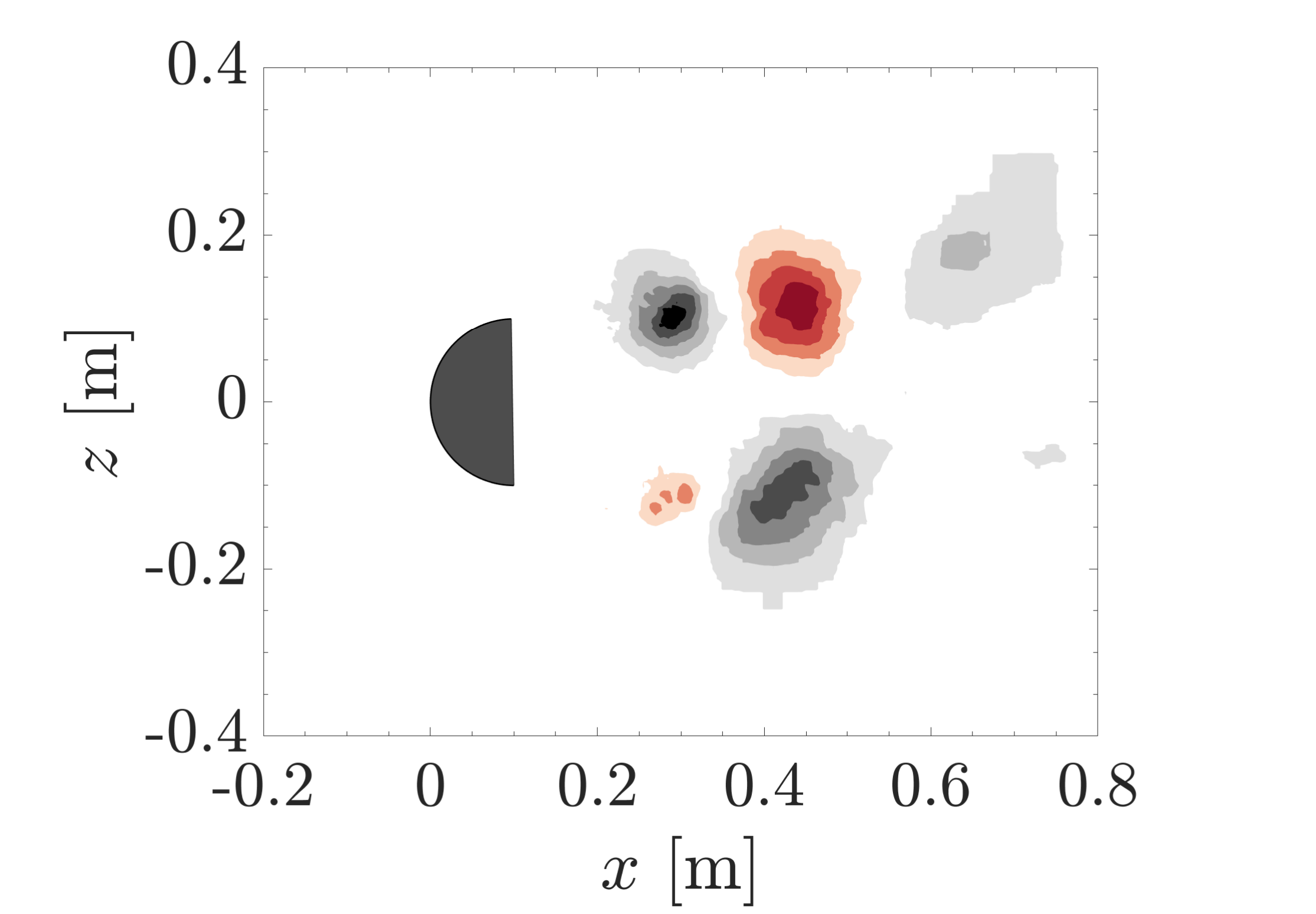}
        }%
                \hspace{-0.5cm}
        \subfigure[]{%
           \includegraphics[width=0.40495\textwidth]{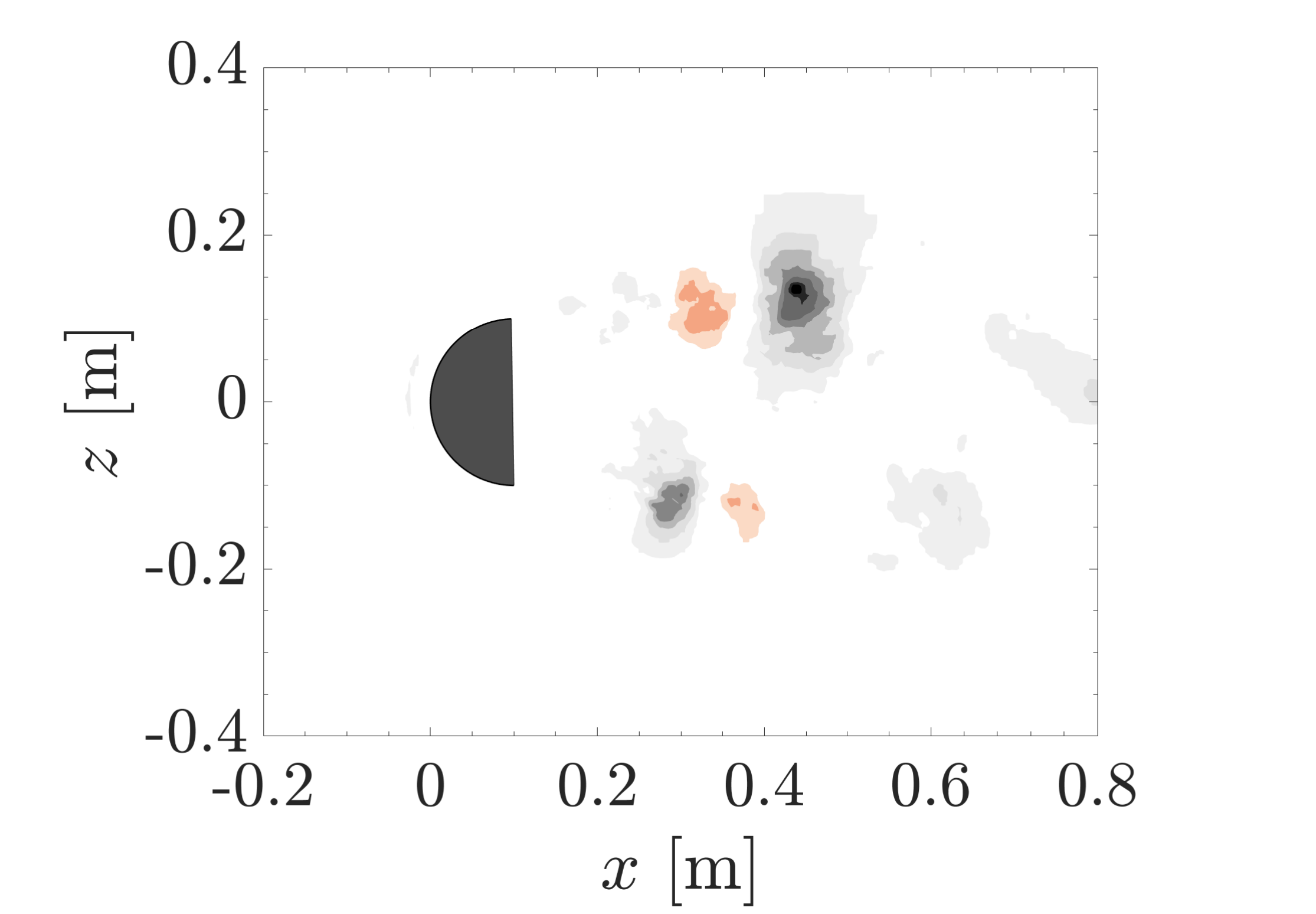}
        }\\
                \hspace{-0.5cm}
        \subfigure[]{%
           \includegraphics[width=0.40495\textwidth]{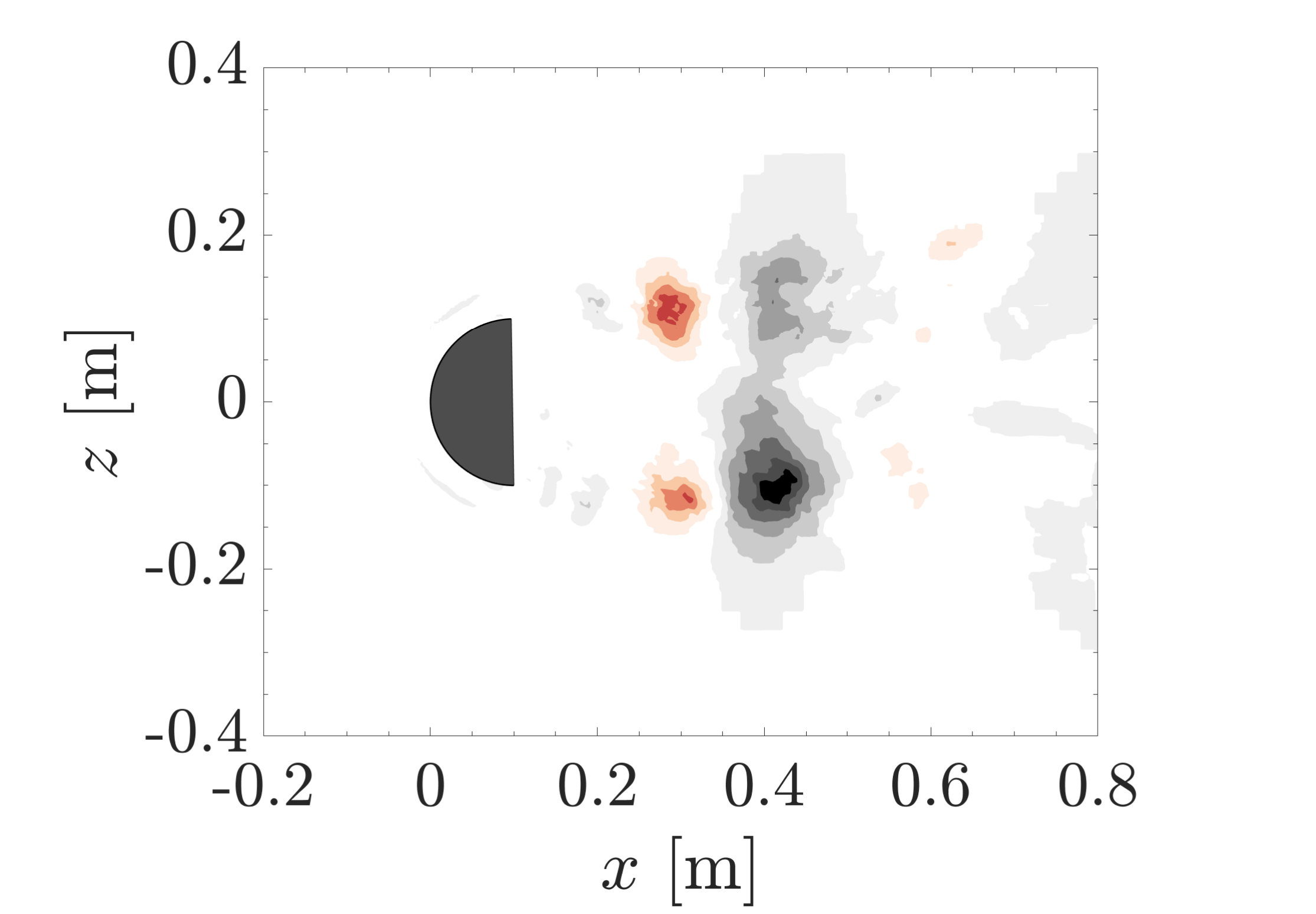}
        }%
                \hspace{-0.5cm}
        \subfigure[]{%
           \includegraphics[width=0.40495\textwidth]{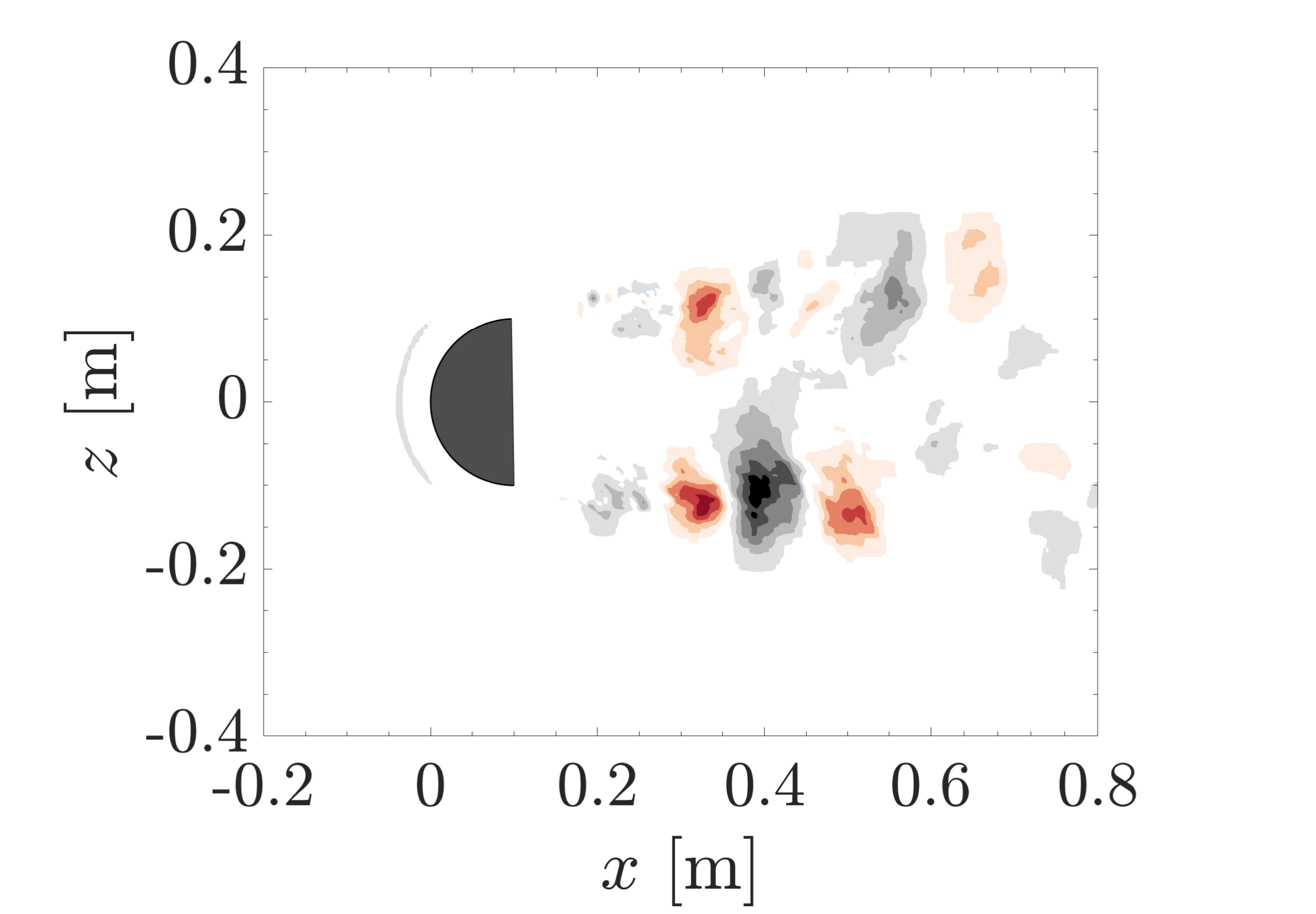}
        }\\
        \subfigure[]{%
           \includegraphics[width=0.40495\textwidth]{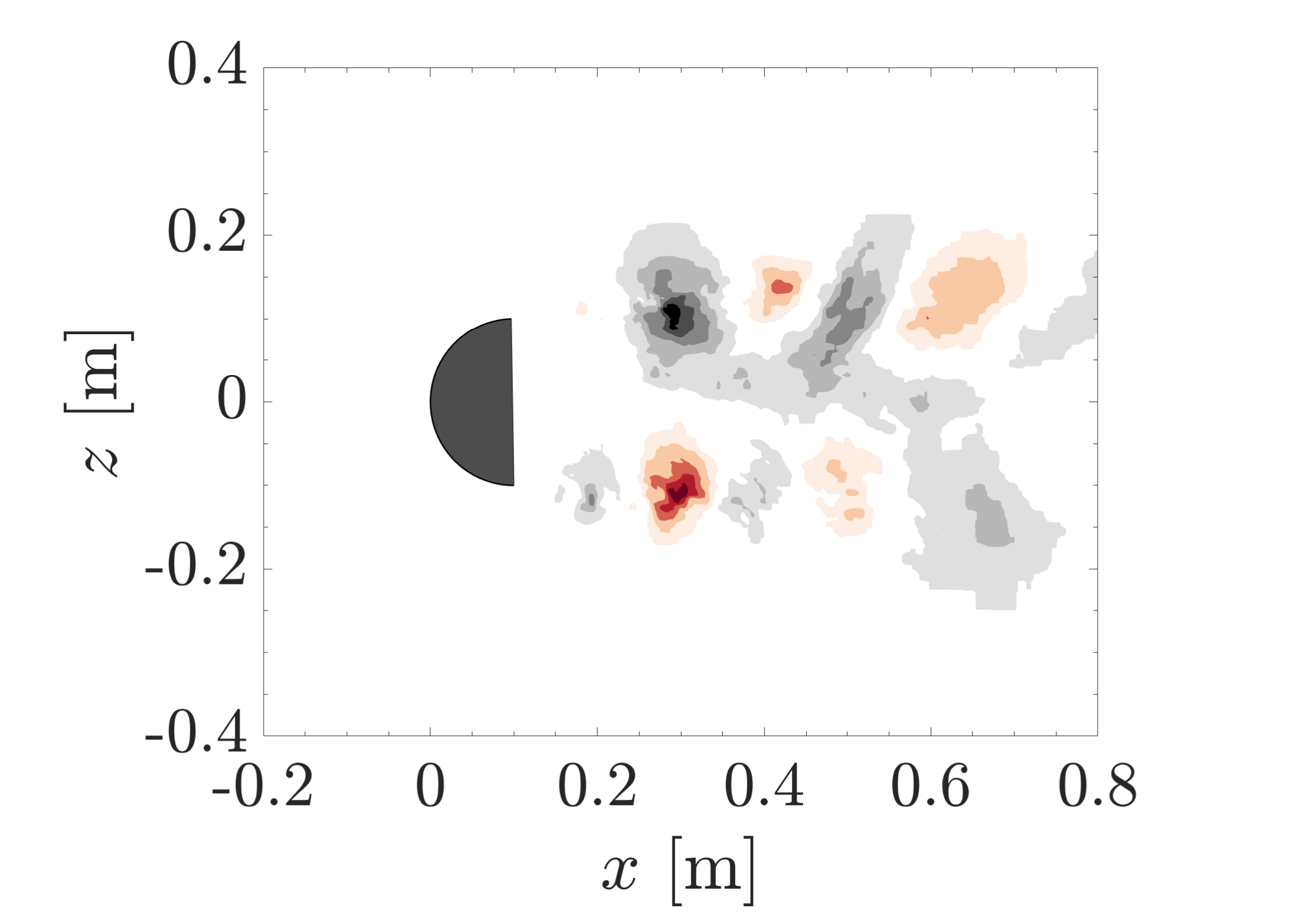}
        }%
                \hspace{-0.5cm}
        \subfigure[]{%
           \includegraphics[width=0.40495\textwidth]{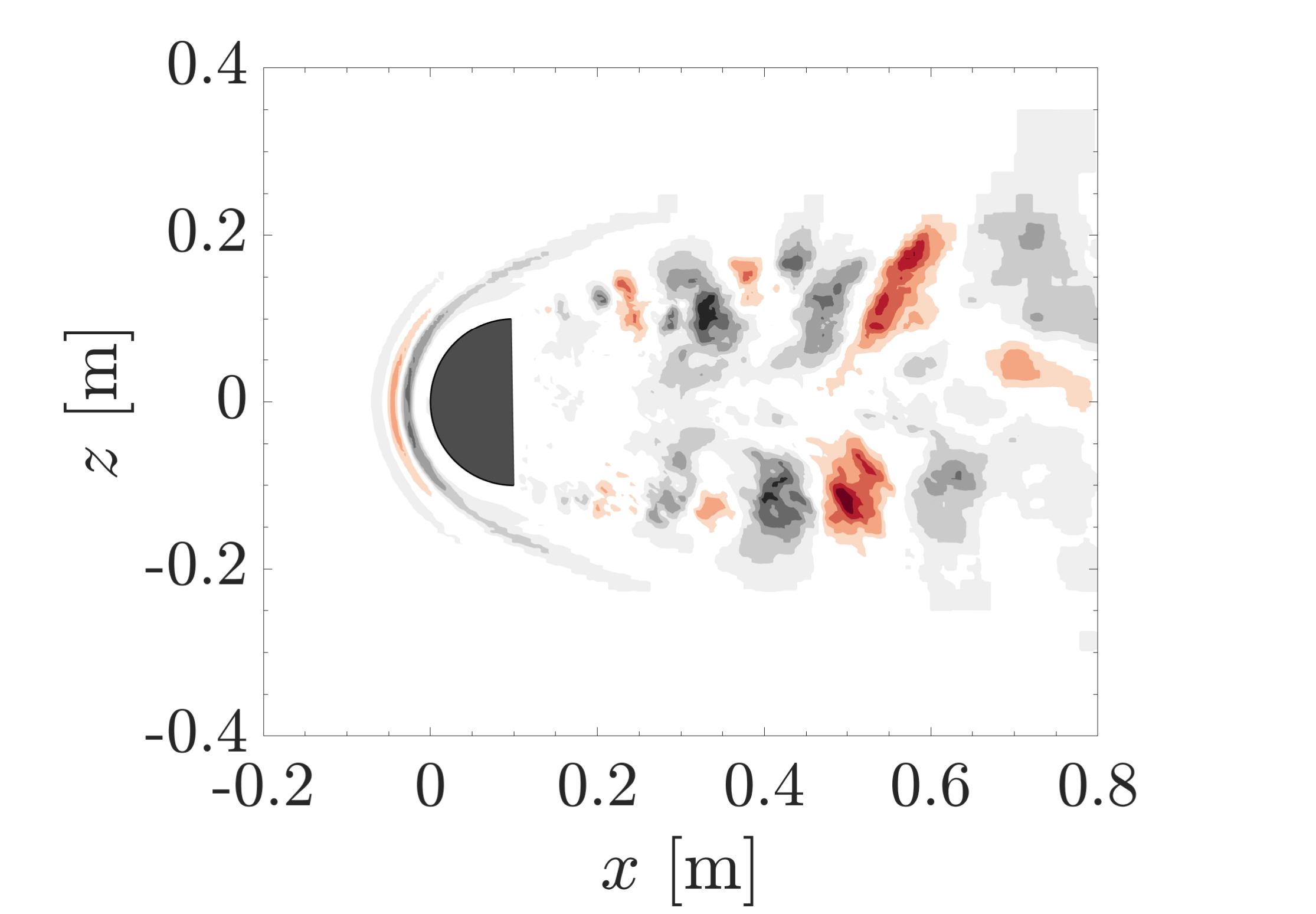}
        }\\
                \hspace{-0.5cm}
        \subfigure[]{%
           \includegraphics[width=0.40495\textwidth]{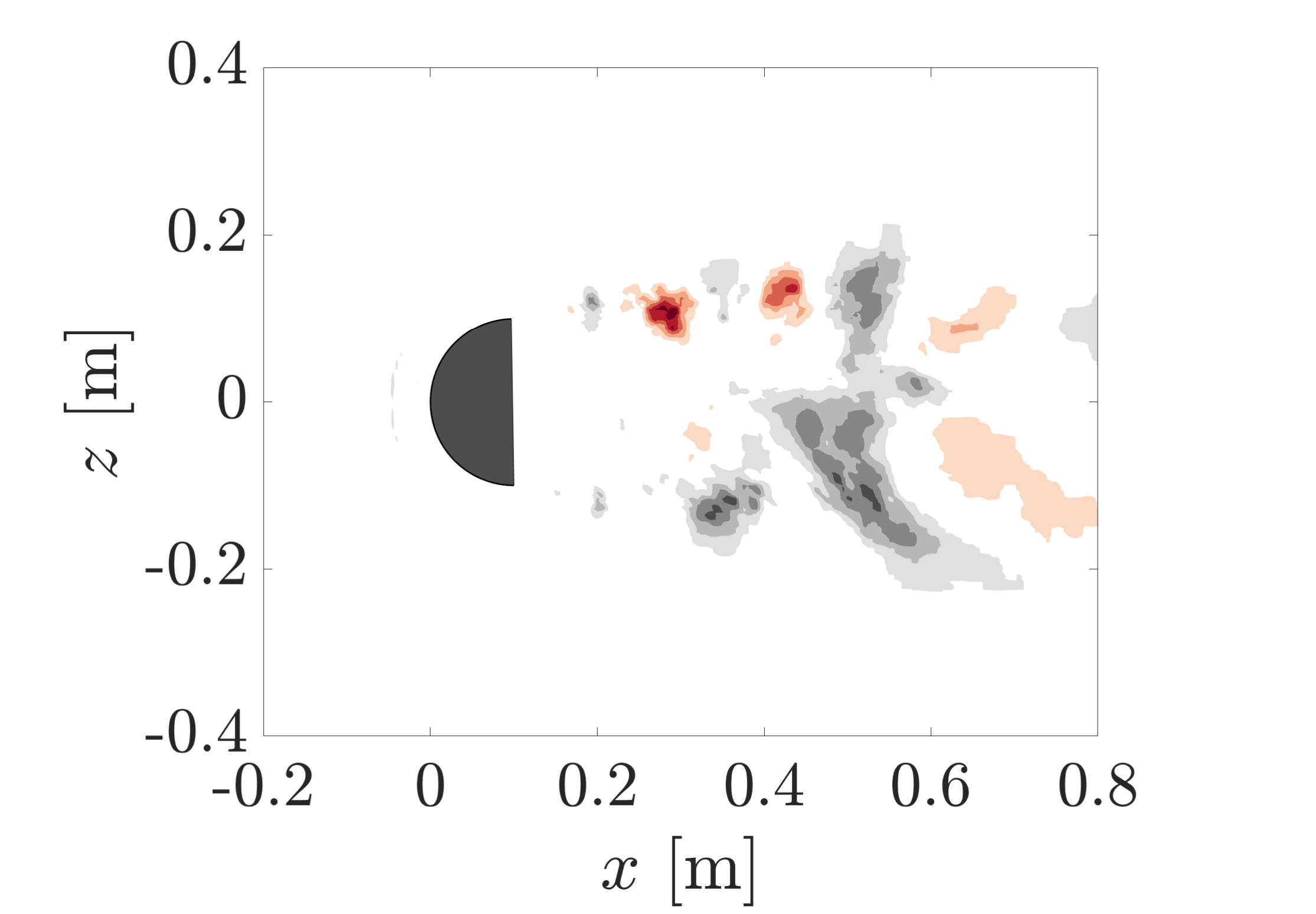}
        }%
                \hspace{-0.5cm}
        \subfigure[]{%
           \includegraphics[width=0.40495\textwidth]{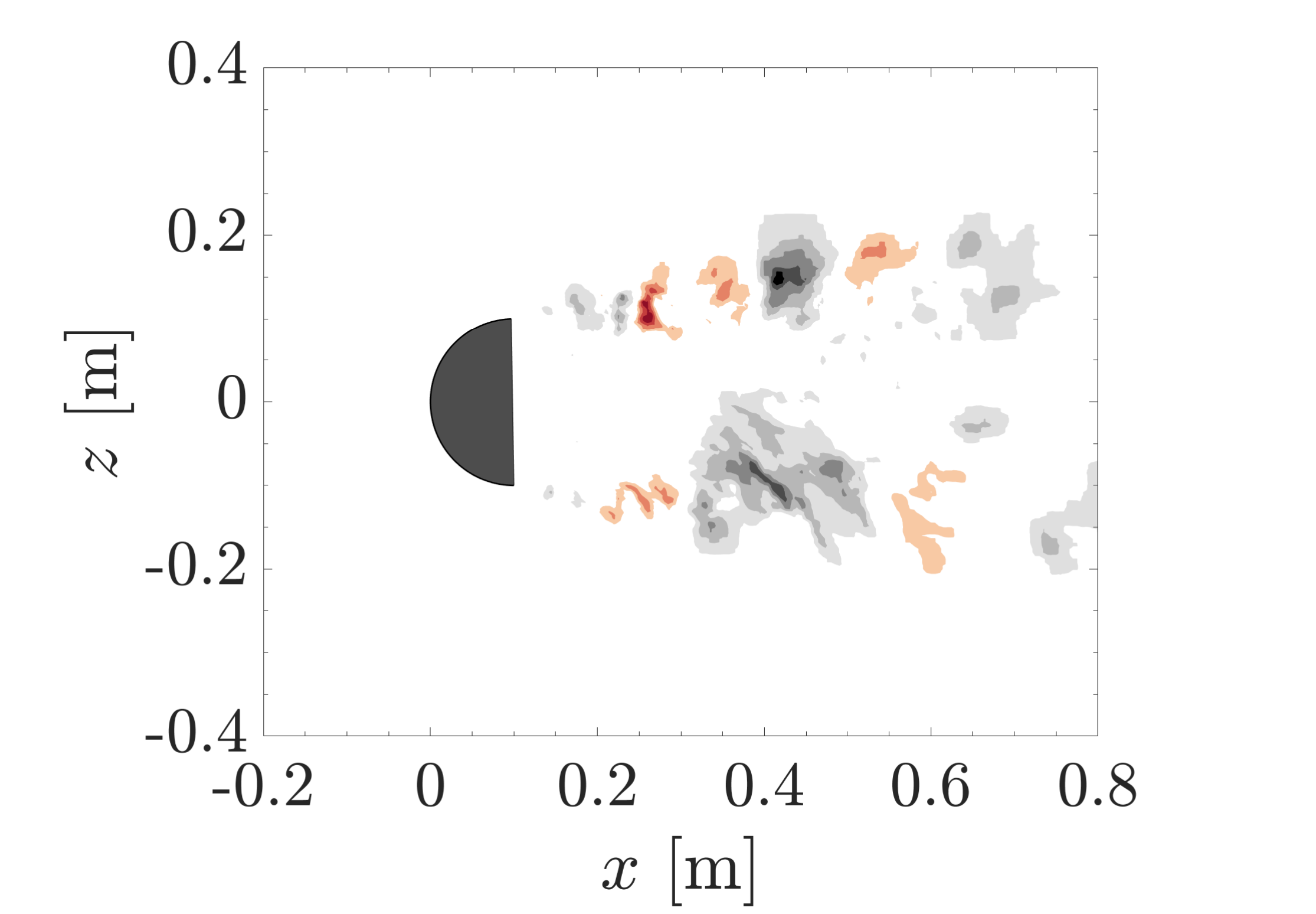}
        }\\
\caption{\small{First four (most-energetic) SPOD modes, ordered from left to right. From top to bottom, at frequencies $f = 62.50\,\mathrm{Hz}$ and $125.00\,\mathrm{Hz}$, respectively.}}
\label{fig-pod-modes}
\end{center}
\end{figure}

The spectral analysis presented in section \ref{sec-sound-pressure-levels} clearly puts in evidence the huge range of temporal and spatial scales typical of turbulent flows. This feature difficults to interpret and overview the vast amount of data produced from high-fidelity CFD simulations, being a key challenge to distinct the deterministic coherent motion from purely stochastic motion. The lack of ability to represents the dynamics of flow structures that evolve coherently in space and time restricts the understanding of flow physics. Among the different alternatives available to deal with this issue, Spectral Proper Orthogonal Decomposion (SPOD) has revealed as an optimal solution. Based on the work of \cite{lib-lumley-j-2007-stochastic-tools-in-turbulence} and proposed in \cite{pap-sieber-m-et-al-2016-spod}, this method has been extensively applied in fluid mechanics and also in ground vehicles, \cite{pap-haffner-y-et-al-2020-spod-ahmed-body-drag-reduction-transient-bimodal-near-wake} and \cite{pap-li-x-et-al-2021-spod-hst} for example. This method allows us to identify energy-ranked modes that each oscillate at a single frequency, are orthogonal to all other modes at the same frequency and, as a set, optimally represent the space-time flow statistics, \cite{pap-haffner-y-et-al-2020-spod-ahmed-body-drag-reduction-transient-bimodal-near-wake}. Figure \ref{fig-pod-analysis}(a) shows the contribution in \% of each mode to the turbulent kinetic energy $k$ of the flow field. It is clear that the first two modes are dominant in terms of energy, the second one still tripling the third mode. The modes are organized in pairs that have similar energy and, as it is observed laterly in \ref{fig-pod-modes}, also similar structure sizes but phase shifted. Figure \ref{fig-pod-analysis}(b) shows the phase portraits (Lissajous figures) of the temporal coefficients of the two most energetic SPOD modes. In the case of a periodic ideal mode with no variation in amplitude and a constant frequency, the phase portrait will show a perfect circle, \cite{pap-rovira-m-et-al-2021-pod-turbulent-jet}. Here it is made out two circles which intersect for a certain period of time. The maxima and minima of one coefficient are slightly skewed from the zeros of the other, but the phase shift is yet closer to $\frac{\pi}{2}$. 

The first four POD modes contain up to $97.7\%$ of the total turbulent kinetic energy, and these modes are considered in fig. \ref{fig-pod-modes} to illustrate the spatial modes at two different frequencies, plotting in $y=0$ plane. The first two modes have a mode shape typical of a vortex shedding phenomena. Indeed, the study of the tonal noise of an actual car side mirror of \cite{pap-werner-m-et-al-2017-tonal-noise-side-mirror-spod} link the first two modes to the long wavelength and low-frequency flapping movement of the shear layer. The second pair of modes (namely third and fourth most energetic modes) describe a traveling vortical structure, \cite{pap-rempfer-d-and-fasel-hf-1994-evolution-coherent-structures-flat-plate-boundary-layer}, \cite{pap-manhart-m-vortex-shedding-hemisphere}. Both pairs of modes show an antisymmetric behaviour.

\section{Conclusions}
\label{sec:summary} 

The second-generation URANS closure STRUCT$-\epsilon$ has been assessed in this work on the simulation of aerodynamic flow around a generic side mirror. Besides, the aeroacoustic capabilities of this turbulence model have also been tested. This new approach aims at advancing the applicability and robustness of hybrid turbulence models by introducing local resolution of unsteady flow structures in areas of timescale overlap, while
not leveraging any grid dependent parameter. This feature is evinced in the mesh sensitivity analysis, where not significant differences are observed when compared the experiments with the simulations using the middle and fine mesh. Nevertheless, the finest mesh get closer to the experimental data and is in good agreement of prediction with that obtained considering incompressible and compressible LES turbulence models, respectively. However, the computational cost (in terms of number of cells) is found to be lower. STRUCT$-\epsilon$ successfully predicts the flow separation, reattachment and vortex unsteadiness observed in the experiment. The turbulence model has also been evaluated in comparison to other hybrid methods, both blending, interfacing or other second-generation URANS turbulence models that use a length scale to define the damping function rather than the turbulent time scale as the STRUCT model does. The response of STRUCT$-\epsilon$ has overachieved the aforementioned turbulence models when these are compared to the experiments, regarding the pressure distribution along the side mirror surface and the sound presure level measured by static and dynamic pressure sensors.

The notable cost reduction, coupled to the robust mesh independence of the model, can support effective aeroacoustic design applications, making it a promising solution for aerodynamic and aeroacoustic optimization.

\section{Acknowledgement}

This work was supported by the Ministerio de Ciencia e Innovaci\'on, the Agencia Estatal de
Investigaci\'on and the FEDER [grant number PID2021-122237OA-I00]. In addition, J. Garc\'ia would
like to thank the Programa de Excelencia para el Profesorado Universitario de la Comunidad de Madrid
for their financial support.

\section{Declarations}

Conflict of interest The authors declare that they have no confict of interest.

\section{Appendix}

\begin{figure}[!h]
\begin{center}
\includegraphics[width=1\textwidth]{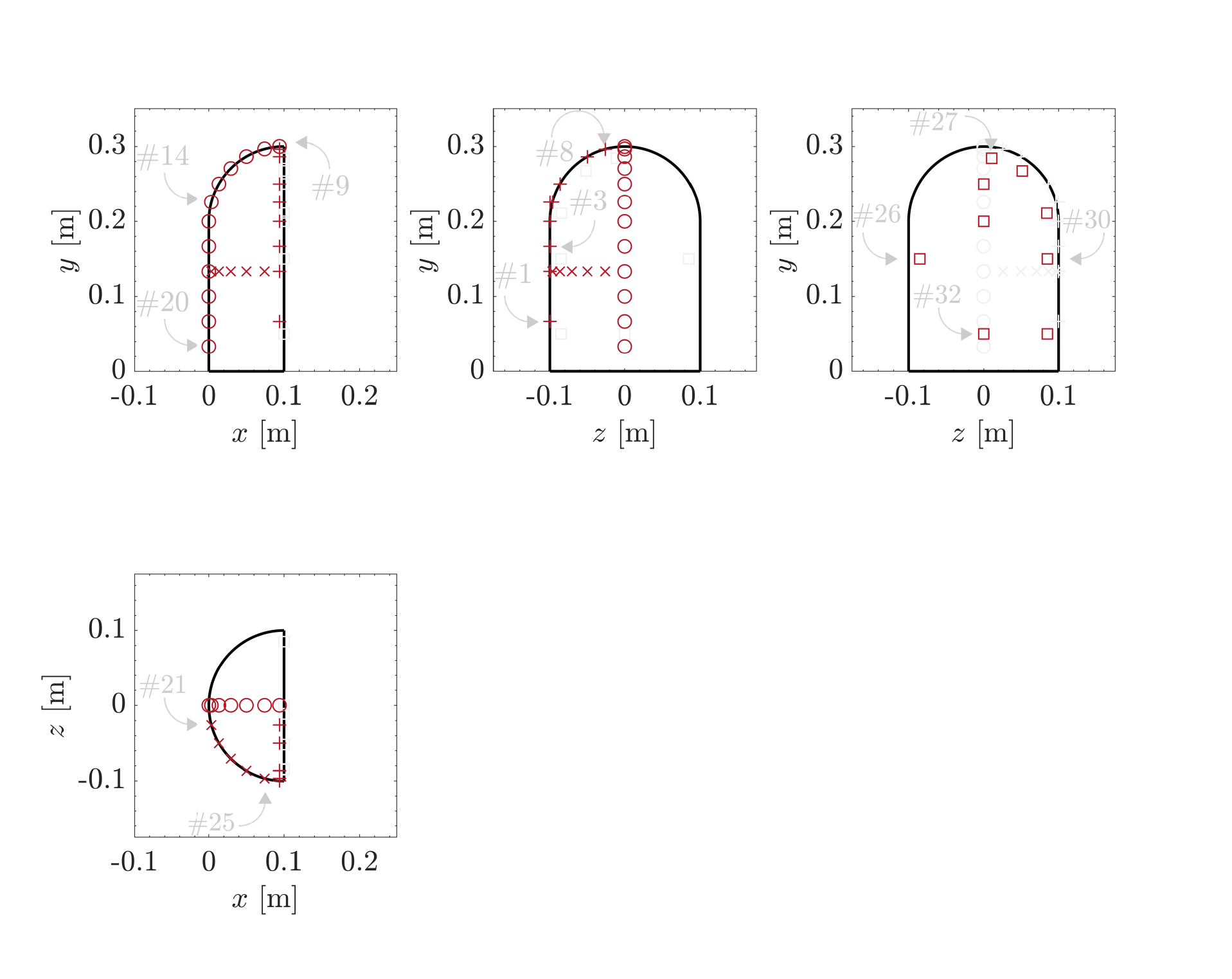}
\caption{\small{From left to right, and from top to bottom, side, front, back and top view of the generic side mirror and the static pressure sensors. Three lines of interest are identified, each one indicated by a different symbol, namely stagnation line ($\Circle$), separation line ($+$) and polar line ($\times$). Sensors located at the rear face of the mirror are plotted with $\square$.}}
\label{fig-position-sensors}
\end{center}
\end{figure}

Figure \ref{fig-position-sensors} shows the locations of the static pressure sensors on the generic side mirror. For the sake of clarity, sensors are grouped into four sets that let studying different flow features: stagnation, separation and recirculation in the rear side. Each of these sets are plotted with a different symbol. The coordinates of the location of these static pressure sensors, as well as the dynamic pressure sensors, are given in table \ref{tab-static-pressure-sensors} and \ref{tab-dynamic-pressure-sensors}, respectively.

\newpage

\newpage
\noindent\begin{minipage}[!ht]{0.6\textwidth}%
              \centering
                \captionof{table}{Static pressure sensors}
                \label{tab-static-pressure-sensors}
                \begin{tabular}{c r r r}
                        \hline
                        \hline
                         Sensor & $x$ [m] & $y$ [m] & $z$ [m]
                        \\\hline
                        \#1  & $0.0940$ & $0.0667$ & $-0.0998$ \\
                        \#2  & $0.0940$ & $0.1333$ & $-0.0998$ \\
                        \#3  & $0.0940$ & $0.1667$ & $-0.0998$ \\
                        \#4  & $0.0940$ & $0.2000$ & $-0.0998$ \\ 
                        \#5  & $0.0940$ & $0.2258$ & $-0.0964$ \\ 
                        \#6  & $0.0940$ & $0.2499$ & $-0.0864$ \\
                        \#7  & $0.0940$ & $0.2864$ & $-0.0499$ \\
                        \#8  & $0.0940$ & $0.2964$ & $-0.0258$ \\
                        \#9  & $0.0940$ & $0.2998$ & $0.0000$ \\ 
                        \#10  & $0.0741$ & $0.2966$ & $0.0000$ \\
                        \#11  & $0.0500$ & $0.2866$ & $0.0000$ \\
                        \#12  & $0.0293$ & $0.2707$ & $0.0000$ \\
                        \#13  & $0.0134$ & $0.2500$ & $0.0000$ \\
                        \#14  & $0.0034$ & $0.2259$ & $0.0000$ \\
                        \#15  & $0.0000$ & $0.2000$ & $0.0000$ \\
                        \#16  & $0.0000$ & $0.1667$ & $0.0000$ \\
                        \#17  & $0.0000$ & $0.1333$ & $0.0000$ \\
                        \#18  & $0.0000$ & $0.1000$ & $0.0000$ \\
                        \#19  & $0.0000$ & $0.0667$ & $0.0000$ \\
                        \#20  & $0.0000$ & $0.0333$ & $0.0000$ \\
                        \#21  & $0.0034$ & $0.1333$ & $-0.0259$ \\
                        \#22  & $0.0134$ & $0.1333$ & $-0.0500$ \\
                        \#23  & $0.0293$ & $0.1333$ & $-0.0707$ \\
                        \#24  & $0.0500$ & $0.1333$ & $-0.0866$ \\
                        \#25  & $0.0741$ & $0.1333$ & $-0.0966$ \\
                        \#26  & $0.1000$ & $0.1500$ & $0.0850$ \\
                        \#27  & $0.1000$ & $0.2843$ & $-0.0111$ \\
                        \#28  & $0.1000$ & $0.2674$ & $-0.0517$ \\
                        \#29  & $0.1000$ & $0.2111$ & $-0.0843$ \\
                        \#30  & $0.1000$ & $0.1500$ & $-0.0850$ \\
                        \#31  & $0.1000$ & $0.0500$ & $-0.0850$ \\
                        \#32  & $0.1000$ & $0.0500$ & $0.0000$ \\
                        \#33  & $0.1000$ & $0.2000$ & $0.0000$ \\
                        \#34  & $0.1000$ & $0.2500$ & $0.0000$ \\
                        \hline
                        \hline                  
                    \end{tabular}
    \end{minipage}%
    \begin{minipage}[t]{0.4\textwidth}%
                \centering
                \captionof{table}{Dynamic pressure sensors}
                \label{tab-dynamic-pressure-sensors}
                \begin{tabular}{c r r r}

                        \hline
                        \hline
                        Sensor & $x$ [m] & $y$ [m] & $z$ [m]\\
                        \hline
                        S111  & $0.1000$ & $0.1167$ & $0.0850$ \\
                        S112  & $0.1000$ & $0.2517$ & $0.0674$ \\
                        S113  & $0.1000$ & $0.2517$ & $-0.0674$ \\
                        S114  & $0.1000$ & $0.1167$ & $-0.0850$ \\ 
                        S116  & $-0.0400$ & $0.0000$ & $0.0000$ \\
                        S117  & $-0.0800$ & $0.0000$ & $0.0000$ \\
                        S118  & $-0.1200$ & $0.0000$ & $0.0000$ \\
                        S119  & $0.2000$ & $0.0000$ & $0.0000$ \\ 
                        S120  & $0.1995$ & $0.0000$ & $-0.1105$ \\
                        S121  & $0.2989$ & $0.0000$ & $-0.1209$ \\
                        S122  & $0.3984$ & $0.0000$ & $-0.1314$ \\
                        S123  & $0.4978$ & $0.0000$ & $-0.1418$ \\ 
                        \hline
                        \hline                       
                    \end{tabular}
    \end{minipage}%

\bibliography{references}{}

\begin{thebibliography}{}

\bibitem[Ask and Davidson,
  2006]{cnf-ask-j-and-davidson-l-2006-des-aeroacoustics-generic-side-mirror}
Ask, J. and Davidson, L. (2006).
\newblock The sub-critical flow past a generic side mirror and its impact on
  sound generation and propagation.
\newblock In {\em 12th AIAA/CEAS Aeroacoustics Conference}, number 2558, pages
  1--20, Cambridge, Massachusetts.

\bibitem[Ask and Davidson,
  2009]{pap-ask-j-and-davidson-l-2009-les-aeroacoustics-generic-side-mirror}
Ask, J. and Davidson, L. (2009).
\newblock A numerical investigation of the flow past a generic side mirror and
  its impact on sound generation.
\newblock {\em Journal of Fluids Engineering}, 31:0611021--12.

\bibitem[Baglietto and Ninokata,
  2007]{pap-baglietto-e-and-ninokata-h-2007-struct}
Baglietto, E. and Ninokata, H. (2007).
\newblock Improved turbulence modeling for performance evaluation of novel fuel
  designs.
\newblock {\em Nuclear Technology}, 158(2):237--248.

\bibitem[Beigmoradi and Vahdati,
  2021]{pap-beigmoradi-s-and-vahdati-m-2019-fractional-factorial-design-hatchback-rear-vehicle-aerodynamic-optimization}
Beigmoradi, S. and Vahdati, M. (2021).
\newblock Multi-objective optimization of a hatchback rear end utilizing
  fractional factorial design algorithm.
\newblock {\em Engineering with Computers}, 1(37):139--153.

\bibitem[Belamri et~al.,
  2007]{cnf-belamri-t-and-menter-f-2007-sas-generic-side-mirror}
Belamri, T., Egorov, Y., and Menter, F. (2007).
\newblock Cfd simulation of the aeroacoustic noise generated by a generic side
  view car mirror.
\newblock In {\em 13th AIAA/CEAS Aeronautics Conference}.

\bibitem[Capizzano et~al.,
  2019]{pap-capizzano-f-et-al-2019-xles-generic-side-mirror}
Capizzano, F., Alterio, L., Russo, S., and de~Nicola, C. (2019).
\newblock A hybrid rans-les cartesian method based on skew-symmetric convective
  operator.
\newblock {\em Journal of Computational Physics}, (390):359--379.

\bibitem[Chen and Li, 2019]{pap-chen-x-and-li-m-2019-ddes-generic-side-mirror}
Chen, X. and Li, M. (2019).
\newblock Delayed detached eddy simulation of subcritical flow past generic
  side mirror.
\newblock {\em Journal of Shanghai Jiao Tong University}, 1(24):107--112.

\bibitem[Chode et~al., 2021]{pap-chode-k-et-al-2021-sbes-generic-side-mirror}
Chode, K.~K., Viswanathan, H., and Chow, K. (2021).
\newblock Noise emitted from a generic side-view mirror with different aspect
  ratios and inclinations.
\newblock {\em Physics of Fluids}, (33):084105.

\bibitem[da~Silva~Pinto and Margnat,
  2020]{pap-goncalves-w-j-and-margnat-f-2020-shape-optimization-pso-noise-bluff-bodies}
da~Silva~Pinto, W. J.~G. and Margnat, F. (2020).
\newblock Shape optimization for the noise induced by the flow over compact
  bluff bodies.
\newblock {\em Computers \& Fluids}, 198:104400.

\bibitem[Demartino and Ricciardelli,
  2017]{pap-demartino-c-and-ricciardelli-f-2017-theory-flow-cylinders}
Demartino, C. and Ricciardelli, F. (2017).
\newblock Aerodynamics of nominally circular cylinders: A review of
  experimental results for civil engineering applications.
\newblock {\em Engineering Structures}, (137):76--114.

\bibitem[Devenport and Simpson,
  1990]{pap-devenport-w-and-simpson-r-1990-bimodal-horseshoe-vortex}
Devenport, W.~J. and Simpson, R.~L. (1990).
\newblock Time-dependent and time-averaged turbulence structure near the nose
  of a wing-body junction.
\newblock {\em Journal of Fluid Mechanics}, (210):23--55.

\bibitem[Egorov et~al., 2010]{pap-egorov-y-et-al-2010-sas-generic-side-mirror}
Egorov, Y., Menter, F., Lechner, R., and Cokljat, D. (2010).
\newblock The scale-adaptive simulation method for unsteady turbulent flow
  predictions. part 2: application to complex flows.
\newblock {\em Flow, Turbulence and Combustion}, 85(1):139--165.

\bibitem[Garc\'ia et~al., 2020]{pap-garcia-j-et-al-2020-struct-freight-trains}
Garc\'ia, J., Munoz-Paniagua, J., Xu, L., and Baglietto, E. (2020).
\newblock A second-generation urans model (struct-$\epsilon$) applied to
  simplified freight trains.
\newblock {\em Journal of Wind Engineering and Industrial Aerodynamics},
  205:104327.

\bibitem[Gilhaus and Hoffmann,
  1998]{ibo-gilhaus-a-and-hoffmann-r-1998-contribution-side-mirror-drag-vehicle}
Gilhaus, A. and Hoffmann, R. (1998).
\newblock {\em Aerodynamics of Road Vehicles}, chapter Directional stability,
  pages 239--310.
\newblock SAE International.

\bibitem[Grahs and Othmer,
  2006]{cnf-grahs-t-and-othmer-c-2006-sas-and-des-generic-side-mirror-optimization}
Grahs, T. and Othmer, C. (2006).
\newblock Evaluation of aerodynamic noise generation: Parameter stud of a
  generic side mirror evaluating the aerocoustic source strength.
\newblock In {\em European Conference on Computational Fluid Dynamics ECCOMAS
  CFD 2006}.

\bibitem[Haffner et~al.,
  2020]{pap-haffner-y-et-al-2020-spod-ahmed-body-drag-reduction-transient-bimodal-near-wake}
Haffner, Y., Borée, J., Spohn, A., and Castelain, T. (2020).
\newblock Mechanics of bluff body drag reduction during transient near-wake
  reversals.
\newblock {\em Journal of Fluid Mechanics}, 894(A14):1--35.

\bibitem[H\"old et~al.,
  1999]{cnf-hold-r-et-al-1999-experiments-side-mirror-acoustics-basic-analysis}
H\"old, R., Brenneis, A., and Eberle, A. (1999).
\newblock Numerical simulation of aeroacoustic sound generated by generic
  bodies placed on a plate: Part i - prediction of aeroacoustic sources.
\newblock In {\em 5th AIAA/CEAS Aeroacoustics Conference}, number AIAA-99-1896,
  Seattle, WA.

\bibitem[Jones and Launder,
  1972]{pap-jones-w-and-launder-b-1972-standard-k-epsilon}
Jones, W.~P. and Launder, B.~E. (1972).
\newblock The prediction of laminarization with a two-equation model of
  turbulence.
\newblock {\em International Journal of Heat and Mass Transfer},
  15(2):301--314.

\bibitem[Kato et~al.,
  2007]{pap-kato-c-et-al-2007-generic-side-mirror-reynolds-effect-yaw-angle}
Kato, C., Murata, O., Kokubo, A., Ichinose, K., Kijima, T., Horinouchi, N., and
  Iida, A. (2007).
\newblock Measurements of aeroacoustic noise and pressure fluctuation generated
  by a door-mirror model placed on a flat plate.
\newblock {\em Journal of Environment and Engineering}, 71(710):278--292.

\bibitem[Kirkil and Constantinescu,
  2015]{pap-kirkil-g-and-constantinescu-g-2015-effect-reynolds-number-horseshoe-vortex-system}
Kirkil, G. and Constantinescu, G. (2015).
\newblock Effects of cylinder reynolds number on the turbulent horseshoe vortex
  system and near wake of a surface-mounted circular cylinder.
\newblock {\em Physics of Fluids}, (27):075102.

\bibitem[Launder and Sharma,
  1974]{pap-launder-b-and-sharma-b-1974-standard-k-epsilon}
Launder, B.~E. and Sharma, B.~I. (1974).
\newblock Application of the energy-dissipation model of turbulence to the
  calculation of flow near a spinning disc.
\newblock {\em Letters in Heat and Mass Transfer}, 1(2):131--137.

\bibitem[Lenci, 2016]{phd-lenci-g-2016-struct}
Lenci, G. (2016).
\newblock {\em A methodology based on local resolution of turbulent structures
  for effective modeling of unsteady flows}.
\newblock PhD thesis, Massachusetts Institute of Technology.

\bibitem[Lenci et~al., 2021]{pap-lenci-g-and-baglietto-e-2021-struct}
Lenci, G., Feng, J., and Baglietto, E. (2021).
\newblock A generally applicable hybrid unsteady reynolds-averaged
  navier-stokes closure scaled by turbulent structures.
\newblock {\em Physics of Fluids}, 33:105117.

\bibitem[Li et~al., 2021]{pap-li-x-et-al-2021-spod-hst}
Li, X.-B., Chen, G., Liang, X.-F., and Li, D.-R. (2021).
\newblock Research on spectral estimation parameters for application of
  spectral proper orthogonal decomposition in train wake flows.
\newblock {\em Physics of Fluids}, (33):125103.

\bibitem[Lokhande et~al.,
  2003]{pap-lokhande-b-et-al-2003-les-generic-side-mirror}
Lokhande, B., Sovani, S., and Xu, J. (2003).
\newblock Computational aeroacoustic analysis of a generic side mirror.
\newblock {\em SAE Transactions}, (2003-01-1698):2175--2184.

\bibitem[Lumley, 2007]{lib-lumley-j-2007-stochastic-tools-in-turbulence}
Lumley, J.~L. (2007).
\newblock {\em Stochastic Tools in Turbulence}.
\newblock Courier Corporation.

\bibitem[Manhart, 1998]{pap-manhart-m-vortex-shedding-hemisphere}
Manhart, M. (1998).
\newblock Vortex shedding from a hemisphere in a turbulent boundary layer.
\newblock {\em Theoretical and Computational Fluid Dynamics}, 12:1--28.

\bibitem[Munoz-Paniagua and Garc\'ia,
  2020]{pap-munozpaniagua-j-and-garcia-j-2020-ga-optimization-hst}
Munoz-Paniagua, J. and Garc\'ia, J. (2020).
\newblock Aerodynamic drag optimization of a high-speed train.
\newblock {\em Journal of Wind Engineering and Industrial Aerodynamics},
  204:104215.

\bibitem[Nakayama and Miyashita,
  2001]{pap-nakayama-a-and-miyashita-k-2001-urans-smooth-topography}
Nakayama, A. and Miyashita, K. (2001).
\newblock Urans simulation of flow over smooth topography.
\newblock {\em International Journal of Numerical Methods for Heat and Fluid
  Flow}, 11(8):723--743.

\bibitem[Otto et~al., 1999]{tec-otto-n-et-al-1999-electric-cars-noise-sources}
Otto, N.~C., Simpson, R., and Wiederhold, J. (1999).
\newblock Electric vehicle sound quality.
\newblock Technical Report 1999-01-1694, SAE Technical Paper.

\bibitem[Paik et~al.,
  2007]{pap-paik-j-et-al-2007-bimodal-dynamics-horseshoe-vortex}
Paik, J., Escauriaza, C., and Sotiropoulos, F. (2007).
\newblock On the bidomal dynamics of the turbulent horseshoe vortex system in a
  wing-body junction.
\newblock {\em Physics of Fluids}, (19):045107.

\bibitem[Papoutsis-Kiachagias et~al.,
  2015]{pap-papoutsiskiachagias-e-et-al-2015-optimization-side-mirror}
Papoutsis-Kiachagias, E.~M., Magoulas, N., Mueller, J., Othmer, C., and
  Giannakoglou, K.~C. (2015).
\newblock Noise reduction in car aerodynamics using a surrogate objective
  function and the continuous adjoint method with wall functions.
\newblock {\em Computers and Fluids}, (122):223--232.

\bibitem[Pereira et~al.,
  2019]{pap-pereira-fs-et-al-2019-cfd-circular-cylinder-re-140000}
Pereira, F.~S., Eca, L., Vaz, G., and Girimaji, S.~S. (2019).
\newblock On the simulation of the flow around a circular cylinder at re =
  140000.
\newblock {\em International Journal of Heat and Fluid Flow}, (76):40--56.

\bibitem[Rempfer and Fasel,
  1994]{pap-rempfer-d-and-fasel-hf-1994-evolution-coherent-structures-flat-plate-boundary-layer}
Rempfer, D. and Fasel, H.~F. (1994).
\newblock Evolution of three-dimensional coherent structures in a flat-plate
  boundary layer.
\newblock {\em Journal of Fluid Mechanics}, 260:351--375.

\bibitem[Robertson et~al.,
  2015]{pap-robertson-e-et-al-2015-cfd-aeroacustical-spectral-analysis-spherocylinder}
Robertson, E., Choudhury, V., Bhushan, S., and Walters, D.~K. (2015).
\newblock Validation of openfoam numerical methods and turbulence models for
  incompressible bluff body flows.
\newblock {\em Computers and Fluids}, 123:122--145.

\bibitem[Roshko,
  1961]{pap-roshko-a-1961-supercritical-reynolds-single-cylinder}
Roshko, A. (1961).
\newblock Experiments on the flow past a circular cylinder at very high
  reynolds number.
\newblock {\em Journal of Fluid Mechanics}, 3(10):345--356.

\bibitem[Rovira et~al., 2021]{pap-rovira-m-et-al-2021-pod-turbulent-jet}
Rovira, M., Engvall, K., and Duwig, C. (2021).
\newblock Proper orthogonal decomposition analysis of the large-scale dynamics
  of a round turbulent jet in counterflow.
\newblock {\em Physical Review Fluids}, 6:014701.

\bibitem[Rung et~al.,
  2002]{cnf-rung-th-et-al-2002-comparison-urans-des-side-mirror}
Rung, T., Eschricht, D., Yan, J., and Thiele, F. (2002).
\newblock Sound radiation of the vortex flow past a generic side mirror.
\newblock In {\em 8th AIAA/CEAS Aeroacoustics Conference \& Exhibit}, number
  AIAA 2002-2549, pages 1505--1514.

\bibitem[Sieber et~al., 2016]{pap-sieber-m-et-al-2016-spod}
Sieber, M., Paschereit, C.~O., and Oberleithner, K. (2016).
\newblock Spectral proper orthogonal decomposition.
\newblock {\em Journal of Fluid Mechanics}, 792:798--828.

\bibitem[Siegert et~al.,
  1999]{cnf-siegert-r-et-al-1999-experiments-generic-side-mirror}
Siegert, R., Schwartz, V., and Reicenberger, J. (1999).
\newblock Numerical simulation of aeroacoustic sound generated by gneric boides
  placed on a plate: Part ii - prediction of radiated sound pressure.

\bibitem[Tosh et~al., 2018]{cnf-tosh-a-et-al-2018-des-generic-side-mirror}
Tosh, A., Caraeni, M., and Caraeni, D. (2018).
\newblock A hybrid computational aeroacoustic method for low speed flows.
\newblock In {\em 2018 AIAA/CEAS Aeroacoustics Conference}, number AIAA
  2018-4096, Atlanta, Georgia.

\bibitem[Werner et~al.,
  2017]{pap-werner-m-et-al-2017-tonal-noise-side-mirror-spod}
Werner, M.~J., Würz, W., and Kraemer, E. (2017).
\newblock Experimental investigation of tonal self-noise emission of a vehicle
  side mirror.
\newblock {\em AIAA Journal}, 55(5):1673--1680.

\bibitem[Xiao et~al.,
  2012]{pap-xiao-z-et-al-2012-numerical-dissipation-massive-separation-tandem-cylinders}
Xiao, Z., Liu, J., Huang, J., and Fu, S. (2012).
\newblock Numerical dissipation effects on massive separation around tandem
  cylinders.
\newblock {\em AIAA Journal}, 50(5):1119--1136.

\bibitem[Xu, 2020]{phd-xu-l-2020-struct}
Xu, L. (2020).
\newblock {\em A Second Generation URANS Approach for Application to
  Aerodynamic Design and Optimization in the Automotive Industry}.
\newblock PhD thesis, Massachusetts Institute of Technology.

\bibitem[Yao and Davidson,
  2018]{pap-yao-hd-and-davidson-l-2018-generic-side-mirror-interior-cavity-noise}
Yao, H.-D. and Davidson, L. (2018).
\newblock Generation of interior cavity noise due to window vibration excited
  by turbulent flows past a generic side-view mirror.
\newblock {\em Physics of Fluids}, (30):036104.

\bibitem[Yu et~al.,
  2021]{pap-yu-l-et-al-2021-fast-transient-fractional-step-method-scheme-generic-side-mirror}
Yu, L., Diasinos, S., and Thornber, B. (2021).
\newblock A fast transient solver for low-mach number aerodynamics and
  aeroacoustics.
\newblock {\em Computers and Fluids}, (214):104748.

\end{thebibliography}

\end{document}